\documentclass[12pt, fullpage]{amsart}
\usepackage{amssymb}
\usepackage{amsmath}
\usepackage{amsfonts}
\usepackage{mathtools}
\usepackage{mathrsfs}
\usepackage{graphicx}
\usepackage{placeins}
\usepackage{color}
\usepackage[onehalfspacing]{setspace}
\usepackage{caption}
\usepackage{subcaption}
\usepackage{natbib}
\usepackage{enumerate}
\usepackage[utf8]{inputenc}
\usepackage{tikz-cd}
\usepackage[charter,cal=cmcal]{mathdesign}
\usepackage{epigraph}
\usepackage{bibunits}
\usepackage[hypertexnames=false,colorlinks=true,citecolor=blue,urlcolor=blue,pdfpagemode=UseNone,pdfstartview=FitH]{hyperref}

\DeclareTextFontCommand{\bi}{%
	\fontseries\bfdefault 
	\itshape
}

\makeatletter
\def\section{\@startsection{section}{1}
	\z@{0.8\linespacing\@plus\linespacing}{.7\linespacing}{\Large}}

\def\subsection{\@startsection{subsection}{2}
	\z@{.5\linespacing\@plus.7\linespacing}{.7\linespacing}{\large}}

\def\subsubsection{\@startsection{subsubsection}{3}
	\z@{.5\linespacing\@plus.7\linespacing}{-.5em}{\normalfont\bfseries}}
\makeatother

\newtheorem{theorem}{Theorem}[section]

\newtheorem{lemma}{Lemma}[section]
\newtheorem{corollary}{Corollary}[section]
\newtheorem{remark}{Remark}[section]

\theoremstyle{definition}
\newtheorem{definition}{Definition}[section]

\theoremstyle{definition}
\newtheorem{assumption}{Assumption}[section]

\theoremstyle{definition}
\newtheorem{example}{Example}[section]

\calclayout

\title{}
\begin{document}

	\vspace*{3ex minus 1ex}
	\begin{center}
		\LARGE \textsc{The Law of Large Numbers for \\Large Stable Matchings}
		\bigskip
			\bigskip
				\bigskip
	\end{center}
	
	\date{%
		\today%
	}
	
	\vspace*{3ex minus 1ex}
	\begin{center}
		Jacob Schwartz$^{a,*}$ and Kyungchul Song$^b$\\
		\textit{University of Haifa and University of British Columbia}
	\end{center}
	
	\thanks{$a$: Department of Economics, University of Haifa, Mount Carmel, Haifa, 3498838, Israel. Email address: jschwartz@econ.haifa.ac.il.\\
    $b$: Vancouver School of Economics, University of British Columbia, BC, V6T 1L4, Canada. Email address: kysong@mail.ubc.ca\\
    $*$ Correspondence to University of Haifa, 199 Aba Houshy Avenue, Department of Economics, Haifa, 3498838, Israel.}

	
	\begin{bibunit}[econometrica]
	\begin{abstract}
		In many empirical studies of a large two-sided  matching market (such as in a college admissions problem), the researcher performs statistical inference under the assumption that they observe a random sample from a large matching market. In this paper, we consider a setting in which the researcher observes either all or a nontrivial fraction of outcomes from a stable matching. We establish a concentration inequality for empirical matching probabilities assuming strong correlation among the colleges' preferences while allowing  students' preferences to be fully heterogeneous. Our concentration inequality yields laws of large numbers for the empirical matching probabilities and other statistics commonly used in empirical analyses of a large matching market. To illustrate the usefulness of our concentration inequality, we prove consistency for estimators of conditional matching probabilities and measures of positive assortative matching.
		
		\medskip
		
		\medskip
		
	\end{abstract}
	
	\maketitle

		\vspace*{5ex minus 1ex}

	\section{Introduction}
	
    Large matching markets have long been a focus of study for empirical researchers and econometricians. (See \cite{Chiappori/Salanie:16:JEL} for a survey in this literature.) For identification and statistical inference, the literature often explicitly or implicitly assumes that the researcher observes a small fraction of matching outcomes drawn by random or stratified sampling from a single large matching market. For example, the literature on two-sided matching with transferable utility performs identification analysis assuming that we observe the distribution of agents' types. Such identification analysis implicitly assumes that that the actual observations observed by the researcher are a random sample from this distribution. (See \cite{Choo/Siow:06:JPE} and \cite{Galichon/Salanie:2022}.) In developing identification and estimation in a large one-to-one matching market with nontransferable utility, \cite{Menzel:2015:Eca} assumes that we observe a random sample of agents from a large limit matching market. \cite{DelBoca/Flinn:14:JET} similarly use an assumption of random sampling from a large market.\footnote{To the best of our knowledge, there are two exceptions to this sampling assumption, which use a finite sample inference approach. First, \cite{Logan/Hoff/Newton:2008:JASA}, \cite{Sorensen:2007:JF}, \cite{Aue/Klein/Ortega:2020:WP} and \cite{He/Sinha/Sun:23:WP} adopted a Bayesian approach to estimate the structural parameters in a two-sided matching market. Second, \cite{Kim/Schwartz/Song/Whang:19:Econometrics} focused on two-sided matching markets with homogeneous preferences on the colleges' side, and adopted the Monte Carlo inference approach of \cite{Dufour:06:JOE} to develop finite sample inference.}
    
    However, in many settings, it is not uncommon for a researcher to observe a nontrivial fraction of the matching outcomes of agents. In fact, many empirical studies exploit administrative data containing information on a large fraction of the participants in a matching market. For example, in their study of the impact of peer characteristics on student outcomes, \cite{Abdulkadiroglu/Angrist/Pathak:2014:ECTA} linked 98.1\% of 7th grade and 99.6\% of 9th grade students participating in a Boston high school matching process in the years 1999-2008 and 2001-2007, respectively, to data on their characteristics and outcomes (see Table C.III in Supplemental note).
	\cite{Kirkeboen/Leuven/Mogstad:16:QJE} were also able to link all the students'  post-secondary applications in Norway from 1998-2004 to socioeconomic characteristics using data from the Norwegian population registry. \cite{Hastings/Neilson/Zimmerman:2013} studied the post-secondary education market in Chile using similarly rich administrative data.  The empirical analysis in \cite{Fack/Grenet/He:AER:2019} examines the matching of middle-school students to  academic-track public high schools in the Parisian southern district using administrative data containing sufficient information to replicate the 2013-2014 academic year matching (see pp. 27-28 of the online appendix associated with the paper).\footnote{There are many other examples of the use of administrative data to study large matching market. See, for example, \cite{Boyd/Lankford/Loeb/Wyckoff:13:JLE}; \citet{Abdulkadiroglu/Agarwal/Pathak:2017:AER}; \cite{Agarwal/Somaini:2018:ECTA}; \cite{Luflade:2018:WP}; \cite{Casamiglia/Fu/Guell:2020:JPE}.} 
 
    In this paper, we consider a large, many-to-one, two-sided matching market with nontransferable utility, where matching outcomes are generated from a stable matching and the researcher observes all or a nontrivial fraction of the matching outcomes for students.  Throughout the paper, we follow \cite{Roth/Sotomayor:90:TwoSidedMatching} and refer to the two-sided matching model as a \textit{college admissions model}, calling one side \textit{students} and the other side \textit{colleges}. The analogy of a colleges admissions model eases the exposition of our paper. However, it is not our goal to develop an empirical model of college admissions or school choice which accommodates institutional details of these market environments in practice. 
    
    In this paper, we view the observations as realized from a \textit{large finite population market} which consists of a finite set of students and a finite set of colleges. In the finite population approach, the population quantities are not defined in terms of a limit continuum market but as part of a finite population matching market from which the data are generated.
    
    Here we assume that the matching mechanism behind the data generation is not known to the researcher except that it yields a stable matching under the preferences of agents. Furthermore, we allow the mechanism to receive reports from students and colleges, not necessarily their preferences directly. In a special case, these reports can coincide with their preferences as in the case of truth-telling strategies. However, in some situations, the assumption of truth-telling may be overly restrictive.\footnote{A notable departure from the assumption of truth-telling is to assume that a school is ranked higher under a student's true preferences whenever the school is ranked higher under the student's stated preferences, a behavior that has been theoretically supported as an undominated strategy in some environments (\cite{Haeringer/Klijn:2009:JET}; \cite{Fack/Grenet/He:AER:2019}). Assumptions of relaxed truth-telling are particularly relevant in situations in which students are either limited in the number of schools they can rank or in which students may choose to omit some schools from their stated preferences (see e.g., the above references) and have been used in combination with the stability assumption in  empirical studies of school choice and centralized two-sided matching (e.g., \cite{Fack/Grenet/He:AER:2019}, \cite{Aue/Klein/Ortega:2020:WP},  \cite{Combe/Tercieux/Terrier:2022:ReStud}).} Our framework allows for a wide range of report maps to cover various strategic settings as we explain below.
    
    Our focus in this paper is on the statistical properties of the empirical matching probability and related statistics. More specifically, let $N=\{1,...,n\}$ and $M = \{1,...,m\}$ be the sets of students and colleges respectively. We consider statistics of the following form:
    \begin{align}\label{generalized statistic}
    	\hat \theta(\boldsymbol{\tau}) = \frac{1}{n} \sum_{i \in N} \sum_{j \in M} \tau_j\left( X_i, Z \right) 1\left\{i\text{ and }j\text{ are matched} \right\}, \quad \boldsymbol{\tau} = (\tau_1,...,\tau_m),
    \end{align}
    where $X_{i}$, $Z_{j}$ are the observed characteristics of student $i$ and college $j$, $\tau_j$ are maps chosen by the researcher, and $Z=(Z_{1},...,Z_m)$. For example, the empirical matching probability for college $j$ measures the fraction of students who have their observed characteristic taking a value in a set, say, $A$, and are matched to college $j$. This probability is expressed as $\hat \theta(\boldsymbol{\tau})$ by choosing $ \boldsymbol{\tau}$ with $\tau_{j'}(X_{i},Z)=1\{X_{i}\in A, j' = j\}$, $j'=1,...,m$. The statistic $\hat \theta(\boldsymbol{\tau})$ can be used to capture other empirical features of the matching. One can, for example, explore matching along observed type categories of students and colleges by choosing $\boldsymbol{\tau}$ with $\tau_{j}(X_{i},Z)=1\{X_i \in A, Z_{j} \in A' \}$, where $A'$ is a set of values for observed college characteristics. Such statistics are useful for investigating positive assortativity of the matching between students and colleges along their observed characteristics.
    
    The main result of this paper establishes a concentration-of-measure phenomenon for  $\hat \theta(\boldsymbol{\tau})$. That is, we establish a finite sample bound $B(t)$ such that\footnote{We focus on the concentration of $\hat \theta(\boldsymbol{\tau})$ around its conditional expectation given colleges' characteristics, $Z, \xi$, to accommodate a model of the matching market in which the number of the colleges is fixed. In such a case, there is no aggregation over colleges' characteristics in large samples, and the concentration of measure arises around the conditional expectation of $\hat \theta_j(\boldsymbol{\tau})$ given $Z, \xi$.}
    \begin{align}
    	\label{concentration inequality}
    	\mathbf{P}\left\{\left| 	\hat \theta(\boldsymbol{\tau}) -  \mathbf{E}[\hat \theta(\boldsymbol{\tau}) \mid Z, \xi] \right| \ge t \mid Z, \xi \right\} \le B(t),
    \end{align}
    for all $t > 0$, where $\xi$ is the collection of unobserved college characteristics. For example, in the case with $\tau_{j'}(X_{i},Z)=1\{X_{i}\in A, j' = j\}$, $j'=1,...,m$, the bound $B(t)$ measures how much the distribution of the empirical matching probability for college $j$ concentrates around its conditional expectation given $Z$ and $\xi$. The bound can also be used to establish the rate of convergence for a law of large numbers. The conditional expectation $ \mathbf{E}[\hat \theta(\boldsymbol{\tau}) \mid Z, \xi]$ here is a finite population quantity that depends on $n$ and $m$, which is in contrast with the matching probability in a limit matching market as $n \rightarrow \infty$. This finite population quantity can be viewed as one obtained by averaging the statistic $\hat \theta(\boldsymbol{\tau})$  across many draws from the distribution of the $n$ students' types in a finite matching market.
    
    As we demonstrate in this paper, the result (\ref{concentration inequality}) yields a number of empirically relevant applications. For instance, in empirical models of large matching, it is common for the identification analysis to assume that the researcher knows the conditional matching probabilities; that is, the probability that a student matches with a specific college when the student has observed characteristic $x$ (e.g., \cite{Diamond/Agarwal:2017:QE} and \cite{He/Sinha/Sun:23:WP}).  As one application of our concentration inequality, we provide conditions under which a kernel-based estimator of a conditional matching probability is consistent. In another other application, we prove consistent estimation of sorting measures (based on Spearman's rho) and distributions of students' characteristics at a college.
    
    There are three main assumptions we rely on to derive the concentration inequality in (\ref{concentration inequality}). First, as mentioned before, we assume that the matching mechanism generates a stable matching under the (true) preferences of agents. The assumption of stable matching is widely used in the empirical matching literature (\cite{Chiappori/Salanie:16:JEL}) and has been justified in both centralized and decentralized two-sided matching environments with non-transferable utility.  \cite{Fack/Grenet/He:AER:2019} discuss the merit of stability in empirical settings in detail. See also the discussions in \cite{Menzel:2015:Eca} and \cite{He/Sinha/Sun:23:WP}, and \cite{Agarwal/Somaini:AnnRev:2020}. 
    
    Second, we assume that the student-specific (both observed and unobserved) types are drawn i.i.d. conditional on the college-specific characteristics. It is not unusual in the literature to assume that the individual utilities are drawn i.i.d.\ from a certain distribution (e.g. \cite{Menzel:2015:Eca}.) 
    
    Third, we assume that the colleges' preferences over students are strongly correlated while allowing students' preferences over  colleges to be fully heterogeneous. While the assumption that college preferences are strongly correlated limits the scope of our paper, such correlation is not necessarily unreasonable in practice. This is especially true in environments in which the role of college preferences is essentially fulfilled by priority indices. For example, in some school choice markets, school priorities  are determined entirely based on common measures of the students' academic performance that do not differ across colleges.\footnote{Many further examples of school choice markets in which the assumption of school-side alignment of priorities is well justified are listed in Table 1 on page 1488 of \citet{Fack/Grenet/He:AER:2019}. (In particular, see \citet{Teo/Sethuraman/Tan:2001:MS}, \citet{Pathak/Sonmez:2013:AER}, \citet{Ajayi:2021:F-JHR}, \citet{PopEleches/Urquiola;2013:AER},  \citet{Artemov/Che/He:2021:JPE}, \citet{Akyol/Krishna:2017:EER}). In Scotland, medical school graduates are assigned to training programs via a centralized matching procedure in which the priority of candidates at the programs is based on a common score (\citet{Irving:2011:MiP}).} In some college admissions markets, the priorities for different choices of colleges and majors  are determined on the basis of a composite index that may differ somewhat across the different options (e.g., \citet{Hastings/Neilson/Zimmerman:2013} and \citet{Kirkeboen/Leuven/Mogstad:16:QJE}). In the market for exam schools in Boston studied in \citet{Abdulkadiroglu/Angrist/Pathak:2014:ECTA},  priorities for students are determined on the basis of a weighted average of the student’s grade-point averages and score on an entrance exam. In the centralized matching of teachers to public secondary schools in France, teachers are  ranked by the central administration using a common, points-based system (\citet{Terrier:2011:MiP}). 
   
    The main method we rely on to derive the concentration inequality is to use a conditional version of McDiarmid's inequality (\cite{McDiarmid:1989:SurveysinCombinatorics}). This inequality is useful in our context, as it enables us to derive a bound for the concentration-of-measure for statistics that involve independent random variables in a complex, nonlinear form. The concentration bound is essentially determined by the bounded difference property that shows how sensitively the statistic responds when one of the input random variables is changed.  We derive the bounded difference condition for a stable matching by drawing heavily on machinery developed in economic theory. \cite{Roth/Vandevate:1990:Ecta} proposed a random process through which an arbitrary matching converges to a stable matching with probability one. \cite{Blum/Roth/Rothblum:1997:JET} and \cite{Blum/Rothblum/2002} developed a re-stabilization operator that takes a matching and produces a stable matching after a finite number of iterations. Similarly, we rely on a re-stabilization operator that transforms a matching into a stable matching, which we use to obtain a bound for the number of the students affected by one student's change of preference. Our bounding the number of such students is related to the rejection chain method that \cite{Kojima/Pathak:09:AER} used in their study of strategic proofness of a large student-optimal stable matching (SOSM) mechanism.
    
    Convergence of empirical matching probabilities has drawn attention in the literature. For example, \cite{Azevedo/Leshno:2016:JPE} and \cite{Che/Kim/Kojima:2019:ECA} showed that a sequence of empirical matching probabilities converges to their counterpart in a continuum economy which is populated by a continuum of students. They assume that colleges' preferences are not over the identities of the students, but over a topological space that the types of students or their distributions take values from. Then a matching is defined between the students' types and the colleges. This setting is used to construct both a sequence of finite economies and a continuum economy where a stable matching is well-defined. Their convergence result bridges between stable matchings from a continuum economy and those from a finite economy. Unlike these papers, we do not consider a continuum limit market in our paper. In our setting, a matching is defined between the identities of the students and those of the colleges, because this is the way the raw data record matching outcomes in practice.
    
    A result more closely related to ours is found in \cite{Menzel:2015:Eca} who proved the convergence of empirical matching probabilities as part of his development of an econometric model of a large one-to-one matching market. (See Corollary 3.1 in \cite{Menzel:2015:Eca}. See also \cite{Peski:2017:JET} for a related contribution in a large stable roommate problem.) There are a few major differences between his result and ours. First, our paper focuses on a many-to-one matching market, where the number of students is greater than that of colleges. Second, as mentioned before, \cite{Menzel:2015:Eca} uses the assumption that the sample involved in the empirical matching probability is drawn by a random sampling or a stratified sampling scheme from a large limit matching market. In contrast, we allow the empirical matching probability to be constructed from the whole or a nontrivial portion of the matching outcomes from a large yet finite matching market.
            
    The work of \cite{Diamond/Agarwal:2017:QE} is also related to our paper. They studied identification and asymptotic inference in many-to-one matching markets - in particular, discovering the value of many-to-one matchings as opposed to one-to-one matchings in identifying the payoff parameters. The main difference between their setting and ours is that they focused on the case of homogeneous preferences on both sides of the market, where the number of students is fixed in proportion to the number of colleges. Furthermore, their asymptotic inference assumes a sampling process where the observed variables are drawn independently from the conditional distribution given utilities. In our setting, the number of the students is allowed to be much larger than the number of colleges, and the preferences of the students and colleges are permitted to be heterogeneous.
    
    The remainder of the paper is organized as follows. In the next section,  we introduce a two-sided, many-to-one matching market drawing on the analogy of a college admissions model, and introduce assumptions on random preferences that define the scope of our paper. In Section 3, we provide a general concentration inequality on functionals of a large matching. We present examples and various concentration-of-measure results. In Section 4, we conclude. In the appendix, we provide the proof of the main results. In the supplemental note, we introduce the basic settings and results from economic theory of matching markets and prove the bounded difference result that is crucial for establishing our concentration inequality.
    
\section{A College Admissions Model}

\subsection{Two-Sided Matching}\label{subsection:Two-Sided Matching}

Throughout the paper, we follow \cite{Roth/Sotomayor:90:TwoSidedMatching} and refer to a generic many-to-one matching model as a \textit{college admissions model}, calling one side \textit{students} and the other side \textit{colleges}. The analogy of a colleges admissions model eases the exposition of our model. However, the reader does not need to assume that our model is intended to reflect all the details specific to this particular matching environment in practice.

A college admissions model consists of the set $N=\{1,...,n\}$ of students and the set $M=\{1,...,m\}$ of colleges. In many situations, colleges are capacity-constrained. For each college $j \in M$, let $q_{j}$ be a positive integer that represents
the quota of college $j$. To accommodate the possibility of unmatched students
and colleges with unfilled positions, we denote $N'=N\cup\{0\}$ and $M'=M\cup\{0\}$,
so that an unmatched student or an unfilled position at a college is viewed as being matched
to $0$. A (many-to-one) \bi{matching}
under the capacity constraint $q=(q_{j})_{j\in N}$ is defined
as a function $\mu:N\rightarrow M'$ such that
$|\mu^{-1}(j)|\le q_{j}$ for each $j\in M$. That is, the matching, $\mu$, is such that the number of students assigned to each college does not exceed the capacity of the college. Throughout the paper, we consider only those matchings that satisfy the capacity constraint. When $\mu^{-1}(j) = \varnothing$ for a college $j$, the college is not matched with any student under $\mu$. Similarly, when $\mu(i) = 0$ for a student $i$, the student is not matched with any college under $\mu$.

In our college admissions model, we let $v_i: M' \rightarrow M'$ denote a bijection that represents student $i$'s strict preference ordering over $M'$. Similarly, we let $w_j:N' \rightarrow N'$ denote a bijection that represents college $j$'s  strict preference ordering over $N'$. For any $j_1,j_2 \in M'$, we write $j_1 \succ_i j_2$ if and only if $v_i(j_1) < v_i(j_2)$, and for any $i_1,i_2 \in N'$, we write $i_1 \succ_j i_2$ if and only if $w_j(i_1) < w_j(i_2)$. If $0 \succ_j i$, this means that college $j$ prefers to be unmatched by any student than to be matched with student $i$, and in this case we say that student $i$ is \bi{unacceptable} to college $j$. Throughout the paper, we will also maintain the assumption that college preferences over sets of students are responsive. Let $\mathbb{V}$ be the set of bijections on $M'$, each of which represents a student's preference over the colleges. Similarly, let $\mathbb{W}$ be the set of bijections on $N'$, representing colleges' preferences over the students. Then the collection of preference profiles, $\boldsymbol{u} = (\boldsymbol{v},\boldsymbol{w})$, is given by 
\begin{align*}
   \mathbb{U} = \mathbb{V}^n\times \mathbb{W}^m,
\end{align*}
where $\boldsymbol{v} = (v_1,...,v_n) \in\mathbb{V}^n$ and $\boldsymbol{w} = (w_1,...,w_m) \in\mathbb{W}^m$. We call any triple $(N,M,\boldsymbol{u})$ a \textbf{\textit{matching market}}.\footnote{Our notation here leaves the quotas of colleges implicit, referring to them only when required.} 

In modeling the predicted outcomes of matching in economics, it is standard in the literature to focus on (pairwise) stable matchings (\cite{Roth/Sotomayor:90:TwoSidedMatching}). We say that a matching $\mu: N \rightarrow M'$ is \bi{stable} under the preference profile $\boldsymbol{u}$, if\medskip

(1) (Individual Rationality) there is no $i\in N$ such that $0\succ_{i}\mu(i)$ and no $j \in M$ such that $0\succ_{j}i'$ for some $i'\in\mu^{-1}(j)$, and 

(2) (Incentive Compatibility) there is no $(i,j)\in N\times M$ such that  $j\succ_{i}\mu(i)$ and either (i) $|\mu^{-1}(j)|<q_{j}$ and $i \succ_{j}0$ or (ii) $|\mu^{-1}(j)|=q_{j}$ and $i\succ_{j}i'$ for some $i'\in\mu^{-1}(j)$. \medskip

Hence, a matching is stable if there is no  student who is currently matched with a college that prefers to be unmatched, no college that prefers to unmatch some student currently matched to it, and no student-college pair not currently matched who can improve over their current match by matching with one another. (Such a pair is called a \bi{blocking pair} to the matching.)

\subsection{Random Preferences with One-Sided Limited Heterogeneity}

In empirical modeling of matching markets, it is common to associate preferences with the agents' observed or unobserved types. In this paper, we allow the type of a student to comprise a vector of observables along with a vector of match qualities capturing the student's unobserved taste for each of the colleges. We model the type of a college in a similar way. Formally, we specify the types as follows: for $i \in N$ and $j \in M$,
\begin{align}
	\label{spec char}
	C_j = (Z_j, \xi_j, \eta_{1j},...,\eta_{nj}), \text{ and } S_i = (X_i, \varepsilon_{i1},...,\varepsilon_{im}),
\end{align}
where $Z_j$ and $X_i$ are observed characteristics specific to college $j$ and student $i$ respectively, and $\eta_{ij}$ refers to college $j$'s match quality with student $i$, $\varepsilon_{ij}$ the student $i$'s match quality with college $j$, and $\xi_j$ the vector of unobserved characteristics of college $j$.

As we make explicit in Assumptions \ref{assump: index utility1} and \ref{assump: index utility2} below, the preferences of colleges and students are determined entirely by the types, $C_j$'s and $S_i$'s. In empirical modeling, we may take $Z_j$ to be a vector of observable characteristics of college $j$, and $X_i$ a vector of observed characteristics of student $i$. For $i\in N $, we also define 
\begin{align*}
	\tilde S_i = (S_i, \eta_{i1},...,\eta_{im}) = (X_i, \varepsilon_{i1},...,\varepsilon_{im},\eta_{i1},...,\eta_{im}),
\end{align*}	
so that $\tilde S_i$ contains both the type of student $i$ as well as the match quality of each college with student $i$. We call $\tilde S_i$ the student $i$'s \bi{quality}. We collect the college-specific quantities $Z$ and $\xi$ and define
\begin{align}
	\label{tilde Z}
	\tilde Z = (Z,\xi),
\end{align}
where $Z = (Z_1,...,Z_m)$ and $\xi = (\xi_1,...,\xi_m)$. Regarding the dependence structure of random quantities, we require only that $\tilde S_i$'s be conditionally independent across $i$'s given $\tilde Z$, which we formalize as follows.
\begin{assumption}
	\label{assump: index utility cond ind} 
	$\tilde{S}_{i}$ are conditionally i.i.d. across $i$'s given $\tilde Z$.
\end{assumption}
We allow the elements of each $\tilde S_i$ to be arbitrarily correlated.  For example, it is reasonable for $\eta_{ij}$ to be correlated with $S_i$, since a college's match quality for a student may depend on the student's type in general. Since our law of large numbers is conditioned on college-specific characteristics $\tilde Z$, we also allow for arbitrary correlation between $\tilde S_i$'s and $\tilde Z$. This permits, for example, correlation between the student's match quality at a college $\varepsilon_{ij}$ with the observed characteristics of the college $Z_j$. However, we require that any two random quantities associated with different students (e.g., two random quantities $X_i$ and $X_j$, or $\varepsilon_{ik}$ and $\eta_{j\ell}$, both pairs associated with two different students $i,j$, $i \ne j$) be conditionally independent given $\tilde Z$.

We relate agents' types to their preference orderings as follows. For the students' preferences over colleges, we allow for full heterogeneity. However, for the colleges' preferences over students, we impose limited heterogeneity in the sense we explain below. We define
\begin{align*}
	S = (S_1,...,S_n), \text{ and } \tilde{S}=(\tilde{S}_1,...,\tilde{S}_n).
\end{align*}
\begin{assumption}
	\label{assump: index utility1}
	For each student $i \in N$, his preference ordering over the colleges, $v_i : M' \rightarrow M'$, is given by 
	\begin{align}
		\label{vi}
		v_i = f_{S_i,\tilde Z},
	\end{align}
    where $f_{S_i,\tilde Z}: M' \rightarrow M'$ is a stochastic map that is measurable with respect to $\sigma(S_i,\tilde Z)$, i.e., the $\sigma$-field generated by $(S_i,\tilde Z)$, such that for any $j_1,j_2 \in M'$,
    \begin{align}
    	\label{eq4}
    	\mathbf{P}\left\{f_{S_i,\tilde Z}(j_1) = f_{S_i,\tilde Z}(j_2) \mid \tilde Z\right\} = 0, \text{ whenever } j_1 \ne j_2.
    \end{align}
\end{assumption}
The assumption on the students' preferences is mild; it assumes that each student's preference is generated by the student's own type and the component of the college types that is unrelated to the students. We do not put any further restriction on the preferences of students. 

The condition (\ref{eq4}) requires that the realized preferences are strict, so that we exclude the case with ties in ranking. While there are research papers which allow for indifferences in preferences in matching markets (e.g., \cite{Erdil/Ergin:2008:AER}, \cite{Erdil/Ergin:2017:JET}; \citet{Abdulkadiroglu/Agarwal/Pathak:2017:AER}), to the best of our knowledge, the assumption of strict preferences remains the most common.\footnote{Here we rely on results that need not hold in the absence of strict preferences, such as the Rural Hospitals Theorem and those results guaranteeing the existence of a unique stable student-optimal matching (see e.g., \cite{Roth/Sotomayor:90:TwoSidedMatching}).}

We introduce a model of college preferences that reflect limited heterogeneity. We model each college's preference to be generated by priority indices over students where each priority index has two components: a vertical component that depends only on students' types and is common across the colleges, and the horizontal component which is different across the colleges.

\begin{assumption}   
	\label{assump: index utility2} 
	(i) Each college $j \in M$ has priority index for a match with student $i \in N$ as:
	\begin{align}
		\label{omega}
		\omega_{ij} = \lambda(S_i) + \sigma_n \eta_{ij},
	\end{align}
    so that $\omega_{ij} > \omega_{i'j}$ if and only if $i \succ_j i'$, where $\eta_{ij}$'s are continuous random variables, and $\sigma_n$ is a positive sequence.
    
    (ii) Each college $j \in M$ has a threshold $c_{j}$ such that $c_j$ is a function of $\tilde Z$ and for each student $i \in N$, $c_{j} > \omega_{ij}$ if and only if $0 \succ_j i$.
\end{assumption}

The sequence $\sigma_n$ in Assumption \ref{assump: index utility2}(i) determines the degree of heterogeneity in the colleges' preferences over the students. Note that if $\sigma_{n}$ were equal to zero, then the college preferences are fully homogeneous, determined by the rank of students according to $\lambda(S_i)$. The degree of heterogeneity of colleges' preferences is captured by the rate at which $\sigma_n \rightarrow 0$. The continuity of $\eta_{ij}$'s is introduced to ensure that the preferences of the colleges are strict with probability one. The priority index generation in (\ref{omega}) allows for a wide range of functional forms for $\lambda$. We do not necessarily require that the colleges observe $\varepsilon_{i1},...,\varepsilon_{im}$. When the colleges do not observe $\varepsilon_{i1},...,\varepsilon_{im}$, this is tantamount to imposing a restriction that the map $\lambda$ does not vary with $\varepsilon_{i1},...,\varepsilon_{im}$. Assumption \ref{assump: index utility2}(ii) says that each college has a threshold as a function of $\tilde Z$ such that when the priority index of a student is below the threshold, the student is unacceptable to the college.

From here on, we say that the preference profile $\boldsymbol{u} = (\boldsymbol{v},\boldsymbol{w})$ is \bi{generated from} $(\tilde S,\tilde Z)$, if the profile is determined from $\boldsymbol{u} = (\boldsymbol{v},\boldsymbol{w})$ according to (\ref{vi}) in Assumption \ref{assump: index utility1} and (\ref{omega}) in Assumption \ref{assump: index utility2}. In this case, the randomness of $(\tilde S,\tilde Z)$ alone is responsible for the randomness of the preference profile $\boldsymbol{u} = (\boldsymbol{v},\boldsymbol{w})$.

The following assumption collects the technical conditions that we require for $\lambda(S_i)$ and $\eta_{ij}$. 

\begin{assumption}   
	\label{assump: index utility3}
	(i) For all $i \in N$ and $j \in M$, and all $t \ge 0$,
	\begin{align*}
		\mathbf{P}\left\{ |\eta_{ij}| > t \mid S, \tilde Z\right\} \le 2 \exp\left( - \frac{t^2}{2} \right).
	\end{align*}
	
	(ii) There exists a bounded interval $B \subset \mathbf{R}$ such that for all $i \in N$,
	\begin{align*}
		\mathbf{P}\left\{ \lambda(S_i) \in B \mid \tilde Z \right\} = 1.
	\end{align*}
    Furthermore, there exists a constant $\overline C>0$ such that for all $i \in N$ and all $t \ge 0$,
    \begin{align*}
    	\sup_{c \in \mathbf{R}} \mathbf{P}\left\{ c - t \le \lambda(S_i) \le c + t \mid \tilde Z \right\} \le \overline C t.
    \end{align*}
\end{assumption}

Assumption \ref{assump: index utility3}(i) is a mild, normalization condition for $\eta_{ij}$, because the scale of the horizontal preference component in the college preferences is captured by the sequence $\sigma_n$. Assumption \ref{assump: index utility3}(ii) requires that $\lambda(S_i)$ be bounded with probability one, and satisfies an anti-concentration condition. The anti-concentration condition is satisfied if $\lambda(S_i)$ is continuous and has a bounded density function.

\subsection{Generation of Matching Outcomes}

The matching outcomes are generated as follows. First, each individual student $i$ reports her rank order list of colleges to a central decision maker, according to a \bi{report map} $\alpha_{s,i}: \mathcal{\tilde S} \times \mathcal{\tilde Z} \rightarrow \mathcal{R}$, where $\mathcal{R}$ denotes a set of rank-order lists of colleges, $\mathcal{\tilde S}$ denotes the support of $\tilde S$, and $\mathcal{\tilde Z}$ denotes the support of $\tilde Z$. Hence each student reports her rank order list to the central decision maker. Our framework allows for a setting where the set $\mathcal{R}$ admits only rank-order lists that meet a certain length limit. (See \cite{Kojima/Pathak:09:AER} who assume such restrictions in studying manipulated reports in matching markets.) Similarly, each college $j$ reports its priority indices of students and a threshold $c_j$ to a central decision maker, according to a report map $\alpha_{c,j}:\mathcal{\tilde S} \times \mathcal{\tilde Z} \rightarrow \Omega^n \times \Xi$, where $\Omega$ denotes the support of the priority index $\omega_{ij}$, and $\Xi$ the support of thresholds $c_{j}$. Thus, each college reports their priority indices for the students and a threshold to the central decision maker. We define the map:
\begin{align*}
	\alpha \equiv (\alpha_{s,1},...,\alpha_{s,n},\alpha_{c,1},...,\alpha_{c,m}): \mathcal{\tilde S} \times \mathcal{\tilde Z} \rightarrow \mathcal{T} \equiv \mathcal{R}^n \times \left(\Omega^{n} \times \Xi \right)^m.
\end{align*}
 The report profile $\boldsymbol{t} = (r_1,...,r_n,\tilde t_1,...,\tilde t_m) \in \mathcal{T}$, $r_i \in \mathcal{R}$ and $\tilde t_j = (\omega_j,c_j) \in \Omega^{n} \times \Xi$, $\omega_j = (\omega_{1j},...,\omega_{nj})$, denotes the combination of the rank order lists $r_i$ submitted by the students, the priority index profile $\omega_j = (\omega_{1j},...,\omega_{nj}) \in \Omega^n$ of colleges over the students, and thresholds $c_j$. Thus, the notation, $\alpha(\tilde S, \tilde Z) =\boldsymbol{t}$, with $\boldsymbol{t} = (r_1,...,r_n,\tilde t_1,...,\tilde t_m) \in \mathcal{T}$, expresses that each student $i$ with type $S_i$ reports her rank-order list $r_i$, and each college $j$ with type $C_j$ reports its priority index vector $\omega_j$ and the threshold $c_j$. It is important to note that we do not require that the reports truthfully reveal the preferences of the students or colleges.

Second, the central decision maker assigns each student to a college or keeps her unmatched, according to a matching mechanism based on the received reported preferences. Thus, we treat a matching as dependent on the reports. We formalize this by treating a matching as indexed by the reports. More specifically, we define a \bi{matching mechanism} to be a matching $\mu(\cdot;\boldsymbol{t}): N \rightarrow M'$, indexed by the report vector $\boldsymbol{t} \in \mathcal{T}$. One can view a matching mechanism as a collection of matchings where each matching is determined once the report $\boldsymbol{t}$ is realized.

Throughout the paper, we assume that the pair of the matching mechanism and the report map, $(\mu,\alpha)$, generates a stable matching. We formalize this into the following assumption.

\begin{assumption}
	\label{assump: stable matching}
	For each $(\tilde s,\tilde z)$ in the support of $(\tilde S,\tilde Z)$, the matching $\mu(\cdot; \alpha(\tilde s, \tilde z))$ is stable under the preference profile generated from $(\tilde s, \tilde z)$.
\end{assumption}

Our framework allows for various information structures. This flexibility is realized through our accommodation of a wide range of report maps, $\alpha$. For each agent, the report map is a map from the agent's information set to the set of possible reports. In the case of a private information setting for students as in \cite{Fack/Grenet/He:AER:2019}, we can place a restriction that each student $i$'s report map $\alpha_{s,i}$ varies only with $S_i$, $Z$, and the priority indices $\omega_{ij}$ by the colleges, $j=1,...,m$. In the case of a setting where all students observe all the other students' preferences and the colleges' priority indices, we may allow each individual student' report map $\alpha_{s,i}$ to vary with the entire profile of students' qualities $\tilde S$ and $\tilde Z$.

Suppose that a matching mechanism $\mu$ and the report map $\alpha$ are given. We assume that the matching $Y_i \in M'$ for each student $i$ is generated as follows:
\begin{align}
	\label{Yi}
	Y_i = \mu(i; \alpha(\tilde S, \tilde Z)).
\end{align}
The researcher observes the matching $Y_i$ for each student $i$, where $Y_i$ is a discrete random variable taking values from $M'$.\footnote{Note that although each matching outcome, $Y_{i}$, depends on the other students and colleges in the market, we choose to omit the dependence of $Y_{i}$ on $n$ and $m$ from our notation for the sake of readability. We will frequently adopt the same practice when notating other quantities, particularly those defined in terms of the matching outcomes.} However, we do not require that the researcher observe the mechanism $\mu$ or the report map $\alpha$.

We introduce an assumption that $(\mu, \alpha)$ is label-free.
\begin{assumption}
	\label{assump: symmetry}
	For any permutation $\pi$ of $\{1,...,n\}$, we have
	\begin{align}
		\label{matching}
		\mu(i; \alpha(\tilde S_\pi, \tilde Z)) = \mu(\pi(i); \alpha(\tilde S, \tilde Z)), \text{ for all } i \in N,
	\end{align}
    with probability one, where $\tilde S_\pi = (\tilde S_{\pi(1)},...,\tilde S_{\pi(n)})$.
\end{assumption}
This assumption is fairly reasonable, especially when for each $i$, $\alpha_{s,i}(\tilde S,\tilde Z) = a(\tilde S_i,\tilde Z)$ for a map $a$ that is the same across students $i$. The assumption says that the matching outcome for student $i$ depends on her numerical label $i$ only through the value of $\tilde S_i$ associated with $i$. If student 1 with quality $\tilde s_1$ is matched with college 1 and student 2 with quality $\tilde s_2$ is matched with college 2, then the condition also means that student 1 with quality $\tilde s_2$ would be matched with college 2 and student 2 with quality $\tilde s_1$ would be matched with college 1. For example, suppose that $n = 4$, and $\pi(1) = 2, \pi(2) = 3, \pi(3) = 4, \pi(4) = 1$. Hence $\mu(i;\alpha(\tilde S_\pi, \tilde Z))$ denotes the match of student $i$, when each student $k$'s quality is $\tilde S_{\pi(k)}$. Then, (writing $\mu(i;\tilde S_\pi) = \mu(i;\alpha(\tilde S_\pi, \tilde Z))$ briefly)
\begin{align*}
	\begin{bmatrix}
		\mu(1;\tilde S_{\pi(1)},\tilde S_{\pi(2)},\tilde S_{\pi(3)},\tilde S_{\pi(4)}) \\
		\mu(2;\tilde S_{\pi(1)},\tilde S_{\pi(2)},\tilde S_{\pi(3)},\tilde S_{\pi(4)}) \\
		\mu(3;\tilde S_{\pi(1)},\tilde S_{\pi(2)},\tilde S_{\pi(3)},\tilde S_{\pi(4)}) \\
		\mu(4;\tilde S_{\pi(1)},\tilde S_{\pi(2)},\tilde S_{\pi(3)},\tilde S_{\pi(4)})
	\end{bmatrix}
	&= \begin{bmatrix}
		\mu(1;\tilde S_{2},\tilde S_{3},\tilde S_{4},\tilde S_{1}) \\
		\mu(2;\tilde S_{2},\tilde S_{3},\tilde S_{4},\tilde S_{1}) \\
		\mu(3;\tilde S_{2},\tilde S_{3},\tilde S_{4},\tilde S_{1}) \\
		\mu(4;\tilde S_{2},\tilde S_{3},\tilde S_{4},\tilde S_{1})
	\end{bmatrix}\\
	&= \begin{bmatrix}
		\mu(2;\tilde S_{1},\tilde S_{2},\tilde S_{3},\tilde S_{4}) \\
		\mu(3;\tilde S_{1},\tilde S_{2},\tilde S_{3},\tilde S_{4}) \\
		\mu(4;\tilde S_{1},\tilde S_{2},\tilde S_{3},\tilde S_{4}) \\
		\mu(1;\tilde S_{1},\tilde S_{2},\tilde S_{3},\tilde S_{4})
	\end{bmatrix}
	= \begin{bmatrix}
		\mu(\pi(1);\tilde S_{1},\tilde S_{2},\tilde S_{3},\tilde S_{4}) \\
		\mu(\pi(2);\tilde S_{1},\tilde S_{2},\tilde S_{3},\tilde S_{4}) \\
		\mu(\pi(3);\tilde S_{1},\tilde S_{2},\tilde S_{3},\tilde S_{4}) \\
		\mu(\pi(4);\tilde S_{1},\tilde S_{2},\tilde S_{3},\tilde S_{4})
	\end{bmatrix}
.
\end{align*}
To see the second equality, note that for example, the student 1 with quality $\tilde S_2$ is relabeled as student 2 with quality $\tilde S_2$. 

An immediate consequence of this assumption is that the observed matches are conditionally exchangeable in $i \in N$ given $\tilde Z$. Let $\Pi$ be the set of all permutations of $\{1,...,n\}$.

 \begin{lemma}
	\label{lemm: exch}
	Suppose that Assumptions \ref{assump: index utility cond ind} and \ref{assump: symmetry} hold, and for each $i \in N$, let $W_i = (\tilde S_{i},Y_i)$. Then, the conditional distribution of $(W_{\pi(1)},...,W_{\pi(n)})$ given $\tilde Z$ is the same across $\pi \in \Pi$.
\end{lemma}

The conditional exchangeability of matches is extremely useful in our context, especially when we consider partial observation of the matching outcomes. Essentially, the conditional exchangeability of matching outcomes allows our results to accommodate a wide range of sampling processes for students in addition to random sampling. We do not require the researcher to know the precise sampling process involved in generating the data. Due to Lemma \ref{lemm: exch}, we can obtain a concentration inequality for a population object which does not depend on the particular sampling process for students used to generate the data. We will give more details later.

\section{The Law of Large Numbers for a Large Stable Matching}

\subsection{The Main Result}

 In this section, we present the concentration inequality that is the main result of our paper. As the random preference profile is generated from $(\tilde S,\tilde Z)$, the randomness of $Y_i$ in (\ref{Yi}) arises solely from that of $(\tilde S,\tilde Z)$. The main challenge in deriving the law of large numbers for the sum of $Y_i$'s over $i \in N$ is that each $Y_i$ is a complex function of common random vector $(\tilde S,\tilde Z)$. Our main result establishes a finite-sample concentration-of-measure for a general statistic that involves $Y_i$'s. From this, we can derive point-wise or uniform law of large numbers for various statistics as we show below.

In order to accommodate an empirical setting with partially observed matches, we follow the approach of \cite{Canen/Schwartz/Song:2020:QE} and consider a generic sampling process which results in a subset $N_Z \subset N$ of students and the subset $M_Z$ of colleges in the sample. For the sets, $N_Z$ and $M_Z$, we make the following assumption.

\begin{assumption}[Sampling Process]
    \label{assump: sampling}
    (i) $N_Z$ is $\sigma(\tilde Z, \zeta)$-measurable and $M_Z$ is $\sigma(\tilde Z)$-measurable, where $\zeta$ is a random vector that is conditionally independent of $\tilde S$ given $\tilde Z$.

    (ii) Students' sampling indicators, $1\{i \in N_Z\}$, are conditionally i.i.d. given $(\tilde Z, \tilde S)$.
\end{assumption}

The randomness in the sampling process is captured by the random vector $\zeta$. We require that it is conditionally independent of $\tilde S$ given $\tilde Z$. In other words, the sampling does not depend on the students' individual characteristics. The subset $N_Z$ of students in the sample can potentially depend on the aggregate characteristics of colleges. As for colleges, we assume that we observe a non-empty subset $M_Z \subset M'$, where $M_Z$ is $\sigma(\tilde Z)$-measurable. For example, we may take $M_Z = \{j \in M: Z_j \in B\}$ for some set $B$, i.e., the set of colleges whose observed characteristics take values in the set $B$. It is important to note that the conditional independence of the sampling indicators $1\{i \in N_Z\}$ does not imply that the matching outcomes for two students, $Y_{i_1}$ and $Y_{i_2}$, are conditionally independent given $\tilde Z$ regardless of whether we condition on that the two students are in the sample or not.
 
The sampling process we consider is general. It covers the scheme of random sampling. Assumption \ref{assump: sampling} also accommodates the case where we observe the entire set of students and colleges. In this case, Assumption \ref{assump: sampling}(ii) is trivially satisfied, because $1\{i \in N_Z\}$'s are simply constants of ones. Let $n_Z = |N_Z|$, i.e., the number of the students in the sample.  Analogously, we define $m_Z = |M_Z|$. The sampling process accommodates the case with $n_Z /n \rightarrow 0$ as $n \rightarrow \infty$ and the case with $n_Z = \alpha n$ for all $n \ge 1$ for some $\alpha \in (0,1]$. 

Let us define a generic form of a statistic:
\begin{align*}
	\hat \theta(\boldsymbol{\tau}) \equiv \sum_{j \in M_Z} \frac{1}{n_Z} \sum_{i \in N_Z} \tau_j\left( X_i, Z \right) 1\{Y_i = j\},
\end{align*}
and the target parameter:
\begin{align}
	\label{def}
	\theta(\boldsymbol{\tau};\tilde Z) \equiv \sum_{j \in M_Z} \mathbf{E}\left[ \tau_j\left( X_i, Z \right) 1\{Y_i = j\} \mid \tilde Z \right], \quad \boldsymbol{\tau} = (\tau_0,\tau_1,...,\tau_m),
\end{align}
for some real functions $\tau_j$ such that $\theta(\boldsymbol{\tau};\tilde Z)$ exists. As we will see, various statistics involving empirical matching probabilities take the form $\hat{\theta}(\boldsymbol{\tau})$ for an appropriate choice of $\tau_{j}$. We emphasize that the population quantity $\theta(\boldsymbol{\tau};\tilde Z)$ depends on the finite matching market, and as such, it depends on $n$ and $m$, although the dependence is left implicit in our notation for simplicity.\footnote{The dependence of the population object on the number of agents in our case can be viewed as arising from a finite population approach. A notable example is the average treatment effect defined as the average of the expected treatment effects, where the expectation is taken with respect to a distribution representing a ``superpopulation''. (See, e.g., \cite{Imbens/Wooldridge:09:JEL} and \cite{Imai/King/Stuart:08:JRSS}.) Such a finite population approach is also used in settings with a large network. (See, e.g., \cite{Aronow/Samii:17:AAS}, \cite{Leung:20:ReStat}, and \cite{He/Song:23:WP}.)}

Due to the conditional exchangeability result in Lemma \ref{lemm: exch} and Assumption \ref{assump: sampling}, $\hat \theta(\boldsymbol{\tau})$ is an unbiased estimator of the target parameter $\theta(\boldsymbol{\tau};\tilde Z)$. 

\begin{lemma}
    \label{lemm: unbiased}
    Suppose that Assumptions \ref{assump: index utility cond ind}, \ref{assump: symmetry} and \ref{assump: sampling} hold. Then,
    \begin{align*}
       \mathbf{E}\left[ \hat \theta(\boldsymbol{\tau}) \mid \tilde Z \right] = \theta(\boldsymbol{\tau};\tilde Z). 
    \end{align*}
\end{lemma}

As for the functions, $\tau_{j}$, we make the following assumption.
\begin{assumption}
	\label{assump: varphi}
	There exists a map $\overline \tau: \mathcal{Z}^m \rightarrow [1, \infty)$ such that for all $x \in \mathcal{X}$ and $z \in \mathcal{Z}^m$, 
	\begin{align*}
		\max_{j \in M'}\left| \tau_{j}(x,z) \right| \le \overline \tau(z),
	\end{align*}
	where $\mathcal{X}$ and $\mathcal{Z}$ denote the sets from which $X_i$ and $Z_j$ take values respectively. 
\end{assumption}

In many applications, it is not hard to find the bound $\overline \tau(z)$. We illustrate this in the simple example involving individual matching probabilities below. See Section \ref{subsec:Examples} for more examples.

\begin{example}[Individual Matching Probabilities]\label{example_individual_matching_probabilities}  We obtain an estimator of the probability of matching with a given college (or being unmatched) as follows. 	For a fixed $j\in M$, we take $M_Z = \{j\}$ and consider
	\begin{align*}
		\hat \theta(\boldsymbol{\tau}) = \frac{1}{n_Z}\sum_{i\in N_Z}1\left\{Y_i = j\right\}, \boldsymbol{\tau} = (\tau_0,\tau_1,...,\tau_m),
	\end{align*}
	where $\tau_{j}\left( X_i, Z \right) = 1$. Thus, Assumption  \ref{assump: varphi} is satisfied with the map $\overline \tau(z)=1$. 
	
	Suppose that one is interested in estimating the fraction of students with characteristic $X_i$ being in some set $A$ among those that are matched to college $j$. Then we may consider
		\begin{align*}
		\hat \theta(\boldsymbol{\tau}) = \frac{1}{n_Z}\sum_{i\in N_Z}1\{X_{i}\in A\}1\left\{Y_i = j\right\},
	\end{align*}
	where $\tau_{j}\left( X_i, Z \right) = 1\{X_{i}\in A\}$.  Here, Assumption  \ref{assump: varphi} is satisfied by taking $\overline \tau(z)=1$. $\square$
\end{example}

The following theorem is our main result. (Recall that $Z$ denotes the vector of observed college-specific  characteristics whereas $\tilde Z$ denotes the vector of both observed and unobserved college-specific  characteristics.)

\begin{theorem}
	\label{thm: concentration inequality}
	Suppose that Assumptions \ref{assump: index utility cond ind}-\ref{assump: varphi} hold. Then, there exist constants $C>0$ and $n_0 \ge 2$ which depend only on the constant $\overline C$ and the set $B$ in Assumption \ref{assump: index utility3} such that whenever $n\ge n_0$, for all $t > 0$, we have
	\begin{align}
		\label{ineq12}
		&\mathbf{P}\left\{ \left| n_Z \hat \theta(\boldsymbol{\tau}) - \mathbf{E}\left[ n_Z \hat \theta(\boldsymbol{\tau}) \mid \tilde Z \right] \right|  \ge  n \pi_Z t \mid \tilde Z \right\}\\ \notag
		&\quad \le 6 \exp\left( - \frac{\displaystyle C n (t^2 \wedge t^{3/2})}{\displaystyle \overline \tau^2(Z) \left( a_n + b_n t \right)}\right) + 2 \exp\left( - \frac{\displaystyle n \pi_Z t^2}{\displaystyle 8 \overline \tau^2(Z)(1 + t)} \right),
	\end{align}
    where $\pi_Z = P\{i \in N_Z \mid \tilde Z\}$ and, with $\tilde \sigma_n = \sigma_n \vee n^{-5/6}$,
    \begin{align*}
    	a_n = n^2 \tilde \sigma_n^2 \ln(n m) + 1 \text{ and } b_n = n \sqrt{n} \tilde \sigma_n.
    \end{align*}
\end{theorem}\medskip		

The bound in Theorem \ref{thm: concentration inequality} is a finite sample bound. As we will see below in Corollary \ref{cor: rate of convergence}, this theorem can be used to derive the rate of convergence of $\hat \theta(\boldsymbol{\tau})-\theta(\boldsymbol{\tau};\tilde Z)$. Due to the finite sample nature of the bound in the theorem, we can see what conditions we require for a sequence of matching markets when we derive the rate of convergence.

\begin{corollary}
	\label{cor: rate of convergence}
	Suppose that Assumptions \ref{assump: index utility cond ind}-\ref{assump: varphi} hold for all $n, m \ge 1$ and $m = g(n)$ for some function $g$. Furthermore, assume that the following conditions hold.\medskip
	
	(i) The constant $\overline C>0$ and the set $B$ in Assumption \ref{assump: index utility3} are independent of $n$.
	
	(ii) There exists an absolute constant $C_1>0$ such that $\overline \tau(Z)  < C_1$ for all $n \ge 1$.
	
	(iii) $\tilde \sigma_n \sqrt{n \ln(n m)} + (n \pi_Z)^{-1/2} \rightarrow_P 0$, as $n \rightarrow \infty$.\medskip

    Then, as $n \rightarrow \infty$, we have
	\begin{align}
		\label{conv3}
		\left|\hat \theta(\boldsymbol{\tau}) - \theta(\boldsymbol{\tau};\tilde Z) \right| = O_P\left( \tilde \sigma_n \sqrt{n \ln(n m)} + (n \pi_Z)^{-1/2} \right).
	\end{align}
\end{corollary}

In regards to Condition (i), note that the constant $\overline C>0$ and the set $B$ are concerned only with the distribution of the colleges' and students' types, not with the matching mechanisms, report maps, or the capacities. Condition (ii) is easily checked as it depends on the choice of the map $\tau_j$ in the statistic. 

The convergence in (\ref{conv3}) shows the rate at which the randomness of $\hat \theta(\boldsymbol{\tau})$, arising from the variations of students' idiosyncratic characteristics, disappears under the conditions stated in the corollary. In other words, when the number of the students in the sample is large enough, and $\sigma_n$ and the number of the colleges in the sample satisfy that
\begin{align}
    \label{cond sigma_n}
    \tilde \sigma_n \sqrt{n \ln(n m)} + (n \pi_Z)^{-1/2} \rightarrow_P 0,
\end{align}
we have 
\begin{align*}
	\hat \theta(\boldsymbol{\tau}) - \theta(\boldsymbol{\tau};\tilde Z) = o_P(1),
\end{align*}
as $n \rightarrow \infty$. That is, particular realizations of the students' idiosyncratic characteristics become more and more irrelevant in determining the value of $\hat \theta(\boldsymbol{\tau})$, and the estimated quantity $\hat \theta(\boldsymbol{\tau})$ gets closer to the population quantity $\theta(\boldsymbol{\tau};\tilde Z)$.

\subsection{Discussion}
\label{sec:DiscussionPH}

To derive the concentration inequality in Theorem \ref{thm: concentration inequality}, we first write 
\begin{align}
    \label{decomp}
    n_Z \hat \theta(\boldsymbol{\tau}) - \mathbf{E}\left[ n_Z \hat \theta(\boldsymbol{\tau})\mid \tilde Z \right] &= n_Z \hat \theta(\boldsymbol{\tau}) - \mathbf{E}\left[ n_Z \hat \theta(\boldsymbol{\tau})\mid \tilde S,\tilde Z \right]\\ \notag
    &\quad  + \mathbf{E}\left[ n_Z \hat \theta(\boldsymbol{\tau})\mid \tilde S,\tilde Z \right] - \mathbf{E}\left[ n_Z \hat \theta(\boldsymbol{\tau})\mid \tilde Z \right].
\end{align}
The first difference on the right hand side is easy to handle because once we condition on $\tilde Z,\tilde S$, $n_Z \hat \theta(\boldsymbol{\tau})$ is a weighted sum of independent Bernoulli random variables, $1\{i \in N_Z\}$. We deal with this difference by using a concentration inequality in \cite{Chung/Lu:AC:2002}. The main challenge is to deal with the second difference, because the match outcomes $Y_i$ are cross-sectionally correlated in a complex form after conditioning on $\tilde Z$. To this end, we use a conditional version of McDiarmid's inequality which is stated as follows:

\begin{lemma}[McDiarmid's Inequality]
	\label{lemm: McDiarmid's}
	Suppose that $W_i$'s are random elements which take values in a space $\mathcal{W}$ and are conditionally independent given a $\sigma$-field, $\mathcal{F}$, and $U$ is a random element that is $\mathcal{F}$-measurable, and takes values from a space $\mathcal{U}$. Let $g:\mathcal{W}^n \times \mathcal{U} \rightarrow \mathbf{R}$ be a measurable map such that for each $i = 1,...,n$, there exists a constant $c_i>0$ satisfying that for all $w_1,....,w_m, w_i' \in \mathcal{W}$, and for all $u \in \mathcal{U}$,
	\begin{align}
		\label{bounded difference}
		|g(w_1,...,w_{i-1},w_i,w_{i+1},...,w_n,u) - g(w_1,...,w_{i-1},w_i',w_{i+1},...,w_n,u)| \le c_i.
	\end{align}

	Then, for all $t>0$,
	\begin{align}
		\mathbf{P}\left\{\left|g(W_1,...,W_n,U) - \mathbf{E}[g(W_1,...,W_n,U) \mid \mathcal{F}]\right| \ge t \mid \mathcal{F}\right\} \le 2 \exp\left( - \frac{2 t^2}{\sum_{i=1}^n c_i^2}\right).
	\end{align}
\end{lemma}
A key step in applying McDiarmid's inequality involves establishing that the matching mechanism of interest obeys the \bi{bounded difference condition} in (\ref{bounded difference}). The bounded difference condition shows how the function $g$ varies as the value of a single argument is arbitrarily perturbed. The inequality shows that the distribution of $g(W_1,...,W_n,U)$ concentrates more around its conditional mean if the bounds $c_i$ in (\ref{bounded difference}) are small.

In our context, we consider the map
\begin{align}
	\label{emp mat prob}
	g(W_1,...,W_n, U) \equiv \sum_{j \in M_Z} \sum_{i \in N} \tau_{j}\left( X_i, Z \right) 1\{ \mu(i;\alpha(\tilde S,\tilde Z)) = j\} \pi_Z = \mathbf{E}\left[n_Z \hat \theta(\boldsymbol{\tau}) \mid \tilde S,\tilde Z \right],
\end{align}
where $W_i = \tilde S_i$ and $U = \tilde Z$, and the last equality follows by Assumption \ref{assump: sampling}. We take $\mathcal{F}$ to be the $\sigma$-field of $\tilde{Z}$. As a crucial first step, we establish a bounded difference result for a stable matching under the assumption that the colleges' preferences over students exhibit a form of limited heterogeneity in terms of maximum rank difference. The maximum rank difference measures the degree of preference heterogeneity among the colleges as we explain below. Given a profile $\boldsymbol{w} = (w_1,...,w_m)$ of college preferences over $N'$, we define the \bi{maximum rank difference} in $\boldsymbol{w}$ by
\begin{align}
	\label{h(w)}
	h(\boldsymbol{w}) & = \max_{j \in M} \max_{(i_{1},i_{2})\in N(\boldsymbol{w})}|w_j(i_{1})-w_j(i_{2})|,
\end{align}
where $N(\boldsymbol{w})$ denotes the set of pairs of students that at least two colleges disagree on their rankings, i.e., 
\begin{align*}
	N(\boldsymbol{w}) = \{(i_1,i_2) \in N' \times N': i_1 \prec_{j_1} i_2, \text{ and } i_1 \succ_{j_2} i_2, \text{ for some } j_1,j_2 \in M \}. 
\end{align*}
(If $N(\boldsymbol{w}) = \varnothing$, we set $h(\boldsymbol{w})= 0$.) Hence $h(\boldsymbol{w})$ represents the maximum rank difference between any two students such that there is a disagreement over the ranking of the two students among colleges. If the preferences in $\boldsymbol{w}$ are homogeneous, then $N(\boldsymbol{w}) = \varnothing$, and $h(\boldsymbol{w}) = 0$. On the other hand, if for example, there exist $w_j$ and $w_k$ in $\boldsymbol{w}$ such that $w_j(i_1) > w_j(i_2)$ if and only if $w_k(i_1)<w_k(i_2)$, then we can have $h(\boldsymbol{w}) = n$ in this case. This can occur, for example, if one college ranks as worst a student who another college ranks as best. Thus, the maximum rank difference $h(\boldsymbol{w})$ measures how ``close'' the preference orderings in $\boldsymbol{w}$ are to each other. If $h(\boldsymbol{w}) = k$, this means that if any two students have rank difference by more than $k$ in any college's preference, all other colleges share the same ordering between the two students. 

With the preference profile $\boldsymbol{u}$ generated from $(\tilde s,\tilde z)$, we rewrite $h(\boldsymbol{w})$ as $h(\tilde s,\tilde z)$ from here on.

\begin{lemma}\label{lemm:bounded_difference_condition} Suppose that $(\tilde s, \tilde z),(\tilde s', \tilde z) \in \mathcal{\tilde S} \times \mathcal{\tilde Z}$ are chosen to satisfy the following three conditions.
	
	(a) $\tilde s$ and $\tilde s'$ differ by $\tilde s_i$ for at most one student $i$, for some $i \in N$.
	
	(b) $\mu(\cdot; \alpha(\tilde s,\tilde z))$ and $\mu(\cdot; \alpha(\tilde s',\tilde z))$ are stable matchings under preference profiles $\boldsymbol{u}$ and $\boldsymbol{u}'$ which are respectively generated from $(\tilde s, \tilde z)$ and $(\tilde s', \tilde z)$.
	
	(c) $h(\tilde s, \tilde z) = h(\tilde s', \tilde z)  = k$, for some $k \in \{0,1,...,n\}$.
	
Then, for any $j  \in M'$,
	\begin{align}
		\label{eq:5pt1ii2-1-1-1}
		\left|\left\{i\in N:1\{\mu(i;\alpha(\tilde s,\tilde z)) = j\} \ne  1\{\mu(i;\alpha(\tilde s',\tilde z)) = j\}\right\} \right|\leq 32(k\vee1) +1.
	\end{align}
\end{lemma}

The remarkable aspect of Lemma \ref{lemm:bounded_difference_condition} is that the bounded difference condition does not impose any restrictions on the report map $\alpha$ other than requiring the resulting matching to be stable. The proof of Lemma \ref{lemm:bounded_difference_condition} is provided in the online supplemental note. It is well known in the literature that under the assumption of strict preferences, the set of two-sided stable matchings has a lattice structure with upper and lower bounds corresponding to two extreme cases of student-optimal and student-worst matchings. Furthermore, it is well known that when the preferences of colleges over students are identical, these two bounds coincide and there is a unique stable matching. Lemma \ref{lemm:bounded_difference_condition} shows a finite sample result, explicitly relating the degree of preference heterogeneity among colleges (as expressed by the maximum rank difference $h(\tilde s,\tilde z)$) to the ``closeness'' of the two bounds. When the maximum rank difference among colleges is bounded by $k$, the student-optimal and student-worst matchings are different at most for $(m+1)(32(k \vee 1) +1)$ students.\footnote{We thank an anonymous referee for pointing out this implication to us.} The role of college preference heterogeneity in our approach are explored further in Example \ref{ex:1} below. 

By applying McDiarmid's inequality together with the bounded difference condition established in Lemma \ref{lemm:bounded_difference_condition}, we obtain a concentration inequality for the second difference on the right hand side of (\ref{decomp}) as follows.

\begin{lemma}\label{lemm:conc_ineq} Suppose that Assumptions \ref{assump: index utility cond ind}-\ref{assump: varphi} hold. Then, for all $t > 0$,
	\begin{align*}
		&\mathbf{P}\left\{ \left| \mathbf{E}\left[ n_Z \hat \theta(\boldsymbol{\tau}) \mid \tilde Z, \tilde S \right] - \mathbf{E}\left[ n_Z \hat \theta(\boldsymbol{\tau}) \mid \tilde Z \right] \right| \ge t \mid \tilde Z\right\}  \le 2 \exp\left(-\frac{ t^2}{2 \displaystyle n \pi_Z^2 \overline \tau^2(Z) \left(32\overline h(\tilde Z)+ 2\right)^2}\right),
	\end{align*}
	where $\pi_Z = P\left\{ i \in N_Z \mid \tilde Z \right\}$ and $\overline h(\tilde z) = \sup_{\tilde s \in \mathcal{\tilde S}} h(\tilde s, \tilde z) \vee 1.$
\end{lemma}

\noindent \textbf{Proof: } For each $i \in N$, let $(\tilde s',\tilde z)$ and $(\tilde s,\tilde z)$ (with the same $\tilde z$) be chosen from the support of $(\tilde S,\tilde Z)$ such that $\tilde s'$ is the same as $\tilde s$ except that its $i$-th component $\tilde s_i'$ is different from the $i$-th component $\tilde s_i$ of $\tilde s$. First, by Assumption \ref{assump: stable matching}, $\mu(\cdot; \alpha(\tilde s',\tilde z))$ and $\mu(\cdot; \alpha(\tilde s,\tilde z))$ are stable matchings. Define a measurable map $\psi_n$ as
\begin{align*}
	\psi_n(\tilde s, \tilde z) = \mathbf{E}\left[n_Z \hat \theta(\boldsymbol{\tau}) \mid (\tilde S, \tilde Z) = (\tilde s, \tilde z) \right].
\end{align*}
Let $y_{i'} = \mu(i'; \alpha(\tilde s,\tilde z))$ and $y_{i'}'= \mu(i'; \alpha(\tilde s',\tilde z))$ for each $i' \in N$. In light of (\ref{emp mat prob}), $\left|\psi_n(\tilde s, \tilde z) - \psi_n(\tilde s', \tilde z) \right|$ is bounded by
\begin{align}
	\label{dev0}
    &2 \overline \tau(z) \pi_Z + \overline \tau(z) \sum_{j \in M_Z} \sum_{i' \in N} \left|1\{y_{i'} = j\} - 1\{y_{i'}' = j\} \right|1\{y_{i'} = j\} \pi_Z\\ \notag
    &\quad \quad + \overline \tau(z) \sum_{j \in M_Z} \sum_{i' \in N} \left|1\{y_{i'} = j\} - 1\{y_{i'}' = j\} \right|1\{y_{i'}' = j\} \pi_Z,
\end{align}    
where $z$ is the observed component of $\tilde z$. By Lemma \ref{lemm:bounded_difference_condition},  
\begin{align*}
	\max_{j \in M'} \sum_{i' \in N} \left|1\{y_{i'} = j\} - 1\{y_{i'}' = j\} \right| \le 32 (h(\tilde s,\tilde z) \vee 1) + 1.
\end{align*}
Hence, by (\ref{dev0}), we have
\begin{align}
	\label{dev}
	\left|\psi_n(\tilde s, \tilde z) - \psi_n(\tilde s', \tilde z) \right| \le 2\pi_Z \overline \tau(z) + 2\pi_Z \overline \tau(z) \left(32\overline h(\tilde z)+1\right),
\end{align}
by Assumption \ref{assump: varphi}.  Since $\tilde S_i$'s are conditionally independent given $\tilde Z$ by Assumption \ref{assump: index utility cond ind}, we obtain the desired bound by Lemma \ref{lemm: McDiarmid's}. $\blacksquare$\medskip

To obtain the result of Theorem \ref{thm: concentration inequality}, we extend Lemma \ref{lemm:conc_ineq} to the case where for any pair of students, there may exist colleges whose rankings over the students disagree, though with small probability for many pairs.

As far as our approach of using McDiarmid’s inequality is concerned, some condition for limited heterogeneity in the college preferences appears inevitable. As we show in the example below, when the disagreement among colleges over the ranking of students is too extensive, a change in preference by a single student can alter the match of every student, which can even render the number of students on the right-hand side of (\ref{eq:5pt1ii2-1-1-1}) to be $n$. Thus, we cannot use the approach based on McDiarmid's inequality to obtain a useful concentration inequality, if we allow an arbitrary degree of preference heterogeneity for colleges.

\begin{example}\label{ex:1} 
	\begin{table}	
		\centering{}\caption{Preferences of Agents for Example \ref{ex:1} }
		\label{example_1}%
		\begin{tabular}{cccccc}
			&  &  &  &  & \tabularnewline
			& Student Preferences &  &  &  & College Preferences \tabularnewline
			$i_{1}$ & $(j_{1},j_{2},j_{3})$ &  &  & $j_{1}$ & $(i_{1},i_{4},i_{2},i_{3},i_{5})$\tabularnewline
			$i_{2}$ & $(j_{2},j_{3},j_{1})$ &  &  & $j_{2}$ & $(i_{1},i_{5},i_{2},i_{3},i_{4})$\tabularnewline
			$i_{3}$ & $(j_{2},j_{3},j_{1})$ &  &  & $j_{3}$ & $(i_{2},i_{3},i_{4},i_{5},i_{1})$\tabularnewline
			$i_{4}$ & $(j_{3},j_{1},j_{2})$ &  &  &  & \tabularnewline
			$i_{5}$ & $(j_{3},j_{2},j_{1})$ &  &  &  & \tabularnewline
			&  &  &  &  & \tabularnewline
		\end{tabular}
	\bigskip
	\end{table}
	Consider a college admissions market with five students, $N=\{i_{1},i_{2},i_{3},i_{4},i_{5}\}$,
	and three colleges, $M=\{j_{1},j_{2},j_{3}\}$ with $q_{j_1}=1$, $q_{j_2}=q_{j_3}=2$.
	Suppose that the preferences $\boldsymbol{u}=(\boldsymbol{v},\boldsymbol{w})$
	are given by Table \ref{example_1}. For example, the preference of student $i_2$ is such that the student considers college $j_2$ the best, $j_{3}$ the second best, and $j_{1}$ the worst. Consider the following stable matching that is obtained from the deferred acceptance algorithm under the given preferences:
	\begin{align*}
		(\mu(i_1),\mu(i_2),\mu(i_3),\mu(i_4),\mu(i_5)) = (j_{1},j_{2},j_{2},j_{3},j_{3}).
	\end{align*}
	Next, suppose that the preferences of agents are instead given by $\boldsymbol{u}'=(\boldsymbol{v}',\boldsymbol{w})$, where $\boldsymbol{v}'$ is defined to be identical
	to $\boldsymbol{v}$, except that we replace the preference of student
	$i_{1}$ with the ordering $(j_{2},j_{3},j_{1})$. The student-optimal matching
	under $\boldsymbol{u}'=(\boldsymbol{v}',\boldsymbol{w})$ is $\mu'=(j_{2},j_{3},j_{3},j_{1},j_{2})$. Thus, as students move from matching $\mu$ to matching $\mu'$ due to one student's preference change, all of the students end up being matched with a different college. $\blacksquare$
\end{example}

The bound in Lemma \ref{lemm:bounded_difference_condition} can be tighter when there are vacancies at the colleges. Indeed, Example \ref{ex:2} below illustrates how the effect of extensive college-preference heterogeneity can be mitigated by the presence of vacancies at the colleges. Since the bounded difference condition must account for a `worst-case scenario' in which such vacancies are absent at the colleges, this example suggests that the bound in Lemma \ref{lemm:bounded_difference_condition} can be conservative in practice when at least some colleges have vacancies. 

\begin{example}\label{ex:2}
	
	Consider again the college admissions market introduced in Example \ref{ex:1}.
	This time, however, suppose that $q_{1}=1$, $q_{2}=3$, $q_{3}=2$.
	That is, college $j_2$ has an additional position. As before, the student
	optimal matching under $\boldsymbol{u}$ is $\mu=(j_{1},j_{2},j_{2},j_{3},j_{3}).$ However, the fact that college $j_{2}$ has a vacant position at $\mu$ implies that the change of preferences from $\boldsymbol{v}$ to $\boldsymbol{v}'$  (as defined in Example \ref{ex:1}) would lead to only student $i_{1}$ changing colleges. That is, the presence of a vacancy at college $j_{2}$ prevents the `cascade' of changes that occurred in the
	previous example. By the same logic, if every college at $\mu$ had one vacant position, the change in preferences from $\boldsymbol{u}$ to \textit{any}  profile $\boldsymbol{u}'$ that differed in the preference of one student would lead to at most one student changing college. $\blacksquare$
	
\end{example}

\subsection{Examples}\label{subsec:Examples}

\subsubsection{Individual Matching Probabilities}
\label{subsubsec: matching prob} 
We revisit Example  \ref{example_individual_matching_probabilities}.  By Lemma \ref{lemm: exch}, the conditional distribution of $(X_i, Y_i)$ given $\tilde Z$ is identical across $i$'s. 
As long as (\ref{cond sigma_n}) is satisfied, by Corollary \ref{cor: rate of convergence}, we have
\begin{align}
	\label{eq212}
	\frac{1}{n_Z}\sum_{i \in N_Z}1\left\{Y_i=j\right\} = \mathbf{P}\left\{Y_i = j \mid \tilde Z\right\} + O_P\left( \tilde \sigma_n \sqrt{n \ln(n m)} + (n \pi_Z)^{-1/2} \right).
\end{align}
Thus, we obtain the rate of convergence for the individual matching probabilities. Similarly, we obtain the same rate of convergence for the second statistic:
\begin{align}
	\label{eq: consist ind matching prob}
	\frac{1}{n_Z}\sum_{i\in N_Z} 1\left\{ X_i \in A, Y_i = j \right\} =  \mathbf{P}\left\{ X_i \in A, Y_i = j \mid \tilde Z\right\} + O_P\left( \tilde \sigma_n \sqrt{n \ln(n m)} + (n \pi_Z)^{-1/2} \right).
\end{align}
\medskip

\subsubsection{Matching Probability on Characteristics}
In many situations, it is of interest to estimate the probability of matching on characteristics. Let $X_i$ and $Z_j$ be as in (\ref{spec char}), so that $X_i$ and $Z_j$ denote student $i$'s own and college $j$'s own characteristics. One might be interested in measuring the fraction of students being matched with a college $j$ with characteristic $Z_j \in A'$ for a set $A'$, when the students have characteristic $X_i$ in $A$. By taking  $\tau_{j}\left( X_i, Z \right) = 1\{X_i \in A, Z_j \in A' \}$ and $M_Z = M'$ in the definition of $\hat{\theta}(\boldsymbol{\tau})$, we obtain
\begin{align}
	\hat{\theta}(\boldsymbol{\tau})=	\frac{1}{n_Z}\sum_{i\in N_Z}\sum_{j\in M'}1\left\{ X_{i}\in A,Z_{j}\in A'\right\}1\{Y_{i}=j\}.
\end{align}
By Corollary \ref{cor: rate of convergence}, we have
\begin{align*}
	\frac{1}{n_Z}\sum_{i\in N_Z}\sum_{j\in M'}1\left\{ X_{i}\in A,Z_{j}\in A',Y_{i}=j\right\} &=\sum_{j\in M'}\mathbf{P}\left\{ X_{i}\in A,Z_{j}\in A',Y_{i}=j\mid \tilde Z\right\} \\
	&\quad + O_P\left( \tilde \sigma_n \sqrt{n \ln(n m)} + (n \pi_Z)^{-1/2} \right).
\end{align*}
\medskip

\subsubsection{Distribution of Characteristics of Students Matched to a College}
It is often of interest to estimate the distribution of characteristics of students matched with a specific college under a stable matching. Define the conditional CDF of students' characteristics conditional on that the student is matched with college $j$:
\begin{align}
	\label{varphi13}
	\hat F_{j}(x \mid \tilde Z) = \frac{ \sum_{i\in N_Z}1\left\{X_i \le x, Y_i = j\right\}}{ \sum_{i \in N_Z} 1\{Y_i = j\}} \text{ and }
	F_j(x \mid \tilde Z) = \mathbf{P}\left\{X_i \le x \mid Y_i = j, \tilde Z\right\},
\end{align}
where $X_i \in \mathbf{R}^d$, and the inequality between vectors is element-wise. Suppose that the set of colleges, $M$, is fixed, i.e., does not depend on $n$, so that the number of colleges $m$ is also fixed. We would like to show that $\hat F_j(x \mid \tilde Z)$ and $F_j( x \mid \tilde Z)$ get closer to each other as $n \rightarrow \infty$.  Let us make the following assumption.

\begin{assumption}
	\label{assump: DF}
	(i) Each random vector, $X_i$, $i \in N$, is either discrete with a finite support that is independent of $n$, or has a continuous conditional distribution function given $\tilde Z$.
	
	(ii) There exist $\epsilon>0$ and a nonempty subset $M(\epsilon) \subset M'$ such that for all $n \ge 1$, and all $j \in M(\epsilon)$, we have $\mathbf{P}\{Y_i = j \mid \tilde Z\} > \epsilon$.
\end{assumption}

Then, we can obtain the uniform convergence of $\hat F_j(\cdot \mid \tilde Z)$ to $F_j(\cdot \mid \tilde Z)$, as $n \rightarrow \infty$.

\begin{corollary}
	\label{cor: unif consistency}
	Suppose that $M$ is a fixed set not depending on $n$, and that Assumptions \ref{assump: index utility cond ind}-\ref{assump: index utility3} and \ref{assump: DF} hold. Suppose further that (\ref{cond sigma_n}) holds. 
	
	Then for all $j\in M(\epsilon)$ with the set $M(\epsilon)$ appearing in Assumption \ref{assump: DF}(ii), as $n \rightarrow \infty$,
	\begin{align*}
		\sup_{x \in \mathbf{R}^d} \left|\hat F_{j}( x \mid \tilde Z) - F_{j}( x \mid \tilde Z)\right| \rightarrow_P 0.
	\end{align*}
\end{corollary}
\medskip

\subsubsection{Consistent Estimation of a Measure of Positive Assortative Matching}
The result of Corollary \ref{cor: unif consistency} can be used to prove consistency of a measure of positive assortative matching. For example, one may want to measure the positive (stochastic) assortative matching between students and colleges along two variables $X_{i,k}$ and $Z_{{Y_i},r}$, where $X_{i,k}$ denotes the $k$-th element of $X_i$ and $Z_{{Y_i},r}$ the $r$-th element of $Z_{Y_{i}}$. One way to measure it is to use a quantity that stems from Spearman's rho defined as follows:\footnote{The Spearman's rho has been proposed or used as a measure of positive assortative matching in the literature. See, e.g., \cite{Gihleb/Lang:2016:NBER}, \cite{Hagedorn/Law/Manovskii:17:Eca}, and \cite{Lochner/Schulz:2021:WP}. The quantity $\rho$ can be viewed as the population version of Spearman's $\rho$ between $X_i$ and $Z_{{Y_i},r}$ after randomly selecting $i$ from the students matched with some college.}
\begin{align*}
	\rho = 12 \int_{\mathbf{R}} \int_{\mathbf{R}} \left(F_{X,Z}(t,s) - F_X(t) F_Z(s)\right) dF_X(t) dF_Z(s),
\end{align*}
where
\begin{align*}
	F_{X,Z}(t,s) &= \mathbf{P}\left\{X_{i,k} \le t, Z_{Y_i,r} \le s \mid \tilde Z, Y_i \ne 0 \right\},\\
	F_X(t) &= \mathbf{P}\left\{X_{i,k} \le t \mid \tilde Z, Y_i \ne 0 \right\}, \text{ and } \\
	F_Z(s) &= \mathbf{P}\left\{Z_{Y_i,r} \le s \mid \tilde Z, Y_i \ne 0 \right\}.
\end{align*}
Note that conditional on $\tilde Z$, $Z_{Y_i}$ is still random, due to the randomness of the students' preferences that affect the matching outcome $Y_i$. 

We can construct the estimator of $\rho$ as follows. First, we define $N_{1,Z} = \{i \in N_Z: Y_i \ne 0\}$, and $n_{1,Z} = |N_{1,Z}|$. Let
\begin{align*}
	\hat \rho = \frac{12}{n_{1,Z}}\sum_{i \in N_{1,Z}} \left(\hat F_{X,Z,-i}(X_{i,k},Z_{{Y_i},r}) - \hat F_{X,-i}(X_{i,k}) \hat F_{Z,-i}(Z_{{Y_i},r})\right),
\end{align*}
where
\begin{align*}
	\hat F_{X,Z,-i}(t,s) &= \frac{1}{n_{1,Z} - 1}\sum_{i' \in N_{1,Z} \setminus \{i\}}1\{X_{i',k} \le t, Z_{Y_{i'},r} \le s \}, \\
	\hat F_{X,-i}(t) &= \frac{1}{n_{1,Z} - 1}\sum_{i' \in N_{1,Z} \setminus \{i\}}1\{X_{i',k} \le t \}, \text{ and }\\
	\hat F_{Z,-i}(s) &= \frac{1}{n_{1,Z} - 1}\sum_{i' \in N_{1,Z} \setminus \{i\}}1\{ Z_{Y_{i'},r} \le s \}.
\end{align*}

Let us make the following assumption which is Assumption \ref{assump: DF}(i) for $X_{i,k}$.

\begin{assumption}
	\label{assump: PA}
	Each random variable, $X_{i,k}$, $i \in N$, is either discrete with a finite support that is independent of $n$, or has a continuous conditional distribution function given $\tilde Z$.
\end{assumption}

Then, using Corollary \ref{cor: unif consistency}, we can show that $\hat \rho$ is consistent for $\rho$.
\begin{corollary}
	\label{cor: PA}
	Suppose that $M$ is a fixed set not depending on $n$, and that Assumptions \ref{assump: index utility cond ind}-\ref{assump: index utility3} and \ref{assump: PA} hold. Suppose further that (\ref{cond sigma_n}) holds, and with probability one,
	\begin{align}
		\label{lower bound}
		\liminf_{n \rightarrow \infty}\mathbf{P}\left\{ Y_i \ne 0 \mid \tilde Z \right\} > 0.
	\end{align}

	Then, as $n \rightarrow \infty$, 
	\begin{align*}
		\hat \rho - \rho \rightarrow_P 0.
	\end{align*}
\end{corollary}

The condition (\ref{lower bound}) is a very mild condition. The failure of this condition means that from some large $n$ on, the probability that no student is matched with any college becomes one.

\subsubsection{Consistent Estimation of Conditional Matching Probabilities}
Suppose that we have a fixed set of colleges $M$ which does not depend on $n$. Let us define the conditional probability of a student $i$ with observed characteristic $x$ matched with college $j$ as follows:
\begin{align}
	p(j \mid x,\tilde Z) = \mathbf{P}\left\{ Y_i = j \mid X_i = x, \tilde Z\right\}.
\end{align}
Suppose that $X_i$ is a continuous random vector in $\mathbf{R}^d$. We consider the following local constant estimator of $p(j \mid x,\tilde Z)$:
\begin{align}
	\hat p(j \mid x,\tilde Z) \equiv \frac{\displaystyle \sum_{i \in N_Z} 1\{Y_i = j\} \mathcal{K}_h\left( X_i - x \right)}
	{\displaystyle \sum_{i \in N_Z} \mathcal{K}_h\left( X_i - x \right)} = \frac{\displaystyle \frac{1}{n \pi_Z} \sum_{i \in N_Z} 1\{Y_i = j\} \mathcal{K}_h\left( X_i - x \right)}
	{\displaystyle \frac{1}{n \pi_Z} \sum_{i \in N_Z} \mathcal{K}_h\left( X_i - x \right)},
\end{align}
where $\mathcal{K}_h(\cdot) = \mathcal{K}(\cdot/h)/h^d$ and $\mathcal{K}$ is a multivariate kernel function on $\mathbf{R}^d$, and $h$ is a bandwidth. 

When we show the consistency of $\hat p(j \mid x,\tilde Z)$, a new challenge (as compared to the standard nonparametric analysis) arises for dealing with the convergence of the following term in the numerator.
\begin{align}
	\frac{1}{n \pi_Z} \sum_{i \in N_Z} 1\{Y_i = j\} \mathcal{K}_h\left( X_i - x \right).
\end{align}
In particular, we would like to show the following:
\begin{align}
	\label{Delta n}
		\Delta_n(x) \equiv \frac{1}{n \pi_Z} \sum_{i \in N_Z} \left( 1\{Y_i = j\} \mathcal{K}_h\left( X_i - x \right) - \mathbf{E}\left[ 1\{Y_i = j\} \mathcal{K}_h\left( X_i - x \right) \mid \tilde Z\right] \right) = o_P(1),
\end{align}
as $n \rightarrow \infty$. For this, we take
\begin{align}
	\hat \theta(\boldsymbol{\tau}) = \frac{1}{n_Z} \sum_{i \in N_Z} 1\{Y_i = j\} \tau_{j}\left( X_i, Z \right), \text{ with } \tau_{j}\left( X_i, Z \right) = \mathcal{K}_h\left( X_i - x \right) h^d,
\end{align}
so that $\Delta_n(x) = n_Z (\hat \theta(\boldsymbol{\tau}) - \mathbf{E}[\hat \theta(\boldsymbol{\tau}) \mid \tilde Z])/ (n \pi_Z h^d)$. Then, it is not hard to see that Assumption \ref{assump: varphi} is satisfied with $\overline \tau(z) = \| \mathcal{K}\|_\infty \vee 1$, where $\|\mathcal{K}\|_\infty = \sup_{x \in \mathbf{R}^d} |\mathcal{K}(x)|$. Therefore, we find from Theorem \ref{thm: concentration inequality} that
\begin{align}
	\label{bd}
   \mathbf{P}\left\{ |\Delta_n(x) | > t \mid \tilde Z \right\}
  \le 6 \exp \left( - \frac{\displaystyle C n ((h^d t)^2 \wedge (h^d  t)^{3/2})}{\displaystyle \overline \tau^2(Z) (a_n + b_n h^d t)} \right) + 2 \exp\left( - \frac{\displaystyle n \pi_Z (h^d t)^2}{\displaystyle 8 \overline \tau^2(Z)(1 + h^d t)} \right).
\end{align}
Hence, if $\tilde \sigma_n h^{-d} \sqrt{n \ln n}  + (n \pi_Z)^{-1/2} h^{-d} \rightarrow_P 0$ as $n \rightarrow \infty$, then we have
\begin{align*}
	\Delta_n(x) = O_P\left(\tilde \sigma_n h^{-d} \sqrt{n \ln n} + (n \pi_Z)^{-1/2} h^{-d}\right). 
\end{align*}

After dealing with the bias part, we can obtain the consistency of the conditional matching probability estimators. Let us present the result formally below.

Let us first introduce smoothness conditions for controlling the bias part.
\begin{assumption}
	\label{assump: kernel est}
     (i) The conditional probability $\mathbf{P}\left\{Y_i = j \mid X_i = x, \tilde Z = \tilde z \right\}$ and the conditional density function of $X_i$ given $\tilde Z=\tilde z$, $f_{X\mid \tilde Z}(x \mid \tilde z)$, both as a function of $x \in \mathbf{R}^d$, are twice continuously differentiable with derivatives bounded uniformly over $j$ and over $\tilde z \in \mathcal{\tilde Z}$.
     
     (ii) The conditional density function $f_{X \mid \tilde Z} (x \mid \tilde z)$ is bounded away from zero on an open ball around $x$ in $\mathbf{R}^d$ uniformly over $\tilde z \in \mathcal{\tilde Z}$. 
	
	(iii) The kernel function $\mathcal{K}$ is symmetric around zero, vanishes outside a compact set, and takes values in $[-1,1]$. 
\end{assumption}

Assumptions \ref{assump: kernel est}(i) and (ii) are an adapted version of standard assumptions used in nonparametric kernel estimation. Using these assumptions and Corollary \ref{cor: rate of convergence}, we can show the following
\begin{align*}
	\frac{1}{n \pi_Z} \sum_{i \in N_Z} \mathcal{K}_h\left( X_i - x \right) &= f_{X \mid \tilde Z}(x \mid \tilde Z) + O_P\left( \frac{1}{\sqrt{n \pi_Z h^d}} + h^2\right), \text{ and }\\
	\frac{1}{n \pi_Z} \sum_{i \in N_Z} \mathbf{E}\left[ 1\{Y_i = j\} \mathcal{K}_h\left( X_i - x \right) \mid \tilde Z\right] &= p(j \mid x,\tilde Z) f_{X \mid \tilde Z}(x \mid \tilde Z) + O_P(h^2),
\end{align*} 
by following the standard arguments. (See Section 2.1 of \cite{Li/Racine:07:NonparamEcon}. Recall that by Assumption \ref{assump: index utility cond ind}, $X_i$'s are conditionally i.i.d. across $i$'s given $\tilde Z$.)

\begin{corollary}
	\label{cor: consistency of matching prob estimators}
	Suppose that Assumptions \ref{assump: index utility cond ind}-\ref{assump: index utility3} and \ref{assump: kernel est} hold. Suppose further that $M$ is a fixed set not depending on $n$, and that $\tilde \sigma_n h^{-d} \sqrt{n \ln n} + (n \pi_Z)^{-1/2} h^{-d} \rightarrow_P 0$, as $n \rightarrow \infty$.
	
	Then, for each $j\in M$ and $x \in \mathbf{R}^d$, as $n \rightarrow \infty$,
	\begin{align}
		\hat p(j \mid x,\tilde Z) = p(j \mid x,\tilde Z) + O_P\left(\tilde \sigma_n h^{-d} \sqrt{n \ln n} + (n \pi_Z)^{-1/2} h^{-d} + h^{-2} \right).
	\end{align}
\end{corollary}

Hence, if $\tilde \sigma_n h^{-d} \sqrt{n \ln n} + (n \pi_Z)^{-1/2} h^{-d} \rightarrow_P 0$ and $h \rightarrow 0$, as $n \rightarrow \infty$, $\hat p(j \mid x,\tilde Z)$ is consistent.

\section{Conclusion}

This paper considers a large two-sided matching market, where a matching between the two sides is stable. In such a situation, it is a non-trivial matter to establish limit theorems for statistics such as empirical matching probabilities as the number of market participants grows. Using the re-equilibration arguments from economic theory, we derive a concentration inequality for various statistics that involve matching outcomes in this environment.

In order to develop inference that does not require random sampling from a large matching, one needs to take into account the dependence structure of the observations carefully. However, this is challenging in large matching markets. It is left to future research to establish limit distribution theory for the large matching setting, where a complex dependence structure arises naturally due to the interdependence among agents in the underlying market. 

\section{Appendix: Proofs}

\noindent \textbf{Proof of Lemma \ref{lemm: exch}: } We write briefly $\mu(i;\tilde S_\pi) = \mu(i;\alpha(\tilde S_\pi, \tilde Z))$ again. Then, note that
\begin{align}
	\label{distr eq}
		[(\tilde S_{1},\mu(1; \tilde S)),...,(\tilde S_{n},\mu(n; \tilde S))] =^d [(\tilde S_{\pi(1)},\mu(1;\tilde S_\pi)),...,(\tilde S_{\pi(n)},\mu(n;\tilde S_\pi))],
\end{align}
where $=_d$ denotes the equality of conditional distributions given $\tilde Z$. The  distributional equality comes from Assumption \ref{assump: index utility cond ind}. To see the  distributional equality, we take a hyper-retangular set $A = A_1 \times ... \times A_n$, where $A_k$ is Borel, for $k=1,...,n$. Then, we can write
\begin{align*}
	&P\left\{ (\tilde S_{1},\mu(1; \tilde S_1,...,\tilde S_n),...,\tilde S_{n},\mu(n; \tilde S_1,...,\tilde S_n)) \in A \mid \tilde Z \right\}\\
	&= P\left\{ (\tilde S_{1},\mu(1; \tilde S_1,...,\tilde S_n)) \in A_1,...,(\tilde S_{n},\mu(n; \tilde S_1,...,\tilde S_n)) \in A_n \mid \tilde Z \right\}\\
	&= P\left\{ (\tilde S_1,...,\tilde S_n) \in B_1,...,(\tilde S_1,...,\tilde S_n) \in B_n \mid \tilde Z \right\},
\end{align*}
for some Borel sets $B_1,...,B_n$. We can approximate each set $B_k$ by a countable union of measurable hyper-rectangles with arbitrary accuracy, and for the last probability, it suffices to focus on the probabilities of the form:
\begin{align*}
	& P\left\{ (\tilde S_1,...,\tilde S_n) \in B_{11} \times... \times B_{1n},...,(\tilde S_1,...,\tilde S_n) \in B_{n1} \times... \times B_{nn} \mid \tilde Z \right\}\\
	&= P\left\{ (\tilde S_{\pi(1)},...,\tilde S_{\pi(n)}) \in B_{11} \times... \times B_{1n},...,(\tilde S_{\pi(1)},...,\tilde S_{\pi(n)}) \in B_{n1} \times... \times B_{nn} \mid \tilde Z \right\},
\end{align*}
for some measurable sets $B_{ij}$ by Assumption \ref{assump: index utility cond ind}. Hence, we have 
\begin{align*}
	&P\left\{ (\tilde S_1,...,\tilde S_n) \in B_1,...,(\tilde S_1,...,\tilde S_n) \in B_n \mid \tilde Z \right\}\\
	&= P\left\{ (\tilde S_{\pi(1)},...,\tilde S_{\pi(n)}) \in B_1,...,(\tilde S_{\pi(1)},...,\tilde S_{\pi(n)}) \in B_n \mid \tilde Z \right\}.
\end{align*}
 We can reverse the arguments back to obtain the probability 
\begin{align*}
	P\left\{ (\tilde S_{\pi(1)},\mu(1; \tilde S_{\pi(1)},...,\tilde S_{\pi(n)}),...,\tilde S_{\pi(n)},\mu(n; \tilde S_{\pi(1)},...,\tilde S_{\pi(n)})) \in A \mid \tilde Z \right\},
\end{align*}
establishing the distributional equality in (\ref{distr eq}). By Assumption \ref{assump: symmetry}, we obtain
\begin{align*}
	[(\tilde S_{\pi(1)},\mu(1;\tilde S_\pi)),..,(\tilde S_{\pi(n)},\mu(n;\tilde S_\pi))]
	=[(\tilde S_{\pi(1)},\mu(\pi(1);\tilde S)),...,(\tilde S_{\pi(n)},\mu(\pi(n);\tilde S))].
\end{align*}
This completes the proof. $\blacksquare$\medskip

\noindent \textbf{Proof of Lemma \ref{lemm: unbiased}: } Note that 
\begin{align*}
    \mathbf{E}\left[ \hat \theta(\boldsymbol{\tau}) \mid \tilde Z \right] &= \sum_{j \in M_Z} \mathbf{E}\left[ \frac{1}{n_Z} \sum_{i \in N_Z} \mathbf{E}[\tau_j(X_i,Z) 1\{Y_i = j\} \mid \tilde Z, \zeta] \mid \tilde Z \right]\\
    &= \sum_{j \in M_Z} \mathbf{E}\left[ \frac{1}{n_Z} \sum_{i \in N} \mathbf{E}[\tau_j(X_i,Z) 1\{Y_i = j\} \mid \tilde Z, \zeta] 1\{ i \in N_Z \} \mid \tilde Z \right],
\end{align*}
because $N_Z$ is $\sigma(\tilde Z,\zeta)$-measurable and $M_Z$ is $\sigma(\tilde Z)$-measurable. Since $\zeta$ is conditionally independent of $\tilde S$ given $\tilde Z$, we have 
\begin{align*}
    \mathbf{E}[\tau_j(X_i,Z) 1\{Y_i = j\} \mid \tilde Z, \zeta] = \mathbf{E}[\tau_j(X_i,Z) 1\{Y_i = j\} \mid \tilde Z].
\end{align*}
By Lemma \ref{lemm: exch}, the last conditional expectation does not depend on $i$. Thus, we obtain the desired result. $\blacksquare$\medskip

The following lemma is a corollary to Lemma 2.1 in \cite{Chung/Lu:AC:2002}.

\begin{lemma}\label{lemm:conc_ineq2} Suppose that Assumptions \ref{assump: index utility cond ind}-\ref{assump: varphi} hold. Then, for all $t > 0$,
	\begin{align}
		\label{ineq1}
		\mathbf{P}\left\{ \left| n_Z \hat \theta(\boldsymbol{\tau}) - \mathbf{E}\left[ n_Z \hat \theta(\boldsymbol{\tau}) \mid \tilde Z, \tilde S \right] \right| \ge t \mid \tilde Z\right\} \le 2 \exp\left(-\frac{ t^2}{\displaystyle 2 \overline \tau^2(Z) n \pi_Z + \left(2 \overline \tau(Z) t/3 \right)}\right).
	\end{align}
\end{lemma}

\noindent \textbf{Proof: } We apply Lemma 2.1 of \cite{Chung/Lu:AC:2002} by letting $p_i \equiv P\{i \in N_Z \mid \tilde Z, \tilde S\} = P\{i \in N_Z \mid \tilde Z\} = \pi_Z$, (with the second equality due to the conditional independence of $\zeta$ and $\tilde S$ given $\tilde Z$) and $a_i = \sum_{j \in M_Z} \tau_j(X_i,Z)1\{Y_i = j\}$, and letting $X_i$ in the lemma be $1\{i \in N_Z\}$ here. Note that 
\begin{align*}
	\sum_{i \in N} a_i^2 p_i &= \sum_{i \in N} \left(\sum_{j \in M_Z} \tau_j(X_i,Z)1\{Y_i = j\}\right)^2 \pi_Z = \sum_{i \in N} \sum_{j \in M_Z} \tau_j^2(X_i,Z)1\{Y_i = j\} \pi_Z  \le \overline \tau^2(Z) n \pi_Z.
\end{align*}
The second equality and last inequality are due to the fact that each student is matched with at most one college. The desired result follows by Lemma 2.1 of \cite{Chung/Lu:AC:2002}. $\blacksquare$\medskip

The lemma below is used to translate the bound in terms of $\overline h(\tilde Z)$ in Lemma \ref{lemm:conc_ineq} into that in terms of $\sigma_n$.

\begin{lemma}\label{lemma:heterogeneity_index_under_additive_preferences}
	Suppose that the preferences of colleges are generated according
	to Assumption \ref{assump: index utility2} with a sequence $\sigma_n>0$. 
	
	Then, for $\overline C>0$ and the set $B$ in Assumption \ref{assump: index utility3}(ii), we have that for all $t>0$,
	\begin{align}
		\label{eq:growth_of_h_index}
		P\left\{ \sigma_n \max_{i \in N,j \in M} |\eta_{ij}| \le t \text{  and  } \overline{h}(\tilde Z) > 12 n \overline C t + 2 n \sup_{c \in B} R_n(c, 6 t) + 1 \right\} = 0,
	\end{align}
    where $\overline{h}(\tilde z)$ is defined in Lemma \ref{lemm:conc_ineq},
    \begin{align*}
    	R_n(c,t) = \frac{1}{n}\sum_{i \in N} \left( f(S_i;c,t) - \mathbf{E} \left[f(S_i;c,t) \mid \tilde Z \right] \right),
    \end{align*}
    and $f(S_i;c,t) = 1\left\{|\lambda(S_i) - c| \le t \right\}.$
\end{lemma}

\noindent \textbf{Proof: } For each $c,t \in \mathbf{R}$, define
\begin{align*}
	N(S;c,t) = \{i \in N: |\lambda(S_i) - c| \le t \}.
\end{align*}
Let $N_2$ be the set of pairs of students in $N$ who are ranked differently
by some colleges. Suppose that $\sigma_n \max_{i \in N,j \in M} |\eta_{ij}| \le t$. Then, whenever $i_1,i_2 \in N$ are such that $\lambda(S_{i_1}) - \lambda(S_{i_2}) > 2 t$, we have $i_1 \succ_j i_2$ for all $j \in M$. Hence, if $\sigma_n \max_{i \in N,j \in M} |\eta_{ij}| \le t$,
\begin{align}
	\label{eq:subset_rho}
	N_2 & \subset \left\{(i_{1},i_{2}) \in N \times N : \left| \lambda(S_{i_2})-\lambda(S_{i_1}) \right| \leq 2 t \right\}.
\end{align}
Take any $(i_1,i_2)$ in the latter set. Then the college $j$'s rank difference between $i_1$ and $i_2$ is bounded by
\begin{align}
	\label{ineq3}
	&\left| \left\{i \in N \setminus \{i_1,i_2\}: i_1 \succ_j i \succ_j i_2\right\} \right| 1\{ i_1 \succ_j i_2\} \\ \notag
	&\qquad + \left| \left\{i \in N \setminus \{i_1,i_2\}: i_2 \succ_j i \succ_j i_1 \right\} \right| 1\{i_2 \succ_j i_1\}+1.
\end{align}
Let us focus on the first term. Note that $i_1 \succ_j i \succ_j i_2$ implies that
\begin{align*}
     0 < \lambda(S_i) - \lambda(S_{i_2}) + \sigma_n (\eta_{ij} - \eta_{i_2 j} ) \le \lambda(S_{i_1}) - \lambda(S_{i_2}) + \sigma_n (\eta_{i_1 j} - \eta_{i_2 j} ) \le 4 t,
\end{align*}
which again implies that $-2 t \le \lambda(S_{i}) - \lambda(S_{i_2})  \le 6 t.$ Hence, on the event that $\lambda(S_{i_2}) \in B$, with $B$ chosen to be the bounded set in Assumption \ref{assump: index utility3}(ii),
\begin{align}
	\label{ineq31}
	\left| \left\{i \in N \setminus \{i_1,i_2\}: i_1 \succ_j i \succ_j i_2 \right\}\right| &\le \left| \left\{i \in N \setminus \{i_1,i_2\}: \left| \lambda(S_{i}) - \lambda(S_{i_2})  \right| \le 6 t \right\}\right|\\ \notag
	&\le \sup_{c \in B} | N(S;c,6 t) |.
\end{align}
We can rewrite
\begin{align*}
	| N(S;c,6 t) | &= \sum_{i \in N} 1\{|\lambda(S_i) - c| \le 6 t\}\\
	 &= \sum_{i \in N} \mathbf{P}\left\{|\lambda(S_i) - c | \le 6 t \mid \tilde Z \right\} + n R_n(c, 6 t) \le 6 n \overline C t + n R_n(c, 6 t),
\end{align*}
where the last inequality follows by Assumption \ref{assump: index utility3}(ii). Let us define the event
\begin{align*}
	A_n(t) = \left\{\sigma_n \max_{i \in N,j \in M} |\eta_{ij}| \le t, \text{ and } \lambda(S_i) \in B \text{ for all } i \in N \right\}.
\end{align*}
Thus, we have shown that on the event $A_n(t)$, for any $(i_1,i_2) \in N_2$, the college $j$'s rank difference between $i_1$ and $i_2$ is bounded by
\begin{align*}
	12 n \overline C t + 2 n  \sup_{c \in B} R_n(c, 6 t) + 1.
\end{align*}
The last bound does not depend on the particular choice $(i_1,i_2)$ from the set $N_2$. Hence on the event  $A_n(t)$, we have
\begin{align*}
	 \overline h(\tilde Z)  \le 12 n \overline C t + 2 n \sup_{c \in B} R_n(c, 6 t) + 1.
\end{align*}
Since $\lambda(S_i) \in B$ for all $i \in N$ with probability one by Assumption \ref{assump: index utility3}(ii),
\begin{align*}
	&P \left\{ \sigma_n \max_{i \in N,j \in M} |\eta_{ij}| \le t, \text{ and } \overline{h}(\tilde Z)  >  12 n \overline C t + 2 n \sup_{c \in B} R_n(c, 6 t) + 1 \right\}\\
	&=P \left(A_n(t) \cap \left\{ \overline{h}(\tilde Z)  >  12 n \overline C t + 2 n \sup_{c \in B} R_n(c, 6 t) + 1 \right\}\right) = 0.
\end{align*}
$\blacksquare$\medskip

Let $B$ be the bounded interval in Assumption \ref{assump: index utility3}(ii), and let $\mathcal{L}(B)$ represent its Lebesgue measure. From here on, without loss of generality, we assume that $\mathcal{L}(B) \ge 1$ and $\overline C \ge 1$. (If $\mathcal{L}(B) < 1$ or $\overline C < 1$, we can replace $\mathcal{L}(B)$ or $\overline C$ by $\mathcal{L}(B) \vee 1$ or $\overline C \vee 1$ respectively below.) Furthermore, let $\tilde \lambda$ be a map such that $\tilde \lambda(S_i) = \lambda(S_i)$ with probability one and $\tilde \lambda$ takes values from the set $B$. For simplicity, from here on, we identify $\lambda$ with $\tilde \lambda$. 

For any sequence $t_n>0$, define
\begin{align*}
	\mathcal{F}(t_n) = \{f(\cdot; c,6t_n): c \in B\},
\end{align*}
where $f(\cdot; c, t)$ is as defined in Lemma \ref{lemma:heterogeneity_index_under_additive_preferences}. Note that if $t_n > \mathcal{L}(B)/6$, then $\mathcal{F}(t_n)$ is the singleton of the constant function one.

\begin{lemma}
	\label{lemm: bracketing}
	For each $0 < \epsilon \le \min\{\sqrt{6 t_n},\mathcal{L}^{1/3}(B)\}$, with $t_n \in (0, \mathcal{L}(B)/6]$, there exist brackets $[f_{L,j},f_{U,j}]$, $j=1,...,n_{[]}(\epsilon)$, that cover $\mathcal{F}(t_n)$ such that for each integer $k \ge 2$, and each $j=1,...,n_{[]}(\epsilon)$,\footnote{The brackets $[f_{L,j},f_{U,j}]$, $j=1,...,n_{[]}(\epsilon)$, refer to pairs of functions $f_{L,j}$ and $f_{U,j}$ such that $f_{L,j} \le f_{U,j}$. We say that they cover $\mathcal{F}(t_n)$ if for all $f \in \mathcal{F}(t_n)$, there exists $j \in \{1,...,n_{[]}(\epsilon)\}$ such that $f_{L,j} \le f \le f_{U,j}$.} 
	\begin{align}
		\label{bound3456}
		\mathbf{E}\left[ \left| f_{L,j}(S_i) - f_{U,j}(S_i) \right|^k \mid \tilde Z \right] \le (5 \overline C)^k \epsilon^2,
	\end{align}
and
\begin{align}
	\label{bound23}
	\ln n_{[]}(\epsilon) \le 1 + \ln \mathcal{L}(B) - 3\ln \epsilon.
\end{align}
\end{lemma}

\begin{figure}[t]
	\begin{center}
		\includegraphics[scale=0.4]{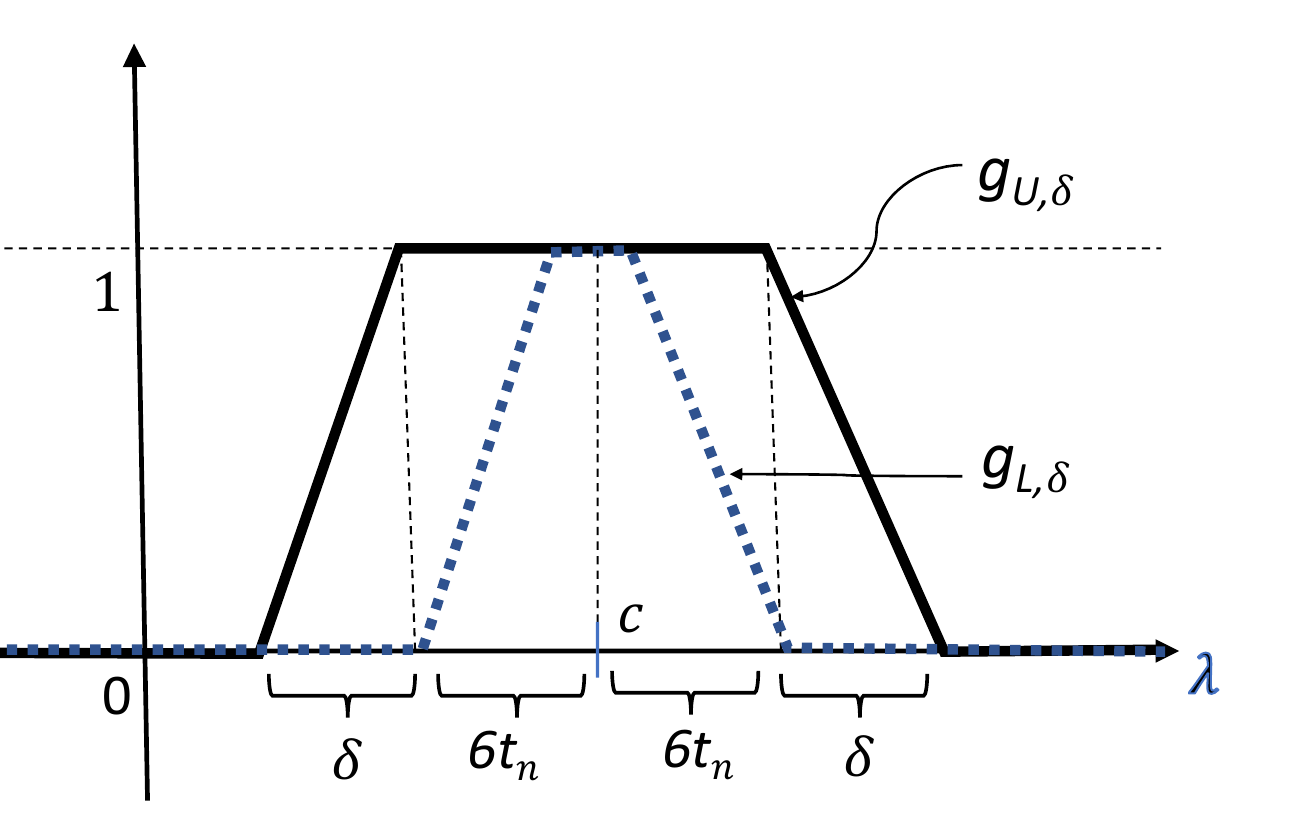}
		\caption{The Shape of $g_{L,\delta}$ and $g_{U,\delta}$}
		\label{fig1}
	\end{center}
\end{figure}

\noindent \textbf{Proof: } The proof adapts part of the arguments in the proof of Proposition A.1 of \cite{Guerre/Sabbah:2012:ET}. First, we take $\delta>0$ such that $\delta \le \mathcal{L}^{2/3}(B)$, and define
\begin{align*}
	z_\delta(\overline \lambda;c) &= \left( 1 - \min\left\{ (\overline \lambda - c -6 t_n)/\delta, 1\right\}\right) \times 1\left\{0 < \overline \lambda - c - 6 t_n\right\} \\
	&\quad + 1\left\{\overline \lambda - c - 6 t_n \le 0 \right\}, \quad \overline \lambda, c \in \mathbf{R}.
\end{align*}
Let
\begin{align*}
	g_{U,\delta}(\overline \lambda;c) &= z_{\delta}(\overline \lambda;c) - z_{\delta}(\overline \lambda + \delta + 12 t_n; c), \text{ and }\\
	g_{L,\delta}(\overline \lambda;c) &= \min\left\{z_\delta(\overline \lambda + \delta; c), z_\delta(-\overline \lambda + 2c + \delta; c) \right\}, \quad \overline \lambda, c \in \mathbf{R},
\end{align*}
and define
\begin{align*}
	f_{U,\delta}(s;c) = g_{U,\delta}(\lambda(s);c), \text{ and } f_{L,\delta}(s;c) = g_{L,\delta}(\lambda(s);c),
\end{align*}
where $\lambda$ is the map in Assumption \ref{assump: index utility2}(i). Then $f_{U,\delta}(s;c)$ and $f_{L,\delta}(s;c)$ are Lipschitz in $c$ with coefficient equal to $\delta^{-1}$. Furthermore,
\begin{align}
	\label{ineq5}
	\mathbf{E}\left[ \left( f_{U,\delta}(S_i;c) - f_{L,\delta}(S_i;c) \right)^2 \mid \tilde Z \right] &\le \mathbf{E}\left[ \left| f_{U,\delta}(S_i;c) - f_{L,\delta}(S_i;c) \right| \mid \tilde Z \right] \\ \notag
	&\le P\left\{ c + 6 t_n - \delta \le \lambda(S_i) \le c + 6 t_n + \delta \mid \tilde Z \right\}\\ \notag
	&\quad + P\left\{ c - 6 t_n - \delta \le \lambda(S_i) \le c - 6 t_n + \delta \mid \tilde Z \right\} \le 2 \overline C \delta, 
\end{align} 
and $f_{L,\delta}(s;c) \le f(s;c,6 t_n) \le f_{U,\delta}(s;c)$. (See Figure \ref{fig1}.) Define
\begin{align*}
	\mathcal{F}_{L,\delta} = \{f_{L,\delta}(\cdot; c): c \in B\}, \text{ and } \mathcal{F}_{U,\delta} = \{f_{U,\delta}(\cdot; c): c \in B\}.
\end{align*}
For any real valued measurable map $f$, we define $\| f \|_{Z,2} = \sqrt{\mathbf{E}[f^2(S_i) \mid \tilde Z]}$. We choose $\epsilon$ such that $0 < \epsilon \le \min\{\sqrt{6 t_n},\mathcal{L}^{1/3}(B)\}$. Since $f_{U,\delta}(s;c)$ and $f_{L,\delta}(s;c)$ are Lipschitz in $c$ with coefficient equal to $\delta^{-1}$ and since $B$ is a bounded interval, it follows
 by Theorem 2.7.11 of \cite{vanderVaart/Wellner:96:WeakConvg} that for any $C_2>0$, there exist $2 C_2\epsilon^3/\delta$-brackets  $[f_{L,a,j},f_{L,b,j}]$, $j=1,...,n_{[]}(\epsilon):=\left\lceil \mathcal{L}(B)/(2 C_2\epsilon^3) \right\rceil$  (with respect to $\| \cdot \|_{Z,2}$) that cover
 $\mathcal{F}_{L,\delta}$, and $2 C_2\epsilon^3/\delta$-brackets $[f_{U,a,j},f_{U,b,j}]$, $j=1,...,n_{[]}(\epsilon)$ (also with respect to  $\| \cdot \|_{Z,2}$) that cover $\mathcal{F}_{U,\delta}$; in other words,  for any pair $f_{U,\delta}(\cdot;c)$ and $f_{L,\delta}(\cdot;c)$, there exists $j\in \{1,...,n_{[]}(\epsilon)\}$  satisfying  
 \begin{align}
	\label{ineqs}
	f_{L,a,j}(\cdot) \le f_{L,\delta}(\cdot;c) \le f_{L,b,j}(\cdot) \le f_{U,a,j}(\cdot) \le f_{U,\delta}(\cdot;c) \le f_{U,b,j}(\cdot).
\end{align}
 Hence, the brackets $[f_{L,j},f_{U,j}] = [f_{L,a,j},f_{U,b,j}]$ for $j=1,...,n_{[]}(\epsilon)$ cover $\mathcal{F}(t_n)$. (The requirement $\delta \le \mathcal{L}^{2/3}(B)$ is fulfilled by the choice of $\epsilon$ so that $\delta = \epsilon^2 \le \mathcal{L}^{2/3}(B)$.) We then set  $C_2 = (\overline C/2)^{1/2}$ and $\delta =\epsilon^2$. Then, we have $n_{[]}(\epsilon)=\lceil \mathcal{L}(B)/((2 \overline C)^{1/2} \epsilon^3) \rceil \le \lceil \mathcal{L}(B)/\epsilon^3 \rceil$ (because $\overline C \ge 1$), and obtain
\begin{align*}
	\ln n_{[]}(\epsilon) \le 1 + \ln \mathcal{L}(B) - 3\ln \epsilon.
\end{align*}
Therefore, we obtain the bound (\ref{bound23}). 

As for the bound (\ref{bound3456}), observe that from (\ref{ineqs}),
\begin{align}
	\label{bound345}
	\mathbf{E}\left[ \left| f_{L,j}(S_i) - f_{U,j}(S_i)\right|^k \mid \tilde Z \right] &= \mathbf{E}\left[\left(f_{U,b,j}(S_i) - f_{L,a,j}(S_i) \right)^k \mid \tilde Z \right]\\ \notag
	&\le \mathbf{E}\left[\left(f_{U,b,j}(S_i) - f_{L,a,j}(S_i)\right)^2 \mid \tilde Z \right],
 \end{align}
because $f_{U,b,j}$ and $f_{L,a,j}$ are bounded between $0$ and $1$ and $k \ge 2$. The last conditional expectation is bounded by
\begin{align*}
	&3\mathbf{E}\left[\left(f_{L,a,j}(S_i) - f_{L,b,j}(S_i)\right)^2 \mid \tilde Z \right]+ 3\mathbf{E}\left[\left(f_{L,b,j}(S_i) - f_{U,a,j}(S_i)\right)^2 \mid \tilde Z \right]\\
	&\quad + 3\mathbf{E}\left[\left(f_{U,a,j}(S_i) - f_{U,b,j}(S_i)\right)^2 \mid \tilde Z \right] \le 3(2 \overline C \epsilon^2 + 2 \overline C \epsilon^2 + 2 \overline C \epsilon^2),
\end{align*}
where the first and third terms $2 \overline C \epsilon^2$ are due to the choice of $(2 \overline C)^{1/2} \epsilon$-brackets with respect to $\| \cdot \|_{Z,2}$ and the middle term $2 \overline C \epsilon^2$ is from (\ref{ineq5}) and (\ref{ineqs}). Because $\left| f_{L,j}(S_i) - f_{U,j}(S_i)\right| \le 1$, we find that for all $k \ge 2$, 
\begin{align}
    \label{bound34}
    \mathbf{E}\left[ \left| f_{L,j}(S_i) - f_{U,j}(S_i)\right|^k \mid \tilde Z \right] \le 18 \overline C \epsilon^2 \le (5 \overline C)^k \epsilon^2,
\end{align}
because $\overline C \ge 1$. We obtain the desired bound in (\ref{bound3456}). $\blacksquare$

\begin{lemma}
	\label{lemm: Rn(c,t)}
	There exist constants $C>0$ and $n_0 \ge 1$ that depend only on the constant $\overline C$ and the set $B$ in Assumption \ref{assump: index utility3}(ii) such that for any $\sigma(\tilde Z)$-measurable random variable $t_n$ satisfying $t_n > n^{-1} \ln n$ for all $n \ge n_0$,
	\begin{align}
        \label{ineq22}
		\mathbf{P}\left\{ \sup_{c \in B} |R_n(c,6t_n)| \ge C t_n \mid \tilde Z \right\} \le 2 \exp\left( - n t_n\right),
	\end{align}
    where $R_n(c,t)$ is as defined in Lemma \ref{lemma:heterogeneity_index_under_additive_preferences}.
\end{lemma}
\medskip

\noindent \textbf{Proof: } Suppose that $t_n > \mathcal{L}(B)/6$. Then, the bound (\ref{ineq22}) trivially holds, because $R_n(c,6t_n) = 0$ for all $c \in B$, with probability one. For the rest of the proof, we assume that $t_n \le \mathcal{L}(B)/6$. Let $C' = 6 \overline C M_1^{-2} $ for simplicity, where 
\begin{align}
	\label{bound M1}
	M_1 = \max \left\{ 5 \overline C, \sqrt{2 \overline C \mathcal{L}(B)} \right\}.
\end{align}
By Assumption \ref{assump: index utility cond ind}, $S_i$'s are conditionally i.i.d.\ across $i$'s given $\tilde Z$. We apply Corollary 6.9 of \cite{Massart:07:ConcentrationInequality}, p.194, to the sum $M_1^{-1} n R_n(c,6t_n)$. The corollary is based on Theorem 6.8 there. It suffices to verify two conditions in Theorem 6.8. (Note that $R_n(c,6t_n)$ remains the same if we replace the set $B$ by a countable dense subset.) First, for any $k \ge 2$,
\begin{align*}
	\mathbf{E}\left[ \left|M_1^{-1} f(S_i;c, 6 t_n)\right|^k \mid \tilde Z \right] &= M_1^{-k} P\left\{ |\lambda(S_i) - c| \le 6 t_n \mid \tilde Z\right\} \\
	&\le C' t_n M_1^{-k+2} \le C' t_n,
\end{align*}
by Assumption \ref{assump: index utility3}(ii) and by $M_1 \ge 1$. So, the first condition in Theorem 6.8 of \cite{Massart:07:ConcentrationInequality} is satisfied by taking $\sigma^2=C't_n$ and $b=1$ there. To verify the second condition of the theorem, we invoke our Lemma \ref{lemm: bracketing}. By (\ref{bound M1}), and choice of $t_n \le \mathcal{L}(B)/6$, we have
\begin{align}
	\label{bound M2}
	\sqrt{C't_n} \le \sqrt{C'\mathcal{L}(B)/6} \le \sqrt{0.5},
\end{align}
and $C' \le 3$, because $M_1 \ge \sqrt{2 \overline C \mathcal{L}(B)}$. Furthermore, $C' t_n \le \min\{6 t_n, \mathcal{L}^{2/3}(B)\}$, because $\mathcal{L}(B) \ge 1$. Using Lemma \ref{lemm: bracketing}, we see that the second condition of Theorem 6.8 of \cite{Massart:07:ConcentrationInequality} holds for $M_1^{-1} n R_n(c,6t_n)$ with $b=1$ again. We apply Corollary 6.9 of \cite{Massart:07:ConcentrationInequality}, p.194, (putting $\varepsilon = 1$ there) to find that for any positive number $x$,
\begin{align*}
	\mathbf{P}\left\{ M_1^{-1} n \sup_{c \in B} |R_n(c,6t_n)| \ge n r_n(x) \mid \tilde Z \right\} \le 2 \exp( - x),
\end{align*}
where 
\begin{align}
	\label{bound14}
	r_n(x) &= \frac{27}{\sqrt{n}} \int_0^{\sqrt{ C' t_n}} \sqrt{ \ln n_{[]}(\epsilon) } d\epsilon + \frac{2}{n}\left(1 + \sqrt{C' t_n}\right)  \ln n_{[]}
	\left(\sqrt{C' t_n}\right)\\ \notag
	&\quad + 7 \left( n^{-1/2} \sqrt{2 C' t_n x} + 2 n^{-1} x\right).
\end{align}
By (\ref{bound23}) in Lemma \ref{lemm: bracketing} and (\ref{bound M2}), and applying the inequality $\sqrt{a+b} \le \sqrt{a} + \sqrt{b}$, $a,b \ge 0$, in the first integral on the right hand side, we have 
\begin{align*}
    r_n(x) &\le \frac{27}{\sqrt{n}} \left( \sqrt{ C' t_n (1 + \ln \mathcal{L}(B))}  + \sqrt{3}\int_0^{\sqrt{C' t_n}} \sqrt{-\ln \epsilon } d\epsilon \right) \\
	&\quad + \frac{2}{n}\left(1 + \sqrt{0.5}\right) \left(1 + \ln \mathcal{L}(B) - 3 \ln \sqrt{C' t_n} \right) + 7 \left( n^{-1/2} \sqrt{x} + 2 n^{-1} x\right).
\end{align*}
Since $\int_0^t \sqrt{-\ln x} dx \le 3t \sqrt{-\ln t}$ for all $0 < t \le \sqrt{0.5}$,  the leading term on the right hand side above is bounded by 
\begin{align*}
    \frac{27}{\sqrt{n}} \left( \sqrt{ C' t_n (1 + \ln \mathcal{L}(B))}  + \frac{3 \sqrt{3} \sqrt{-C' t_n (\ln(C') + \ln(t_n))}}{\sqrt{2}} \right).
\end{align*}
Taking $x = n t_n$, and noting that $t_n \ge n^{-1} \ln n$, we can find $n_0$ and constants $C_1,C_2>0$ which depend only on $\overline C$ and the set $B$ such that for all $n \ge n_0$,
\begin{align*}
	r_n(x) &\le C_1\left(n^{-1/2} \sqrt{-t_n \ln(t_n \wedge 1)} - n^{-1} \ln(t_n \wedge 1) + t_n\right) \le C_2 t_n.
\end{align*}
Thus, we obtain the desired result. $\blacksquare$\medskip

\noindent \textbf{Proof of Theorem \ref{thm: concentration inequality}: } First, we bound 
\begin{align*}
	&\mathbf{P}\left\{ \left| n_Z \hat \theta(\boldsymbol{\tau}) - \mathbf{E}\left[ n_Z \hat \theta(\boldsymbol{\tau}) \mid \tilde Z \right] \right|  \ge  n \pi_Z t \mid \tilde Z \right\} \\
    &\quad \le \mathbf{P}\left\{ \left| n_Z \hat \theta(\boldsymbol{\tau}) - \mathbf{E}\left[ n_Z \hat \theta(\boldsymbol{\tau}) \mid \tilde Z, \tilde S \right] \right|  \ge \frac{n \pi_Z t}{2} \mid \tilde Z \right\} \\
	&\quad \quad + \mathbf{P}\left\{ \left| \mathbf{E}\left[ n_Z \hat \theta(\boldsymbol{\tau}) \mid \tilde Z, \tilde S \right] - \mathbf{E}\left[ n_Z \hat \theta(\boldsymbol{\tau}) \mid \tilde Z \right] \right|  \ge \frac{n \pi_Z t}{2} \mid \tilde Z \right\} = H_{n,1} + H_{n,2}, \text{ say}.
\end{align*}
By Lemma \ref{lemm:conc_ineq2}, we have 
\begin{align}
	\label{H_{n,1}}
	H_{n,1} \le 2 \exp\left(-\frac{ n \pi_Z t^2/4}{\displaystyle 2 \overline \tau^2(Z) + \left(\overline \tau(Z) t/3 \right)}\right)
    \le 2 \exp\left(-\frac{ n \pi_Z t^2/4}{\displaystyle 2 \overline \tau^2(Z) (1 + t)}\right),
\end{align}
because $\overline \tau(Z) \ge 1$.

Let us turn to $H_{n,2}$. We let $n'$ be the smallest positive integer $n$ such that $n^{-5/6} \sqrt{\ln n} -  (\sqrt{2} n)^{-1}> n^{-1} \ln n$. We take $n_0$ to be the maximum between $n'$ and the integer $n_0$ in Lemma \ref{lemm: Rn(c,t)}. From here on, we take $n \ge n_0$. We choose $t_n$ to be a $\sigma(\tilde Z)$-measurable, positive sequence of random variables. We will determine the precise value of $t_n$ later. Let $A_{n,1}$ be the event under which $\sigma_n |\eta_{ij}| \le t_n$ for all $i \in N$ and $j \in M$, and let $A_{n,2}$ be the event
\begin{align*}
	\left\{\sup_{c \in B} |R_n(c, 6 t_n)| \le C t_n\right\},
\end{align*}
for the constant $C>0$ that appears in Lemma \ref{lemm: Rn(c,t)}. Let $A_n = A_{n,1} \cap A_{n,2}$. Also define the event
\begin{align*}
	E_n =  \left\{\overline h(\tilde Z) \leq (12 \overline C + 2C) n t_n + 1\right\}.
\end{align*}
Then, $P(A_n \cap E_n^c) = 0$, because $P(A_n \cap E_n^c)$ is bounded by the probability in (\ref{eq:growth_of_h_index}). For simplicity, define 
\begin{align*}
	D_n = \mathbf{E}\left[ n_Z \hat \theta(\boldsymbol{\tau}) \mid \tilde Z, \tilde S \right] - \mathbf{E}\left[ n_Z \hat \theta(\boldsymbol{\tau}) \mid \tilde Z \right].
\end{align*}
By Lemmas \ref{lemm:conc_ineq} and \ref{lemma:heterogeneity_index_under_additive_preferences}, we have
\begin{align}
	\label{ineqB}
	\mathbf{P}\left\{\left\{|D_n| \ge \frac{n \pi_Z t}{2} \right\} \cap A_n \mid \tilde Z \right\} &= \mathbf{P}\left\{\left\{|D_n| \ge \frac{n \pi_Z t}{2} \right\} \cap A_n \cap E_n \mid \tilde Z \right\}  \\ \notag
	&\le 2 \exp\left(-\frac{\tilde  C n t^2}{\displaystyle \overline \tau^2(Z) \left(n^2 t_n^2 + 1\right)} \right),
\end{align}
where $\tilde C > 0$ is a constant that depends only on $\overline C$ and set $B$ in Assumptions \ref{assump: index utility3}(ii). Without loss of generality, we assume that $\tilde C \le 1$. (If $\tilde C > 1$, we simply replace $\tilde C$ by $\tilde C \wedge 1$ to obtain the same bound.) 

On the other hand, by Assumptions \ref{assump: index utility3}(i),
\begin{align}
	\label{bound}
	\mathbf{P}\left\{ A_{n,1}^c \mid \tilde Z \right\} &\le 2 n m \exp\left( - \frac{t_n^2}{2\sigma_n^2}\right) = 2 \exp\left( - \left( \frac{t_n^2}{2\sigma_n^2} - \ln\left( n m \right) \right)\right)\\ \notag
	&\le 2 \exp\left( - \left( \frac{t_n^2}{2 \tilde \sigma_n^2} - \ln\left( n m \right) \right)\right),
\end{align}
where we recall $\tilde \sigma_n = \sigma_n \vee n^{-5/6}$. Since $\tilde \sigma_n \ge n^{-1}$, $2 \tilde \sigma_n^2 n^2 \ln(n m) \ge 1$, because $n \ge n_0$, $n_0 \ge 2$ and $m \ge 1$. We set $t_n^2$ to be
\begin{align}
	\label{eqn2}
	 t_n^2 =  \frac{2 \tilde \sigma_n^2 n^2 \ln(n m)  - 1 + \sqrt{ \left(2 \tilde \sigma_n^2 n^2 \ln(n m)  - 1 \right)^2 + 4n^2\left( 2 \tilde \sigma_n^2 \ln(n m) + n \tilde \sigma_n^2 \tilde t^2 \right)}}{2 n^2},
\end{align}
where $\tilde t = \sqrt{2 \tilde C} t / \overline \tau(Z)$, so that we have
\begin{align}
	\label{eqn}
	\frac{t_n^2}{2\tilde \sigma_n^2} - \ln(n m) = \frac{\tilde C n t^2}{\overline \tau^2(Z) \left(n^2 t_n^2 + 1 \right)}.
\end{align}
Since $\sqrt{a + b} \le \sqrt{a} + \sqrt{b}$ for all $a,b \ge0$, we have
\begin{align}
	\label{bound232}
	t_n^2 \le (t_n')^2 \equiv 2 \tilde \sigma_n^2 \ln(n m) + \frac{\sqrt{2} \tilde \sigma_n \sqrt{\ln(n m)}}{n} + \frac{ \tilde \sigma_n \tilde t}{\sqrt{n}}.
\end{align}
Therefore, from (\ref{bound}) and (\ref{eqn}),
\begin{align}
	\label{ineq52}
	\mathbf{P}\left\{ A_{n,1}^c \mid \tilde Z \right\} \le 2 \exp\left(  - \frac{\tilde C n t^2}{\overline \tau^2(Z)  \left(n^2 t_n^2 + 1\right)} \right).
\end{align}

Since $\tilde \sigma_n \ge n^{-5/6}$, we have from (\ref{eqn2}) that
\begin{align}
	\label{ineq32}
    t_n^2 \ge \frac{ \tilde \sigma_n \tilde t}{\sqrt{n}} = \frac{\sqrt{2\tilde C} \tilde \sigma_n t}{ \overline \tau(Z) \sqrt{n}} \ge \frac{\sqrt{\tilde C} t}{\overline \tau^{4/3}(Z) n^{4/3}} \ge \frac{\tilde C^2 t}{\overline \tau^{4/3}(Z) n^{4/3}},
\end{align}
where the second to the last inequality uses the fact that $\overline \tau(Z) \ge 1$ and the last inequality is due to the choice of $\tilde C \le 1$. This implies that 
\begin{align}
	\label{bd321}
	n t_n \ge \frac{\tilde C^3 n t^{3/2}}{\overline \tau^2(Z) n^2 t_n^2} \ge \frac{\tilde C^3 n t^{3/2}}{\overline \tau^2(Z) \left(n^2 (t_n')^2 + 1\right)}.
\end{align}
From (\ref{eqn2}), we have 
\begin{align*}
	t_n^2 \ge \frac{1}{2n^2}\left( 2 \tilde \sigma_n^2 n^2 \ln (n m) - 1 \right),
\end{align*}
and hence (using $(a + b)^2 \ge a^2 + b^2$ for all $a,b \ge 0$)
\begin{align*}
	t_n \ge \tilde \sigma_n \sqrt{\ln(nm)} - (\sqrt{2} n)^{-1} \ge n^{-5/6} \sqrt{\ln(nm)} - (\sqrt{2} n)^{-1} > n^{-1} \ln n,
\end{align*}
for all $n \ge n_0$, due to the choice of $n_0$. By Lemma \ref{lemm: Rn(c,t)}, for all $n \ge n_0$,
\begin{align}
	\label{bounds21}
	\mathbf{P}\left\{ A_{n,2}^c \mid \tilde Z \right\} \le 2 \exp(- n t_n) \le 2\exp\left( -\frac{\tilde C^3 n t^{3/2}}{\overline \tau^2(Z) \left(n^2 (t_n')^2 + 1\right)} \right),
\end{align}
by (\ref{bd321}). Hence, collecting the bounds in (\ref{ineqB}), (\ref{ineq52}) and (\ref{bounds21}), we find that 
\begin{align}
	\label{H_{n,2}}
	H_{n,2} = \mathbf{P}\left\{|D_n| \ge \frac{n \pi_Z t}{2} \mid \tilde Z  \right\} &\le \mathbf{P}\left\{\left\{|D_n| \ge \frac{n \pi_Z t}{2} \right\} \cap A_n \mid \tilde Z \right\} + \mathbf{P}\left\{A_{n,1}^c \mid \tilde Z \right\} + \mathbf{P}\left\{A_{n,2}^c \mid \tilde Z \right\}\\ \notag
	&\le 6 \exp\left(-\frac{\tilde  C^3 n (t^2 \wedge t^{3/2})}{\displaystyle \overline \tau^2(Z) \left(n^2 (t_n')^2 + 1\right)} \right),
\end{align}
because $\tilde C \le 1$. Now, observe that 
\begin{align*}
	n^2(t_n')^2 + 1 &= 2 n^2 \tilde \sigma_n^2 \ln(n m) + \sqrt{2} n \tilde \sigma_n \sqrt{\ln(n m)} + n \sqrt{n} \tilde \sigma_n \tilde t + 1 \\
	&\le 4 n^2 \tilde \sigma_n^2 \ln(n m) +1 + n \sqrt{n} \tilde \sigma_n \tilde t \\
	&\le 4 n^2 \tilde \sigma_n^2 \ln(n m) +1 + \sqrt{2 \tilde C} n \sqrt{n} \tilde \sigma_n t = (4 \vee \sqrt{2 \tilde C})(a_n + b_n t),
\end{align*}
where the first inequality follows because 
\begin{align*}
	n \tilde \sigma_n \le n^2 \tilde \sigma_n^2 (n^{-1} \tilde \sigma_n^{-1}) \le n^2 \tilde \sigma_n^2 n^{-1} n^{5/6} \le n^2 \tilde \sigma_n^2,
\end{align*}
and the second inequality uses the definition of $\tilde t = \sqrt{2 \tilde C} t/\overline \tau(Z)$ and the fact that $\overline \tau(Z) \ge 1$. We take $C = \tilde C^3 /(4 \vee \sqrt{2 \tilde C})$ for the constant $C>0$ in the theorem and combining the bound (\ref{H_{n,2}}) with that in (\ref{H_{n,1}}), we obtain the desired result. $\blacksquare$\medskip

\noindent \textbf{Proof of Corollary \ref{cor: rate of convergence}:} First, note that 
\begin{align*}
    \mathbf{E}[(n_Z - n \pi_Z) \hat \theta(\boldsymbol{\tau}) \mid \tilde Z] = \mathbf{E}[n_Z - n \pi_Z \mid \tilde Z] \mathbf{E}[ \hat \theta(\boldsymbol{\tau}) \mid \tilde Z]= 0,
\end{align*}
because $n_Z - n \pi_Z$ and $\hat \theta(\boldsymbol{\tau})$ are conditionally independent given $\tilde Z$. Hence, we have 
\begin{align*}
    |n \pi_Z \hat \theta(\boldsymbol{\tau}) - \mathbf{E}[n \pi_Z \hat \theta(\boldsymbol{\tau}) \mid \tilde Z]| &\le |n_Z \hat \theta(\boldsymbol{\tau}) - \mathbf{E}[ n_Z \hat \theta(\boldsymbol{\tau}) \mid \tilde Z] | \\
    &\quad \quad + |n_Z - n \pi_Z| |\hat \theta(\boldsymbol{\tau})| + |\mathbf{E}[(n_Z - n \pi_Z) \hat \theta(\boldsymbol{\tau}) \mid \tilde Z]|\\
    &\le |n_Z \hat \theta(\boldsymbol{\tau}) - \mathbf{E}[ n_Z \hat \theta(\boldsymbol{\tau}) \mid \tilde Z] | + |n_Z - n \pi_Z| \overline \tau(Z), 
\end{align*}
because $|\hat \theta(\boldsymbol{\tau})| \le \overline \tau(Z)$. 

We take a large number $M_2 \ge 1$ and let 
\begin{align*}
    \nu_n = \tilde \sigma_n \sqrt{n\ln(n m)} + (n \pi_Z)^{-1/2}.
\end{align*}
Fix $\epsilon>0$ such that $\epsilon M_2 \le 1$. Then, $P\{\nu_n > \epsilon\} \rightarrow 0$ as $n \rightarrow \infty$, by Condition (iii) of the corollary. Let $E_n = \{\nu_n \le \epsilon\}$. From the previous result, we obtain that
\begin{align*}
    P\left\{ |\hat \theta(\boldsymbol{\tau}) - \mathbf{E}[ \hat \theta(\boldsymbol{\tau}) \mid \tilde Z]| > 2M_2 \nu_n \mid \tilde Z \right\} \le I_{n,1} + I_{n,2},
\end{align*}
where 
\begin{align*}
    I_{n,1} &= P\left\{ |n_Z \hat \theta(\boldsymbol{\tau}) - \mathbf{E}[ n_Z \hat \theta(\boldsymbol{\tau}) \mid \tilde Z]| > n \pi_Z M_2 \nu_n \mid \tilde Z \right\}, \text{ and }\\
    I_{n,2} &= P\left\{ \overline \tau(Z) |n_Z - n \pi_Z|  > n \pi_Z M_2 \nu_n \mid \tilde Z \right\}.
\end{align*}
In light of Lemma \ref{lemm: unbiased}, it suffices to show that $\mathbf{E}[(I_{n,1} + I_{n,2})1_{E_n}] \rightarrow 0$ as $n \rightarrow \infty$ and then $M_2 \rightarrow \infty$. We will show that $\mathbf{E}[I_{n,1}1_{E_n}] \rightarrow 0$ and $\mathbf{E}[I_{n,2}1_{E_n}] \rightarrow 0$ separately.

As for $I_{n,1}$, we apply Theorem \ref{thm: concentration inequality} with $t = M_2 \nu_n$. Note that $t \le 1$ on the event $E_n$ by the choice of $\epsilon$. Furthermore, 
\begin{align}
	\label{eq3}
	\frac{n \nu_n^2}{a_n + b_n M_2 \nu_n} &= \frac{ \tilde \sigma_n^2 n^2 \ln(nm) + 2\tilde \sigma_n n \sqrt{n \ln(nm)} n^{-1/2} \pi_Z^{-1/2} + (\pi_Z)^{-1}}{\tilde \sigma_n^2  n^2 \ln(nm) + 1 + n \sqrt{n} \tilde \sigma_n M_2 \{\tilde \sigma_n \sqrt{n \ln(nm)} + (n\pi_Z)^{-1/2}\}}\\ \notag
    & \ge \frac{ \tilde \sigma_n^2 n^2 \ln(nm) + \tilde \sigma_n n \sqrt{\ln(nm)} \pi_Z^{-1/2} + 1}{(1 + M_2) \tilde \sigma_n^2  n^2 \ln(nm) + 1 + M_2 n \tilde \sigma_n \pi_Z^{-1/2}} \ge \frac{1}{2 M_2}.
\end{align}
 By Theorem \ref{thm: concentration inequality}, we can take a large enough $n_0$ such that for all $n \ge n_0$, on the event $E_n$ (so that we have $M_2 \nu_n \le 1$),
\begin{align*}
    I_{n,1} &\le 6 \exp\left( - \frac{C M_2^2 n \nu_n^2}{\overline \tau^2(Z)(a_n + b_n M_2 \nu_n)}\right) + 2 \exp\left( - \frac{\displaystyle n \pi_Z M_2^2 \nu_n^2}{\displaystyle 8 \overline \tau^2(Z)(1 + M_2 \nu_n)} \right)\\
    &\le 6 \exp\left( - \frac{C M_2}{2C_1^2} \right) + 2 \exp\left( - \frac{\displaystyle n \pi_Z M_2^2 \nu_n^2}{\displaystyle 16 C_1^2}\right) \\
	&\le 6 \exp\left( - \frac{C M_2}{2C_1^2} \right) + 2\exp\left( - \frac{M_2^2}{16 C_1^2} \right),
\end{align*}
where the second inequality follows from (\ref{eq3}) and the third inequality follows because $n \pi_Z \nu_n^2 \ge 1$ and $\overline \tau(Z) < C_1$. Therefore, $\mathbf{E}[I_{n,1}1_{E_n}] \rightarrow 0$, as $n \rightarrow \infty$ and then $M_2 \rightarrow \infty$.

Let us turn to $I_{n,2}$. By Lemma 2.1 of \cite{Chung/Lu:AC:2002}, we have on the event $E_n$,
\begin{align*}
    I_{n,2} &\le 2 \exp\left( - \frac{n^2 \pi_Z^2 M_2^2 \nu_n^2 /\overline \tau^2(Z)}{2 n \pi_Z + 2 \left(n \pi_Z M_2 \nu_n / (3 \overline \tau(Z))\right)}\right) \\
	&= 2 \exp\left( - \frac{n \pi_Z M_2^2 \nu_n^2}{2 \left(\overline \tau^2(Z) + \left( \overline \tau(Z) M_2 \nu_n / 3\right)\right)}\right) \le 2 \exp\left( - \frac{M_2^2}{2(C_1^2 + (C_1/3))}\right),
\end{align*}
where the last inequality follows because $M_2 \nu_n \le 1$ on the event $E_n$, $n \pi_Z \nu_n^2 \ge 1$, and $\overline \tau(Z) < C_1$. Hence, $\mathbf{E}[I_{n,2}1_{E_n}] \rightarrow 0$, as $n \rightarrow \infty$ and then $M_2 \rightarrow \infty$. $\blacksquare$\medskip

\noindent \textbf{Proof of Corollary \ref{cor: unif consistency}: } For simplicity, we focus on the case with $X_i \in \mathbf{R}$, $i \in N$. The proof is similar for a general case, involving hyperrectangles in place of intervals. Also, the proof for the case with $X_i$ being discrete is straightforward. We focus on the case where $X_i$ has a continuous conditional distribution function given $\tilde Z$. The proof modifies the proof of Lemma 2.11 of \cite{vanderVaart:98:AsympStat}. Since 
\begin{align*}
	\frac{1}{n_Z} \sum_{i \in N_Z} \left( 1\{Y_i = j\} - \mathbf{P}\left\{Y_i = j \mid \tilde Z\right\} \right) = o_P(1),
\end{align*}
 by Corollary \ref{cor: rate of convergence}, it suffices to show that
\begin{align}
	\label{conv32}
	\frac{1}{n_Z} \sum_{i \in N_Z} \left( 1\{X_i \le x, Y_i = j\} - \mathbf{P}\left\{X_i \le x, Y_i = j \mid \tilde Z\right\} \right) = o_P(1),
\end{align}
uniformly over $x \in \mathbf{R}$. For any $\sigma(\tilde Z)$-measurable random variable $W$, we have
\begin{align}
	\label{op}
	\frac{1}{n_Z} \sum_{i \in N_Z}  \left( 1\{X_i \le W, Y_i = j\} - \mathbf{P}\left\{X_i \le W, Y_i = j \mid \tilde Z\right\} \right) = o_P(1),
\end{align}
as shown in (\ref{eq: consist ind matching prob}). Since the conditional distribution function of $X_i$ given $\tilde Z$ is continuous, for any $\epsilon>0$, and any $\sigma(\tilde Z)$-measurable random variable $W$, there exists a $\sigma(\tilde Z)$-measurable random variable $\eta>0$ such that
\begin{align*}
	\mathbf{P}\left\{ W - \eta \le X_i \le W + \eta, Y_i = j  \mid \tilde Z \right\} \le \mathbf{P}\left\{ W - \eta \le X_i \le W + \eta \mid \tilde Z \right\} \le \epsilon.
\end{align*}
Fix $\epsilon>0$ and choose $\sigma(\tilde Z)$-measurable random variables $W_1,...,W_{k-1}$ such that $-\infty = W_0 < W_1 < W_2 < ... < W_k = \infty$, such that for all $\ell = 0,1,2,...,k$,
\begin{align*}
	\mathbf{P}\left\{X_i \le W_{\ell+1}, Y_i = j \mid \tilde Z \right\} - \mathbf{P}\left\{X_i \le W_{\ell}, Y_i = j \mid \tilde Z \right\} \le \epsilon.
\end{align*}
Then, for $W_{\ell-1} \le x \le W_{\ell}$, 
\begin{align*}
	&\frac{1}{n_Z} \sum_{i \in N_Z}  \left( 1\{X_i \le W_{\ell-1}, Y_i = j\} - \mathbf{P}\left\{X_i \le W_{\ell-1}, Y_i = j \mid \tilde Z\right\} \right) - \epsilon\\
	&\le \frac{1}{n_Z} \sum_{i \in N_Z} \left( 1\{X_i \le x, Y_i = j\} - \mathbf{P}\left\{X_i \le x, Y_i = j \mid \tilde Z\right\} \right) \\
	&\le \frac{1}{n_Z} \sum_{i \in N_Z} \left( 1\{X_i \le W_{\ell}, Y_i = j\} - \mathbf{P}\left\{X_i \le W_{\ell}, Y_i = j \mid \tilde Z\right\} \right) + \epsilon.
\end{align*}
Hence,
\begin{align*}
	&\sup_{x \in \mathbf{R}} \left| \frac{1}{n_Z} \sum_{i \in N_Z}  \left( 1\{X_i \le x, Y_i = j\} - \mathbf{P}\left\{X_i \le x, Y_i = j \mid \tilde Z\right\} \right) \right|\\
	&\le  \max_{0 \le \ell \le k} \left| \frac{1}{n_Z} \sum_{i \in N_Z}  \left( 1\{X_i \le W_{\ell}, Y_i = j\} - \mathbf{P}\left\{X_i \le W_{\ell}, Y_i = j \mid \tilde Z\right\} \right)\right| + \epsilon = \epsilon+o_P(1),
\end{align*}
as $n \rightarrow \infty$, by (\ref{op}). By sending $\epsilon$ to zero, we obtain the desired result. $\blacksquare$\medskip

\noindent \textbf{Proof of Corollary \ref{cor: PA}: } We define
\begin{align*}
	\hat F_{X,Z}(t,s) &= \frac{1}{n_{1,Z}}\sum_{i' \in N_{1,Z}} 1\{X_{i',k} \le t, Z_{Y_{i'},r} \le s\}, \text{ and } \\
	\hat F_{Z}(s) &= \frac{1}{n_{1,Z}}\sum_{i' \in N_{1,Z}} 1\{Z_{Y_{i'},r} \le s\}.
\end{align*}
Note that
\begin{align*}
	\hat F_{X,Z}(t,s) = \frac{n_Z}{n_{1,Z}} \sum_{j \in M} \frac{1}{n_Z}\sum_{i' \in N_Z} 1\{X_{i',k} \le t, Z_{j,r} \le s\}1\{Y_{i'} = j\}.
\end{align*}
Furthermore,
\begin{align}
	\label{conv2}
	\frac{n_{1,Z}}{n_Z} = \frac{1}{n_Z}\sum_{i \in N_Z} 1\{Y_i \ne 0\} =  \frac{1}{n_Z}\sum_{i \in N_Z} \mathbf{P}\{Y_i \ne 0 \mid \tilde Z\} + o_P(1) = \mathbf{P}\{Y_i \ne 0 \mid \tilde Z\} + o_P(1),
\end{align}
as $n \rightarrow \infty$. The second equality follows from (\ref{eq212}) because $M$ is independent of $n$. The last equality follows because the conditional distribution of $Y_i$ given $\tilde Z$ is identical across $i$'s by Lemma \ref{lemm: exch}. On the other hand, by Corollary \ref{cor: unif consistency}, for each $j \in M$,
\begin{align*}
	\frac{1}{n_Z}\sum_{i' \in N_Z} 1\{X_{i',k} \le t, Y_{i'} = j\} = \mathbf{P}\left\{X_{i',k} \le t, Y_{i'} = j \mid \tilde Z \right\} + o_P(1),
\end{align*}
uniformly over $t \in \mathbf{R}$, as $n \rightarrow \infty$. Since, conditional on $\tilde Z$, the indicator $1\{Z_{j,r} \le s\}$ is treated as constant, we have
\begin{align*}
	\sup_{t,s \in \mathbf{R}}\left|\hat F_{X,Z}(t,s) - F_{X,Z}(t,s) \right| = o_P(1).
\end{align*}
Because $n_Z = n\pi_Z (1 + o_P(1))$ by Lemma \ref{lemm:conc_ineq2} and for any $M >0$, $P\{n \pi_Z > M \} \rightarrow 1$ as $n \rightarrow \infty$ by (\ref{cond sigma_n}), 
\begin{align*}
	\max_{i \in N_Z} \sup_{t,s \in \mathbf{R}}\left|\hat F_{X,Z,-i}(t,s)  - \hat F_{X,Z}(t,s)\right| = o_P(1).
\end{align*}
Hence,
\begin{align*}
	\max_{i \in N_Z} \sup_{t,s \in \mathbf{R}}\left|\hat F_{X,Z,-i}(t,s)  - F_{X,Z}(t,s)\right| = o_P(1),
\end{align*}
as $n \rightarrow \infty$. Similarly, $\max_{ i \in N_Z} \sup_{t \in \mathbf{R}}\left|\hat F_{X,-i}(t)  - F_{X}(t)\right| = o_P(1)$, as $n \rightarrow \infty$. On the other hand,
\begin{align*}
	\sup_{t \in \mathbf{R}} \left|\hat F_{Z}(t) - F_Z(t) \right| &\le \sum_{j \in M}  \sup_{t \in \mathbf{R}} 1\{Z_{j,r} \le t\}\left|\frac{1}{n_{1,Z}}\sum_{i' \in N_{1,Z}}\left(1\{Y_{i'} = j\} - \mathbf{P}\left\{ Y_{i'} = j \mid \tilde Z  \right\}\right)\right|\\
	&= \sum_{j \in M}  \sup_{t \in \mathbf{R}} 1\{Z_{j,r} \le t\}\left|\frac{1}{n_{1,Z}}\sum_{i' \in N_Z}\left(1\{Y_{i'} = j\} - \mathbf{P}\left\{ Y_{i'} = j \mid \tilde Z  \right\}\right)\right|,
\end{align*}
where the last equality follows because $j$'s are chosen from $M$. The last term is bounded by
\begin{align*}
   \frac{n_Z}{n_{1,Z}} \sum_{j \in M}  \left|\frac{1}{n_Z}\sum_{i' \in N_Z}\left(1\{Y_{i'} = j\} - \mathbf{P}\left\{ Y_{i'} = j \mid \tilde Z  \right\}\right)\right| = o_P(1),
\end{align*}
as $n \rightarrow \infty$. The last equality follows by (\ref{conv3}) and (\ref{conv2}), and the condition (\ref{lower bound}) in Corollary \ref{cor: PA}, and due to $n_Z/n_{1,Z} = O_P(1)$ by (\ref{conv2}) and (\ref{lower bound}). Collecting these results, we conclude that
\begin{align*}
	\max_{i \in N_Z} \sup_{t,s \in \mathbf{R}} \left|\hat F_{X,Z,-i}(t,s) - \hat F_{X,-i}(t) \hat F_{Z,-i}(s) -\left(F_{X,Z}(t,s) - F_{X}(t) F_{Z}(s) \right) \right| = o_P(1).
\end{align*} 
Since $|\hat \rho - \rho|$ is bounded by 12 times the left hand side term, the desired result follows. $\blacksquare$\medskip

\noindent \textbf{Proof of Corollary \ref{cor: consistency of matching prob estimators}: } Define
\begin{align*}
	\hat g(x) = \frac{1}{n \pi_Z} \sum_{i \in N_Z} 1\{Y_i = j\} \mathcal{K}_h\left( X_i - x \right).
\end{align*}
Also, let
\begin{align*}
     g(x) &= p(j \mid x,\tilde Z) f_{X \mid \tilde Z}(x \mid \tilde Z) \text{ and } \hat f_{X \mid \tilde Z} (x \mid \tilde Z) = \frac{1}{n \pi_Z} \sum_{i \in N_Z} \mathcal{K}_h\left( X_i - x \right).
\end{align*}
Since $X_i$'s are conditionally i.i.d. given $Z$ by Assumption \ref{assump: index utility2}, and 
\begin{align*}
	\mathbf{E}\left[ \mathcal{K}_h\left( X_i - x \right) 1\{i \in N_Z\} \mid \tilde Z \right] 
	= \mathbf{E}\left[ \mathcal{K}_h\left( X_i - x \right) \mid \tilde Z \right] \pi_Z,
\end{align*}
(due to the conditional independence of $\zeta$ and $\tilde S$ given $\tilde Z$ in Assumption \ref{assump: sampling}), we have
\begin{align*}
	\hat f_{X \mid \tilde Z} (x \mid \tilde Z) - f_{X \mid \tilde Z} (x \mid \tilde Z) = O_P\left( \frac{1}{\sqrt{n \pi_Z h^d}} + h^2\right),
\end{align*}
using standard arguments of kernel density estimation. Hence
\begin{align*}
	\hat p(j \mid x,\tilde Z) - p(j \mid x,\tilde Z) &= \frac{\hat g(x) - g(x)}{\hat f_{X \mid \tilde Z}( x \mid \tilde Z)} + g(x)\left( \frac{1}{\hat f_{X \mid \tilde Z}(x \mid \tilde Z)} - \frac{1}{f_{X \mid \tilde Z}(x \mid \tilde Z)} \right)\\
	   &= \frac{\hat g(x) - g(x)}{f_{X \mid \tilde Z}( x \mid \tilde Z)}\left( 1 + o_P(1) \right) + O_P\left( \frac{1}{\sqrt{n \pi_Z h^d}} + h^2\right).
\end{align*}
Using the standard arguments in dealing with the bias part, we find that
\begin{align*}
	\hat g(x) - g(x) = \Delta_n(x) + O_P(h^2),
\end{align*}
where $\Delta_n(x)$ is as defined in (\ref{Delta n}). The desired result follows from (\ref{bd}), because $m$ is fixed. $\blacksquare$

\section{Acknowledgements}

We would like to thank Li Hao, Wei Li, Shunya Noda, and the participants at the seminars at Caltech, UBC, University of Haifa and Western University for valuable comments and questions. We thank the Co-Editor, the Associate Editor and two anonymous referees for valuable comments and criticisms. All errors are ours. Song acknowledges that this research was supported by Social Sciences and Humanities Research Council of Canada [grant number 435-2020-0204].\bigskip 

\putbib[LLN]
\end{bibunit}

\begin{bibunit}[econometrica]

\pagebreak
\renewcommand\thesection{\Alph{section}}

\begin{center}
	\Large \textsc{Supplemental Note to ``The Law of Large Numbers for Large Stable Matchings''}
	\bigskip
\end{center}

\date{%
	\today%
}

\setcounter{section}{0}
\setcounter{subsection}{0}
\setcounter{equation}{0}
\setcounter{lemma}{0}

\vspace*{7ex minus 1ex}
\begin{center}
	Jacob Schwartz and Kyungchul Song\\
	\textit{University of Haifa and University of British Columbia}
	\bigskip
\end{center}

The supplemental note is devoted to the proof of the bounded difference result (Lemma \ref{lemm:bounded_difference_condition}) in the main text. Let us present a brief summary of our proof strategy. We first introduce a re-stabilization operator which we use to transform any matching obtained by unmatching one student from a stable matching into another stable matching. Using the re-stabilization operator, we establish a bounded difference condition for a student-optimal stable matching (see Lemma \ref{lemm:BDC sosm} below),  where the bound depends on the maximum rank difference $h(\boldsymbol{w})$ defined in (\ref{h(w)}) in the main text. Next, using the Rural Hospital Theorem and the fact that the student-optimal stable matching is the college-worst stable matching, we establish a bound for the number of students matched with different colleges between a student-optimal stable matching and any stable matching (see Lemma \ref{lemm:bounded_between_two_stable_preliminary} below.) As in the case of the first bound, the second is also expressed in terms of $h(\boldsymbol{w})$. By combining these two bounds, we obtain the desired bounded difference condition for any stable matching in the market.

\section{Re-stabilization Operator and Its Properties}

We introduce a \bi{re-stabilization operator} that transforms an unstable matching in one market into a stable matching in another, by \textit{repeatedly satisfying a blocking pair}.\footnote{The definition in the one-to-one matching case is well known (e.g., \cite{Blum/Roth/Rothblum:1997:JET}): given an unstable matching $\mu'$ with a blocking pair $(i,j)$ we say that a matching $\mu$ \bi{is obtained from $\mu'$ by satisfying the blocking pair $(i,j)$} if $i$ and $j$ are matched to each other in $\mu$, their mates (if any) in $\mu'$ are unmatched in $\mu$ and the status of the remaining matched agents is unchanged. In our many-to-one setup, the notion of satisfying a blocking pair relevant for our purposes will be made precise later when we define a special operator (see Definition \ref{T} below). } \citet{Roth/Vandevate:1990:Ecta} showed that given an arbitrary matching, a sequence of matchings in which each is obtained from the previous matching by satisfying blocking pairs is guaranteed to converge to a stable matching when blocking pairs may be chosen randomly at each step in the sequence.  Rather than re-stabilizing arbitrary matchings, we focus on small perturbations to the outcomes of stable matchings caused by a change in the type of a single student. Therefore, we will consider a re-stabilization operator that takes in a matching that is already ``close'' to being stable.

We begin with some key definitions that we use repeatedly later.\footnote{Closely related definitions have been used. In particular, the notion of doctor quasi-stable matchings discussed in \cite{Wu/Roth:2018:GEB}. The matchings in the second item of the definition can be viewed as a many-to-one version of the matchings studied by \citet{Blum/Rothblum/2002} in the one-to-one case.}
\begin{definition}\label{def:college-envy-free}
	Given a matching $\mu$, we say that 
	\begin{enumerate}
		\item $\mu$ is \bi{individually rational} if there is no $i\in N$ or $j \in M$ such that $0\succ_{i}\mu(i)$ and $0\succ_{j}i'$ for some $i'\in\mu^{-1}(j)$.
		\item $\mu$ is \bi{envy-free} if it is individually rational and any blocking pair, if it exists, involves an unmatched student in $\mu$.
		\item $\mu$ is \bi{1-envy-free} if it is (i) envy free and (ii) either $\mu$ is stable or there exists one and only student who belongs to every blocking pair of $\mu$. 
	\end{enumerate}
\end{definition}
Given a market $(N,M,\boldsymbol{u})$, let $\mathcal{S}(N,M,\boldsymbol{u})$ be the set of stable matchings, and let $\mathcal{E}_1(N,M,\boldsymbol{u})$ be the set of matchings that are 1-envy-free. For each $i \in N$, with each map $\mu_{i}:N\backslash\{i\}\rightarrow M'$, we associate a map $g_{i}(\mu_{i})(\cdot):N\rightarrow M'$ defined by
\begin{align}
	g_{i}(\mu_{i})(i') = \left\{ \begin{array}{ll}
		\mu_{i}(i'),& \text{ if } i'\neq i,\\
		0,& \text{ if } i' = i.
	\end{array}
	\right.
\end{align}
The map $g_i(\mu_i)$ is a matching on $N$ constructed from $\mu_i$ by matching student $i$ to $0$. 

Given a preference profile $\boldsymbol{u}$, for each student $i \in N$, we define
\begin{align*}
	\boldsymbol{u}_{-i} = (\boldsymbol{v}_{-i},\boldsymbol{w}_{-i}),
\end{align*}
where $\boldsymbol{v}_{-i}$ is constructed by removing $v_i$ from $\boldsymbol{v}$, and $\boldsymbol{w}_{-i}$ is constructed by replacing each college's preference $w_j$ by the bijection $w_j^i: N'\setminus \{i\} \rightarrow N'\setminus \{i\}$ such that $w_j(i_1) > w_j(i_2)$ if and only if $w_j^i(i_1) > w_j^i(i_2)$ for all $i_1,i_2 \in N'\setminus \{i\}$. In other words, the preference profile $\boldsymbol{u}_{-i}$ is obtained by ``eliminating'' the student $i$ from the market.

Let $(N\backslash\{i\},M,\boldsymbol{u}_{-i})$ be the matching market derived from $(N,M,\boldsymbol{u})$ by eliminating the student $i$. The following remark follows immediately from our definitions. 
\begin{remark}\label{student_optimal_one_envy_free} For every $\boldsymbol{u}\in\mathbb{U}$ and $i\in N$, \label{rem}$\mu_{i} \in\mathcal{S}(N\backslash\{i\},M,\boldsymbol{u}_{-i})$
	if and only if $g_i(\mu_i)\in\mathcal{E}_1(N,M,\boldsymbol{u})$.  
\end{remark}

Given an envy-free matching $\mu$, a pair $(i,j)$ is a \bi{student-maximal blocking pair} for $\mu$ if $(i,j)$ is a blocking pair for $\mu$ and $j$ is student $i$'s most preferred college among those with whom he can form a blocking pair for $\mu$. Our re-stablization operator iteratively satisfies student-maximal blocking pairs.\footnote{The operator we propose is a straightforward adaptation of the correcting procedures described in \citet{Blum/Rothblum/2002} (itself a special case of   \cite{Blum/Roth/Rothblum:1997:JET}), adapted to a many-to-one setup. The approach is also similar to \citet{Biro/Cechlarova/Fleiner:208:IJGT}, who discuss algorithms for stabilizing matching markets when a single agent is added to a market that is presumed stable in the absence of the
	additional agent. Note that \cite{Wu/Roth:2018:GEB} showed that in many-to-one markets in which no students have justified envy, stable matchings can be obtained as fixed points of a lattice operator that generalizes the college-optimal deferred acceptance algorithm.} 

Before defining the operator, we introduce some further notation. Define $B_{\mu}$ to be the set of all blocking pairs to a matching $\mu$ for the market $(N,M,\boldsymbol{u})$. Let the set of student-maximal blocking pairs for a matching $\mu$ be
\begin{align*}
	B_{\mu}' & =\{(i',j')\in B_{\mu}:j'\succ_{i'}j'',\text{ for all }(i',j'')\in B_{\mu} \text{ such that } j' \ne j''\}.
\end{align*}
Note that $B_{\mu}' = \varnothing$ if and only if $B_{\mu} = \varnothing$. In the case that $\mu\in\mathcal{E}_1(N,M,\boldsymbol{u})$, $B_{\mu}'$ is either empty, or contains exactly one blocking pair. In the case that  $B_{\mu}\neq\varnothing$, we define $j^{*}(i)$ to be a college such that $B_{\mu}' = \{(i,j^{*}(i))\}$.

\begin{definition}\label{T} 
	For any $\mu\in \mathcal{E}_1(N,M,\boldsymbol{u})$, we define the operator $T: \mathcal{E}_1(N,M,\boldsymbol{u})\rightarrow \mathcal{E}_1(N,M,\boldsymbol{u})$ as follows.\medskip
	
	Suppose $B_{\mu}=\varnothing$. Then we take $T(\mu)=\mu$.
	
	Suppose $B_{\mu}\neq\varnothing$. Then $B_\mu'$ is a singleton, say, $\{(i,j^*(i))\}$ and we set $T(\mu)(i)=j^{*}(i)$. For each $i' \ne i$, let us denote $j' = \mu(i')$. We set $T(\mu)(i')$ as follows:
	
	\quad Case 1: $j'\neq j^{*}(i)$. Then $T(\mu)(i')=j'$. 
	
	\quad Case 2: $j'=j^{*}(i)$. Then,
	\begin{align}
		T(\mu)(i') = \left\{ \begin{array}{ll}
			j',& \text{ if } |\mu^{-1}(j')|<q_{j'},\\
			j',& \text{ if }  |\mu^{-1}(j')|=q_{j'}, \text{ and } i''\prec_{j'}i' \text{ for some }i''\in\mu^{-1}(j')\\
			0,& \text{ if }  |\mu^{-1}(j')|=q_{j'}, \text{ and } i'\prec_{j'}i'' \text{ for all }i''\in\mu^{-1}(j')\backslash \{i'\}.
		\end{array}
		\right.
	\end{align}
\end{definition}\medskip

From the definition of $T$, it is clear that any $\mu\in\mathcal{S}(N,M,\boldsymbol{u})$
satisfies $T(\mu)=\mu$; i.e., any stable matching is a fixed point
of $T$. The sets $B_{\mu}$, $B_{\mu}'$, and the operator $T$ certainly
	depend on $\boldsymbol{u}$, but we will often suppress these from
	our notation for simplicity. It is also true that any fixed point $\mu$ of $T$ is a stable matching. Note also that $T$ maps from $\mathcal{E}_1(N,M,\boldsymbol{u})$ to itself,
since for any $\mu\in\mathcal{E}_1(N,M,\boldsymbol{u})$, $T(\mu)$
is either stable or blockable by at most one student that is unmatched.

The next result, Lemma \ref{DA_as_monotone_operator}, shows that
repeated iterations of the operator $T$ yield a stable matching when the input is a 1-envy-free matching with respect
to one student. For this, it is convenient to introduce a partial order $\succsim$ over matchings. Suppose that $\succ_j^*$ is college $j$'s preference ordering over \textit{groups of} students. The following definition is from Definition 5.2. on page 128 of \cite{Roth/Sotomayor:90:TwoSidedMatching}.
\begin{definition}\label{def:responsiveness}
	The preference relation of a college $j$, $\succ_{j}^*$, over sets
	of students is \textbf{\textit{responsive}} to the preference over
	individual students if, whenever $\mu_{1}^{-1}(j)=(\mu_{2}^{-1}(j)\cup\{i_{1}\})\backslash\{i_{2}\}$
	for $i_{2}\in\mu_{2}^{-1}(j)$ and $i_{1}\notin\mu_{2}^{-1}(j)$,
	then $\mu_{1}^{-1}(j)\succ_{j}^*\mu_{2}^{-1}(j)$ if and only if $i_{1}\succ_{j}i_{2}$.
\end{definition}
We assume that $\succ_j^*$ is responsive to $\succ_j$. For any pair of matchings $\mu_{1}$ and $\mu_{2}$, we write $\mu_{1}\succsim \mu_{2}$
if and only if for all $j\in M$, either $\mu_{1}^{-1}(j)\succ_j^* \mu_{2}^{-1}(j)$
or $\mu_{1}^{-1}(j)=\mu_{2}^{-1}(j)$. 

\begin{lemma}\label{DA_as_monotone_operator} For each $\mu\in\mathcal{E}_1(N,M,\boldsymbol{u})$, the following is satisfied.
	
	(i) $T(\mu)\succsim\mu$.
	
	(ii) There is a finite sequence of matchings, $\mu_{1},\mu_{2},...,\mu_{r}$, with
	$\mu_{r}\in\mathcal{S}(N,M,\boldsymbol{u})$, where $\mu_0 = \mu$, and $\mu_{r'}=T(\mu_{r'-1})$
	for each $r'=1,...,r$.
\end{lemma}

\noindent \textbf{Proof: }If $B_{\mu}'$ is empty, the matching $\mu$ 
is stable and we have $T(\mu)=\mu,$ satisfying both (i) and (ii). If
$B_{\mu}'$ is not empty, it contains exactly one blocking pair, say, $(i,j^{*}(i))$,
and $T$ assigns student $i$ to college $j^{*}(i)$. In the case
that $j^{*}(i)$ has no vacancies under $\mu$, its worst student
under $\mu$, say, $i'$, is made unmatched, and in the case that $j^{*}(i)$ has a vacancy under $\mu$, then all the other students remain in their colleges. Thus, $T(\mu)\succsim\mu$,
since $T$ either affects no colleges, or leaves exactly one college,
$j^{*}(i)$, strictly better off while leaving the remaining colleges
unaffected.\footnote{By strictness and responsiveness of college preferences and the fact
	that $(i,j^{*}(i))$ is a blocking pair for $\mu$, we have either
	(i) $|\mu^{-1}(j^{*}(i))|=q_{j}$ and $j^{*}(i)$ strictly prefers
	$(\mu^{-1}(j^{*}(i))\cup\{i\})\backslash\{i'\}$ to $\mu^{-1}(j^{*}(i))$
	for some $i'\in\mu^{-1}(j^{*}(i))$ or (ii) $|\mu^{-1}(j^{*}(i))|<q_{j}$
	and $j^{*}(i)$ strictly prefers $\mu^{-1}(j^{*}(i))\cup\{i\}$ to
	$\mu^{-1}(j^{*}(i))$.} Since there are a finite number of student-college pairs, repeated iterations
of $T$ from any $\mu\in\mathcal{E}_1(N,M,\boldsymbol{u})$ are guaranteed
to converge to a fixed point which is a stable matching, after finite iterations. $\blacksquare$

\medskip{}

Lemma \ref{DA_as_monotone_operator} shows that for any $\mu\in\mathcal{E}_1(N,M,\boldsymbol{u})$,
repeated iterations of $T$ lead to a stable matching of the market
$(N,M,\boldsymbol{u})$, in finite iterations. Furthermore, the output of repeated iterations of $T$ is uniquely determined by the given choice of $\mu\in\mathcal{E}_1(N,M,\boldsymbol{u})$, since there is at most one student-maximal blocking pair after each iteration of the operator.  It is convenient to develop notation for the stable output of
repeated iterations of $T$ in terms of an input matching. Given any
$\mu\in\mathcal{E}_1(N,M,\boldsymbol{u})$, we denote $T^{*}(\mu)\equiv\mu_{r}$
where  $\mu_{0}=\mu,\mu_{1},\mu_{2},...,\mu_{r}$ is the finite sequence
of matchings with $\mu_{r}\in\mathcal{S}(N,M,\boldsymbol{u})$, where
$\mu_{r'}=T(\mu_{r'-1})$ for each $r'=1,...,r$.

The following is a consequence of what is called Rural Hospital Theorem. 

\begin{lemma}
	\label{lem:taxonomy_1_vacancy_or_no_filled}
	Let $\mu\in\mathcal{S}(N,M,\boldsymbol{u})$	and $\mu'\in\mathcal{E}_1(N,M,\boldsymbol{u})$. Then either of the following two cases must hold.
	
	\noindent Case 1:  For every college $j\in M$, 
	\begin{align}\label{eq:same_number_of_college_positions_filled_every_college}
		&|\mu^{-1}(j)|=|\mu'^{-1}(j)|
	\end{align}
	\noindent Case 2: There exists one and only one college, $j^{*}\in M$, such that
	\begin{align}\label{eq:one_and_only_one_college_fills_extra}
		&|\mu^{-1}(j^{*})|=|\mu'^{-1}(j^{*})|+1,
	\end{align}
	and, for every college $j\in M\backslash \{ j^{*}\}$,
	\begin{align}\label{eq:everybody_else_the_same}
		&|\mu^{-1}(j)|=|\mu'^{-1}(j)|.
	\end{align}	
\end{lemma}
\noindent \textbf{Proof:}  Let $T^{*}(\mu')$ be the matching obtained by iterations of $T$ starting from $\mu'$. Since $\mu$ is stable and $T^{*}(\mu')$ is stable by Lemma \ref{DA_as_monotone_operator}, the set of college positions filled under $\mu$ and $T^{*}(\mu')$ is identical by Theorem 5.12 of \cite{Roth/Sotomayor:90:TwoSidedMatching}. By the arguments in the proof of Lemma \ref{DA_as_monotone_operator}, either of the following two cases hold.

Case 1: Every college has the same number of matched students between $\mu$ and $\mu'$.

Case 2: One and only one college $j^{*}$ fills an additional position as we move from $\mu'$ to $\mu$, whereas all the other colleges have the same number of matched students between $\mu'$ and $\mu$. $\blacksquare$\medskip 

Lemma \ref{lem:taxonomy_1_vacancy_or_no_filled} immediately implies the following corollary on the cardinality of sets, $N_{j,1}$ and $N_{j,0}$ defined as
\begin{align}
	\label{N_{j1} N_{j0}}
	&		N_{j,1} \equiv\mu^{-1}(j)\backslash\mu'^{-1}(j)\text{ and }
	N_{j,0} \equiv\mu'^{-1}(j)\backslash\mu^{-1}(j),
\end{align}
where $\mu\in\mathcal{S}(N,M,\boldsymbol{u})$	and $\mu'\in\mathcal{E}_1(N,M,\boldsymbol{u})$.

\begin{corollary}\label{cor:relative_size_Nj0Nj1}  Let  $\mu\in\mathcal{S}(N,M,\boldsymbol{u})$	and $\mu'\in\mathcal{E}_1(N,M,\boldsymbol{u})$, and $N_{j,1}$ and $N_{j,0}$ be as defined in (\ref{N_{j1} N_{j0}}).
	
If Case 1 holds under Lemma \ref{lem:taxonomy_1_vacancy_or_no_filled}, then for every college $j\in M$,
	\begin{align}\label{vacancy_or_no_filled_samae_Size}
		|N_{j,1}|=|N_{j,0}|.
	\end{align}

If Case 2 holds under Lemma \ref{lem:taxonomy_1_vacancy_or_no_filled}, then for each $j\in M$,
	\begin{align}\label{eq:one_and_only_one_college_fills_extra_NJ}
		&|N_{j,1}|=|N_{j,0}|+1\{j=j^{*}\},
	\end{align}
where $j^*$ is the college in Lemma \ref{lem:taxonomy_1_vacancy_or_no_filled}.
\end{corollary}

\noindent \textbf{Proof: }
Let $j\in M$. Since $\mu^{-1}(j)$ and $\mu'^{-1}(j)$ are finite, 
\begin{align}
	& |N_{j,1}|=|\mu^{-1}(j)\backslash\mu'^{-1}(j)|=|\mu^{-1}(j)|-|\mu'^{-1}(j)\cap\mu^{-1}(j)|,\text{ and }\label{eq:setdiffcardin1}\\
	& |N_{j,0}|=|\mu'^{-1}(j)\backslash\mu^{-1}(j)|=|\mu'^{-1}(j)|-|\mu'^{-1}(j)\cap\mu^{-1}(j)|.\label{eq:setdiffcardin2}
\end{align}

First, suppose that we are in Case 1 under Lemma \ref{lem:taxonomy_1_vacancy_or_no_filled}. Then, $|\mu^{-1}(j)|=|\mu'^{-1}(j)|$, so that by (\ref{eq:setdiffcardin1}) and (\ref{eq:setdiffcardin2}) we have $|N_{j,1}|=|N_{j,0}|$. Next, suppose we are in Case 2 under Lemma \ref{lem:taxonomy_1_vacancy_or_no_filled}. Then $|N_{j,1}|=|N_{j,0}|+1$, if $j = j^*$, whereas if $j\in M\backslash \{j^{*}\}$, we have $|N_{j,1}|=|N_{j,0}|$ as before. $\blacksquare$\medskip

\section{Related One-To-One Markets}
When preferences of agents are strict (as is the case in our setup under our assumptions), a unique student optimal stable matching exists and can be realized through the Deferred Acceptance (DA) mechanism proposed by \cite{Gale/Shapley:1962}. 

The main result of this section is Lemma \ref{lemm:equilibration_theorem} below, which is essentially a many-to-one version of Theorem 5.2 of \cite{Blum/Roth/Rothblum:1997:JET} adapted to our setup.\footnote{A special case of the result also appears as the second item of Theorem 2.3 in \citet{Blum/Rothblum/2002}. } It says that applying iterations of $T$ in a given market to the student-optimal stable matching (SOSM) associated with the market altered to exclude any one student, we obtain the SOSM in the original market with the student included. To proceed, we introduce a notion of one-to-one
matching markets that are analogous to the many-to-one matching markets we have dealt
with so far. Our definitions follow Section 5.2 of \cite{Roth/Sotomayor:90:TwoSidedMatching}, but we summarize the main details here for convenience.

Given a many-to-one market $(N,M,\boldsymbol{u})$, we define the corresponding one-to-one market  $(N,\bar{M},\boldsymbol{\bar{u}})$ as follows. First, the set of colleges $\bar{M}$ in the one-to-one market is obtained by ``splitting'' each college $j\in M$ into $q_{j}$ positions, $c_{j,1},...,c_{j,q_{j}}$, where each position of $j$ has the same preferences over students as college $j$.  Students' preferences over the positions in the one-to-one market are such that each student $i$ prefers a position of college $j$ to a position of college $j'$ in the one-to-one market  if and only if $i$ prefers college $j$ to college $j'$ in the many-to-one market. Moreover, when comparing any two positions of the same college $j$, each student is assumed simply to prefer the position with the smaller index. (So, all students have the same preference ordering between positions in the same college.) For example, each student considers $c_{j,1}$ the best position of college $j$, $c_{j,2}$ to be the second best position of $j$, and so on.

Next, we define a matching $\bar{\mu}$ in a market $(N,\bar{M},\boldsymbol{\bar{u}})$ to be a
pair of maps $(\bar{\mu}_{N},\bar{\mu}_{M})$, $\bar{\mu}_{N}: N\rightarrow \bar{M}\cup\{0\}$, $\bar{\mu}_{M}:M\rightarrow \bar{N}\cup\{0\}$ such that for all $i\in N$ and $j\in M$, $\bar{\mu}_{N}(i)=j$ if and only if $\bar{\mu}_{M}(j)=i$. Under the assumption of strict preferences, we then obtain the following one-to-one correspondence between matchings for the market $(N,M,\boldsymbol{u})$ and matchings for the market  $(N,\bar{M},\boldsymbol{\bar{u}})$. 
A matching $\mu$ for market $(N,M,\boldsymbol{u})$ which matches college $j\in M$ with students $\mu^{-1}(j)$, corresponds to the matching $\bar{\mu}=(\bar{\mu}_{N},\bar{\mu}_{M})$ for market $(N,\bar{M},\boldsymbol{\bar{u}})$ in which the students  in $\mu^{-1}(j)$ are matched in the order they occur in the college's preferences, with the ordered positions of $j$ that appear in $\bar{M}$. Thus, if $i$ is college $j$'s most preferred student in $\mu^{-1}(j)$ then $\bar{\mu}_{N}(i)=c_{j,1}$ and $\bar{\mu}_{M}(c_{j,1})=i$, and so on. Following \cite{Roth/Sotomayor:90:TwoSidedMatching}, we call $(N,\bar{M},\boldsymbol{\bar{u}})$ the \bi{related market}. The notion of simple matchings in \citet{Sotomayor:1996:GEB} defined below is useful for our purpose.

\begin{definition} Given a matching $\bar{\mu}=(\bar{\mu}_{N},\bar{\mu}_{M})$ in 
	market $(N,\bar{M},\boldsymbol{\bar{u}})$, we say that
	
	(1) $\bar{\mu}$ is \bi{individually rational} if there is no $i\in N$ or $j\in\bar{M}$ such that $0\succ_{i}\bar{\mu}_{N}(i)$
	or $0\succ_{j}\bar{\mu}_{M}(j)$,
	
	(2) $\bar{\mu}$ is \bi{simple} if it is individually rational and any $(i,j)\in N\times\bar{M}$ such that $j\succ_{i}\bar{\mu}_{N}(i)$
	and $i\succ_{j}\bar{\mu}_{M}(j)$, satisfies that $\bar{\mu}_{N}(i)=0$,\footnote{That is,  $\bar{\mu}$ is simple if it is individually rational and any blocking pair, if it exists, involves an unmatched student in  $\bar{\mu}$.} and
	
	(3) $\bar{\mu}$ is \bi{1-simple} if it is: (i) simple and (ii) either $\bar{\mu}$ is stable or there exists one and only one student $i\in N$ who belongs to every blocking pair of $\bar{\mu}$.
	\par
\end{definition}\label{def:simple}
The following result and its proof are similar to Proposition 2.2 of \citet{Wu/Roth:2018:GEB}.

\begin{lemma}\label{lem:prop_2pt2_wu_and_roth} Let $\mu$ be a 1-envy-free matching in $(N,M,\boldsymbol{u})$ (in the sense of Definition \ref{def:college-envy-free}). Then its corresponding matching $\bar{\mu}$ in $(N,\bar{M},\boldsymbol{\bar{u}})$ is 1-simple. \end{lemma}

\noindent \textbf{Proof:} First we show that if $\mu$ is envy-free, then the corresponding matching is simple. Let $\mu$ be envy-free in $(N,M,\boldsymbol{u})$. Suppose by contradiction that its corresponding matching $\bar{\mu}$ is not simple in $(N,\bar{M},\boldsymbol{\bar{u}})$. We assume that $\bar{\mu}$ is individually rational, as otherwise the contradiction is immediate. Then there is a blocking pair $(i,j)\in N\times\bar{M}$
for $\bar{\mu}$ with $\bar{\mu}_{N}(i)=j'\neq0$. Suppose that $j$ and $j'$ are positions of distinct colleges in $M$.\footnote{Note that for such a blocking pair to exist, $j$ and $j'$ cannot be positions of the same college under $M$; if $j$ is the better position, then the college fills it with a preferred student; if $j$ is the worse position then $i$ does not prefer $j$ to $j'$.} Then we obtain the contradiction that $\mu$ is not envy-free, since $(i,j)$ is a blocking pair for $\mu$, yet $i$ is matched to a college under $\mu$.  

Now let $\mu$ be 1-envy-free in $(N,M,\boldsymbol{u})$. Suppose
by contradiction that its corresponding matching $\bar{\mu}$ is simple but not 1-simple in $(N,\bar{M},\boldsymbol{\bar{u}})$. Then there are more than one students forming a blocking pair to $\bar{\mu}$. Each of those students forms a blocking pair to $\mu$. This contradicts that $\mu$ is 1-envy-free. $\blacksquare$\medskip

Thus, whenever $\mu$ is stable and $\mu'$ is 1-envy free, $\bar{\mu}$ is stable and $\bar{\mu}'$ is 1-simple in the related market, by Lemma 5.6 of \citet{Roth/Sotomayor:90:TwoSidedMatching} and Lemma \ref{lem:prop_2pt2_wu_and_roth}. 

The following lemma is Lemma A.2 of \cite{Blum/Roth/Rothblum:1997:JET}, translated into our notation. The lemma is a version of Knuth's Decomposition Lemma (see Corollary 2.2 1 of \citet{Roth/Sotomayor:90:TwoSidedMatching}).

\begin{lemma}
	\label{lemm: decomp}
	Let $\bar{\mu}=(\bar{\mu}_{N},\bar{\mu}_{M})$
	and $\bar{\mu}'=(\bar{\mu}'_{N},\bar{\mu}'_{M})$ be the stable and
	1-simple matchings in $(N,\bar{M},\boldsymbol{\bar{u}})$ respectively. Suppose that $s'$ is a student who does not belong to any blocking pair for $\bar \mu'$, and $s$ is a student who does not belong to any blocking pair for $\bar \mu$. Then the following statements hold:
	
	(i) $\bar \mu_N(s') \succ_{s'} \bar \mu_N'(s') \equiv c'$ if and only if $s' \equiv \bar \mu_M'(c') \succ_{c'} \bar \mu_M(c')$.
	
	(ii) $\bar \mu_N'(s) \succ_{s} \bar \mu_N(s) \equiv c$ if and only if $s \equiv \bar \mu_M(c) \succ_{c} \bar \mu_M'(c)$.
\end{lemma}

We use this lemma to obtain the following result which could be viewed as an adaptation of Lemma 5.25 of \cite{Roth/Sotomayor:90:TwoSidedMatching}. The difference is that our setting involves two matchings where one is a 1-envy-free matching rather than a stable matching. This means that  the set of positions filled by colleges may not be identical between the two matchings. For the sake of full transparency, we provide detailed arguments in the proof. Let $\mu$ and $\mu'$ be matchings with  corresponding one-to-one matchings $\bar{\mu}=(\bar{\mu}_{N},\bar{\mu}_{M})$ and $\bar{\mu}'=(\bar{\mu}_{N}',\bar{\mu}_{M}')$. For any college $j\in M$ and college position $c\in \bar{M}$ of $j$, we write $\bar{\mu}_{M}(c)\succeq_{j}\bar{\mu}_{M}'(c)$ if either $\bar{\mu}_{M}(c)\succ_{j}\bar{\mu}_{M}'(c)$ or $\bar{\mu}_{M}(c)=\bar{\mu}_{M}'(c)$.

\begin{lemma}
	\label{lem:RS5pt25}
	Let $\mu\in\mathcal{S}(N,M,\boldsymbol{u})$
	and $\mu'\in\mathcal{E}_1(N,M,\boldsymbol{u})$, and let $\bar{\mu}=(\bar{\mu}_{N},\bar{\mu}_{M})$
	and $\bar{\mu}'=(\bar{\mu}'_{N},\bar{\mu}'_{M})$ be the stable and
	1-simple matchings corresponding to $\mu$ and $\mu'$ in the related
	one-to-one market. Suppose that for some college $j \in M$ and one of its positions $c$,
	\begin{align}
		\label{ineq2}
		\bar{\mu}_{M}(c) &\succ_{j}\bar{\mu}_{M}'(c).
	\end{align}
	Then, $\bar{\mu}_{M}(c')\succeq_{j}\bar{\mu}_{M}'(c')$
	for all the positions $c'$ of the college $j$.
\end{lemma}

\noindent \textbf{Proof:}
Case 1 under  Lemma \ref{lem:taxonomy_1_vacancy_or_no_filled} can be dealt with using the proof of Lemma 5.25 of \cite{Roth/Sotomayor:90:TwoSidedMatching} using Lemma \ref{lemm: decomp} in place of the decomposition lemma.

We focus on Lemma \ref{lem:taxonomy_1_vacancy_or_no_filled} under Case 2. Then the position vacant under $\mu'$ and filled under $\mu$ should be $c_{j,\ell}$ with $\ell = |\mu^{\prime -1}(j)|+1 = |\mu^{-1}(j)|$, and $\bar \mu_M(c_{j,\ell}) \succ_j \bar \mu_M'(c_{j,\ell}) = 0$ by the individual rationality of $\bar \mu$. Furthermore, if $|\mu^{-1}(j)|<q_{j}$, the positions of college $j$ must be vacant at indices $|\mu^{-1}(j)|+1,...,q_{j}$ under both $\bar{\mu}$ and $\bar{\mu}'$. 

To prove the lemma, we assume that (\ref{ineq2}) holds for some position $c$. Without loss of generality, assume that there exists $i \in \{1,...,|\mu^{-1}(j)|\}$ such that
\begin{align*}
	\bar{\mu}_{M}(c_{j,\ell}) &= \bar{\mu}_{M}'(c_{j,\ell}), \text{ for all } \ell = 1,...,i-1,\\
	\bar{\mu}_{M}(c_{j,i}) &\succ_j \bar{\mu}_{M}'(c_{j,i}).
\end{align*}
We show that $\bar{\mu}_{M}(c_{j,\ell})\succ_{j}\bar{\mu}_{M}'(c_{j,\ell})$ for all $\ell=i+1,...,|\mu^{-1}(j)|$.\footnote{The case where $\bar{\mu}_{M}(c_{j,\ell}) = \bar{\mu}_{M}'(c_{j,\ell}), \text{ for all } \ell = 1,...,i-1,$ and $\bar{\mu}_{M}(c_{j,i}) \prec_j \bar{\mu}_{M}'(c_{j,i})$ is excluded, because otherwise, we can use the same arguments to prove that $\bar{\mu}_{M}(c_{j,\ell})\prec_{j}\bar{\mu}_{M}'(c_{j,\ell})$ for all $\ell=i+1,...,|\mu^{-1}(j)|$, but this contradicts the assumption that (\ref{ineq2}) holds for some position $c$.} For this, we follow the arguments in the proof of Lemma 5.25 of \cite{Roth/Sotomayor:90:TwoSidedMatching}. Suppose by contradiction that for some $\ell=i,...,|\mu^{-1}(j)|-1$,
\begin{align}
	\label{statements1}
	&\bar{\mu}_{M}(c_{j,\ell})\succ_{j}\bar{\mu}_{M}'(c_{j,\ell}), \text{ and }\\ \label{statements2}
	&\bar{\mu}_{M}'(c_{j,\ell+1})\succeq_{j}\bar{\mu}_{M}(c_{j,\ell+1}).
\end{align}
Since $c_{j,\ell}$ with $\ell=i,...,|\mu^{-1}(j)|-1$ is filled both under $\bar \mu$ and $\bar \mu'$, we have $s_{\ell}'\equiv\bar{\mu}_{M}'(c_{j,\ell}) \in N$, and hence $s_{\ell}'$ does not belong to any blocking pairs of $\bar{\mu}'$ (because $\bar{\mu}'$ is 1-simple.) 

Since the lower-indexed positions are filled by better students at any 1-simple matching, we have from (\ref{statements2}):
\begin{align}
	\label{eq2}
	s_{\ell}'=\bar{\mu}_{M}'(c_{j,\ell}) & \succ_{j}\bar{\mu}_{M}'(c_{j,\ell+1})\succeq_{j}\bar{\mu}_{M}(c_{j,\ell+1}).
\end{align}

Next, since (\ref{statements1}) implies that $c_{j,\ell}=\bar \mu_{N}'(s_\ell')\ne \bar \mu_{N}(s_\ell')$, we must have either 
\begin{align}\label{slprime_preference}
c_{j,\ell} = \bar{\mu}_{N}'(s_{\ell}') \succ_{s_{\ell}'}\bar{\mu}_{N}(s_{\ell}')\text{ or }\bar{\mu}_{N}(s_{\ell}') \succ_{s_{\ell}'}\bar{\mu}_{N}'(s_{\ell}')= 	c_{j,\ell}.
\end{align}
However, the preference of $s_{\ell}'$ satisfies the first relation. To see this, suppose instead that it satisfies the second relation in (\ref{slprime_preference}). Then by Lemma \ref{lemm: decomp}(i), we must have $s'_{\ell}\equiv \bar \mu_M'(c_{j,\ell}) \succ_j \bar \mu_M(c_{j,\ell})$, which violates (\ref{statements1}).  We conclude that $c_{j,\ell}=\bar{\mu}_{N}'(s_{\ell}') \succ_{s_{\ell}'}\bar{\mu}_{N}(s_{\ell}')\ne c_{j,\ell+1}$, where the fact that $\bar{\mu}_{N}(s_{\ell}')\ne c_{j,\ell+1}$ is by (\ref{eq2}). This implies that
\begin{align}
	\label{eq}
	c_{j,\ell+1}\succ_{s_{\ell}'} \bar{\mu}_{N}(s_{\ell}'),
\end{align}
since $c_{j,\ell+1}$ immediately follows $c_{j,\ell}$ in the strict preference of $s_{\ell}'$ over all the positions in colleges.

From (\ref{eq}) and (\ref{eq2}), the student-college pair $(s_{\ell}',c_{j,\ell+1})$ blocks $\bar{\mu}$, contradicting the
stability of $\bar \mu$, and of $\mu$ via Lemma 5.6 of \citet{Roth/Sotomayor:90:TwoSidedMatching}. $\blacksquare$\medskip

The following lemma is a many-to-one version of Theorem A6 of \citet{Blum/Rothblum/2002}.

\begin{lemma}\label{lem:thmA6BRR} For any $\mu' \in\mathcal{E}_1(N,M,\boldsymbol{u})$, there is no stable matching $\mu''$ for market $(N,M,\boldsymbol{u})$ satisfying that $\mu''\ne T^{*}(\mu')$ and yet 
	\begin{align}
		T^{*}(\mu') \succsim \mu''\succsim  \mu'.
	\end{align}
\end{lemma}

\noindent \textbf{Proof: } The proof uses some additional notation. Given two many-to-one matchings $\mu_{1}$ and $\mu_{2}$ with  corresponding one-to-one matchings $\bar \mu_{1}=(\bar{\mu}_{1,N},\bar{\mu}_{1,M})$ and $\bar \mu_{2}=(\bar{\mu}_{2,N},\bar{\mu}_{2,M})$,   we write  $\bar{\mu}_{1}\geq\bar{\mu}_{2}$ to denote $\bar{\mu}_{1,M}(j)\succ_{j}\bar{\mu}_{2,M}(j)$ or $\bar{\mu}_{1,M}(j)=\bar{\mu}_{2,M}(j)$ for all colleges $j$.\footnote{Thus $\bar{\mu}_{1}\geq\bar{\mu}_{2}$ represents the statement that all colleges weakly prefer $\bar{\mu}_{1}$ to $\bar{\mu}_{2}$.}

Let us prove the lemma. Let $\mu = T^{*}(\mu')$ and let $\mu'$ be the matching in the lemma. Let $\bar \mu'$ be the 1-simple matching corresponding to $\mu'$. By Theorem A6 of \citet{Blum/Rothblum/2002}, the matching $\bar{\mu}=T^{*}(\bar{\mu}')$ is the college-worst stable matching weakly preferred by the colleges to $\bar{\mu}'$.\footnote{Recall Remark \ref{student_optimal_one_envy_free}. Note that when $\bar \mu'$ is an 1-simple matching in a market, it is a stable matching for the market once any student that forms a blocking pair is eliminated from the market.} We now argue that $\mu$ is the college-worst stable matching weakly preferred by colleges to $\mu'$.  
Suppose by contradiction that there is a stable many-to-one matching $\mu''$ satisfying $\mu\succsim\mu''\succsim\mu'$ with $\mu''\neq\mu$. We assume that $\mu\ne \mu'$ (as otherwise, the contradiction is immediate).  Since $\mu\ne \mu'$ and $\mu$ is stable, it follows by the properties of $T^{*}$ that $\mu'$ must be an unstable 1-envy free matching. Hence, $\mu'' \ne \mu'$. Next, let $\bar{\mu}''$ denote the stable one-to-one matching corresponding to $\mu''$. Since $\mu$ and $\mu''$ are stable with $\mu \neq \mu''$, it follows by Lemma 5.25 of \cite{Roth/Sotomayor:90:TwoSidedMatching} that $\bar{\mu}\geq\bar{\mu}''$.  Similarly, since $\mu''$ is stable and  $\mu'$ is 1-envy free with  $\mu''\succsim\mu'$ and $\mu''\neq \mu'$, we obtain $\bar{\mu}''\geq\bar{\mu}'$ from Lemma \ref{lem:RS5pt25}.    Hence, $\bar{\mu}\geq\bar{\mu}''\geq\bar{\mu}'$, contradicting the requirement that $\bar{\mu}$ is the college-worst stable matching weakly preferred to $\bar{\mu}'$.  $\blacksquare$\medskip

For the rest of the proofs, we take $\mu^{*}: N \times \mathbb{U} \rightarrow M'$ to be a matching mechanism such that for all $\boldsymbol{u} \in \mathbb{U}$, the matching $\mu^{*}(\cdot, \boldsymbol{u})$ is the SOSM for market $(N,M,\boldsymbol{u})$. For any $i \in N$, we take $\mu_{i}^*(\cdot,\boldsymbol{u}_{-i})$ to be the SOSM for market $(N\backslash\{i\},M,\boldsymbol{u}_{-i})$.

\begin{lemma} \label{lemm:equilibration_theorem} For any $\boldsymbol{u}\in\mathbb{U}$ and $i\in N$, we have
	\begin{align*}
		\mu^{*}(\cdot,\boldsymbol{u})= & T^*(g_i(\mu_i^*(\cdot,\boldsymbol{u}_{-i}))).
	\end{align*}
\end{lemma}

The proof of Lemma \ref{lemm:equilibration_theorem} draws on Corollary A7 of \citet{Blum/Rothblum/2002}. See
also Theorem 4.3 of \cite{Blum/Roth/Rothblum:1997:JET}, and Theorem 3.12 of \cite{Wu/Roth:2018:GEB}.\medskip

\noindent \textbf{Proof: } Since $\mu^{*}(\cdot,\boldsymbol{u})$ is the college-worst stable
matching by Corollary 5.30 of \cite{Roth/Sotomayor:90:TwoSidedMatching}, we have
\begin{align*}
	T^*(g_i(\mu_i^*(\cdot, \boldsymbol{u}_{-i}))) \succsim \mu^*(\cdot,\boldsymbol{u}).
\end{align*}
On the other hand, by Theorem 5.34 of \cite{Roth/Sotomayor:90:TwoSidedMatching}, $\mu^{*}(\cdot,\boldsymbol{u})\succsim g_{i}(\mu_{i}^*(\cdot,\boldsymbol{u}_{-i}))$, so that
\begin{align*}
	T^*(g_i(\mu_i^*(\cdot, \boldsymbol{u}_{-i}))) \succsim \mu^*(\cdot,\boldsymbol{u}) \succsim g_{i}(\mu_{i}^*(\cdot,\boldsymbol{u}_{-i})).
\end{align*}
By Lemma \ref{lem:thmA6BRR}, we must have $\mu^{*}(\cdot,\boldsymbol{u})=T^{*}(g_{i}(\mu_{i}^*(\cdot,\boldsymbol{u}_{-i})))$. $\blacksquare$\medskip

\section{Bounded Difference Condition for Stable Matching Mechanisms}

\subsection{Bounded Difference Condition for SOSM Mechanisms}
\begin{lemma}
	\label{lemm: Nj1 Nj0}
	Let for $j \in M$ and $i \in N$,
	\begin{align}
		\label{Nj1 Nj0}
		N_{j,1} \equiv\mu^{*-1}(j,\boldsymbol{u})\backslash\mu_{i}^{*-1}(j,\boldsymbol{u}_{-i}), \text{ and }
		N_{j,0} \equiv\mu_{i}^{*-1}(j,\boldsymbol{u}_{-i})\backslash\mu^{*-1}(j,\boldsymbol{u}).
	\end{align}
	Then for any $j$ with $|\mu_{i}^{*-1}(j,\boldsymbol{u}_{-i})|=q_{j}$,
	\begin{align}
		\left|N_{j,1}\right|=\left|N_{j,0}\right|\label{eq:equality-1}.
	\end{align}
\end{lemma}	

\noindent \textbf{Proof: } By Lemma \ref{DA_as_monotone_operator}, $|\mu_{i}^{*-1}(j,\boldsymbol{u}_{-i})|=q_{j}$ implies that $|\mu^{*-1}(j,\boldsymbol{u})|=q_{j}$, since all students in $\mu_{i}^{*-1}(j,\boldsymbol{u}_{-i})$ are acceptable to $j$ under
preferences $\boldsymbol{u}$ (i.e., a college $j$ with no vacancy unmatches a student at an iteration of the operator if and only if
college $j$ forms a student-maximal blocking pair with some other student). Thus, the desired result follows. $\blacksquare$\medskip

From here on, we will also write $T(\cdot;\boldsymbol{u})$ and $T^*(\cdot;\boldsymbol{u})$
when the distinction between the preferences in the underlying market is important.

\begin{lemma}
	\label{lem:change_at_college}  For each $j\in M$ and $i\in N$, we have that for any $\boldsymbol{u} = (\boldsymbol{v},\boldsymbol{w}) \in\mathbb{U}$ with $h(\boldsymbol{w}) =k$ for some $k = 0,1,...,n$,
	\begin{align*}
		\left|N_{j,1}\right| \leq 4(k\vee 1),\text{ and } \left|N_{j,0}\right| \leq  4(k\vee 1),
	\end{align*}
	where $N_{j,1}$ and $N_{j,0}$ are defined as in (\ref{Nj1 Nj0}) above and $h(\boldsymbol{w})$ as in (\ref{h(w)}) in the main text.
\end{lemma}

\noindent \textbf{Proof}: Let $\mu_{0},\mu_{1},...,\mu_{r'}$ be matchings, generated  by $\mu_{r}=T(\mu_{r-1}; \boldsymbol{u})$ for each $r=1,...,r'$, with $\mu_{0}=g_{i}(\mu_{i}^*(\cdot,\boldsymbol{u}_{-i}))$. By Lemma \ref{DA_as_monotone_operator}, $r'$
is finite and $\mu_{r'}=T^{*}(\mu_{0}; \boldsymbol{u})\in\mathcal{S}(N,M,\boldsymbol{u})$.
However, by Lemma \ref{lemm:equilibration_theorem}, we also have
that $T^{*}(\mu_{0}; \boldsymbol{u})(\cdot)=\mu^*(\cdot,\boldsymbol{u})$.
Note that by the definition of $T$, any $j$ with $|\mu_{i}^{*-1}(j,\boldsymbol{u}_{-i})|<q_{j}$
can be involved in at most one iteration of $T$.\footnote{Moreover, by Lemma \ref{lemm:equilibration_theorem}, there is at most one such vacancy-filling college, as any iteration of $T$ that fills a vacancy is a stable matching, and hence, must be the final iteration of $T$.}  Hence the first
bound holds trivially with the value of one for any college with a
vacancy under $\mu_{i}^*(\cdot,\boldsymbol{u}_{-i})$. The second bound
holds with the value of zero for any such college. Therefore, for
the remainder of the proof, we show that for every $j\in M$ with $|\mu_{i}^{*-1}(j,\boldsymbol{u}_{-i})|=q_{j}$,
\begin{align}\label{eq:displaced_and_new}
	\left|N_{j,1}\right|\leq 4k^{*} \text{ and }\left|N_{j,0}\right| & \leq 4k^{*},
\end{align}
where we write $k^{*}\equiv k\vee 1$. In showing this, we use the following implication of Lemma \ref{DA_as_monotone_operator}: for each $j\in M$,
\begin{align}
	i_{1}\succ_{j}i_{2},\text{ for all }(i_{1},i_{2})\in N_{j,1}\times N_{j,0}.\label{eq:displaced_student_is_worse_than_any_new_student-1}
\end{align}

First, by Lemma \ref{lemm: Nj1 Nj0}, $|N_{j,1}| = |N_{j,0}|$. To prove (\ref{eq:displaced_and_new}), suppose by contradiction that $|N_{j,0}|= t$ and $|N_{j,1}|= t$ for some integer $t>4k^{*}$. Let us enumerate
\begin{align*}
	N_{j,1} = \{a_{1},a_{2},...,a_{t-1},a_{t}\}, \text{ and } N_{j,0} = \{b_{1},b_{2},...,b_{t-1},b_{t}\},
\end{align*}
so that a smaller index indicates that the change to the match of the student with the index took place at an earlier iteration of $T$.  First, we establish the following three facts.\medskip

\noindent \textbf{Fact 1: } For each $s=1,...,4k^{*}+1$, $a_{s} \succ_j b_{4k^{*}+1} \succ_j ... \succ_j b_1$.\medskip

\noindent \textbf{Proof: } The ordering among the $b_{s}$'s is by the definition of $T$ (i.e., students less preferred by $j$ are dropped in earlier iterations of $T$).\footnote{Note that by transitivity of college preferences, if a student $i'$ is unmatched by a college $j'$ on some iteration of $T$,  $i'$ can never be rematched to $j'$ on a later iteration of $T$. This also implies that once a student $a_{s}\in N_{j,1}$ is matched to college $j$ on some iteration of $T$, $a_{s}$ is never unmatched from college $j$ on a subsequent iteration.} The fact that college $j$ prefers any of the $a_{s}$'s to any of the $b_{s}$'s is by  (\ref{eq:displaced_student_is_worse_than_any_new_student-1}). $\square$ \medskip

For each $s=1,...,k^{*}$, let $c_s^b$ denote the student unmatched from some college (not $j$) which then matches $b_s$, and let $c_{s}^a$ denote the student matched to some college (not $j$) which then unmatches $a_{s+1}$ (before $a_{s+1}$ matches with college $j$.) Then the following fact is a consequence of Lemma \ref{DA_as_monotone_operator}.\medskip 

\noindent \textbf{Fact 2: } For each $s=1,...,k^{*}$,

(i) $c_{s}^b$ is not ranked higher than $b_{s}$ by more than $k^{*}$ positions under $\succ_{j}$, and

(ii) $c_{s}^a$ is not ranked lower than $a_{s+1}$ by more than $k^{*}$ positions under $\succ_{j}$.\medskip

\noindent \textbf{Proof: } Recall that by the condition that $h(\boldsymbol{w}) = k$, if the rank difference between two students is more than $k^{*}$ in some college's preference, every college agrees with the ranking between the two students. Hence violation of (i) or (ii) implies that for some $s=1,...,k^{*}$, some college $j'$ is made worse off in an iteration of $T$, violating Lemma \ref{DA_as_monotone_operator}. $\square$\medskip

For notational brevity, we will occasionally write $a\succ_j b \succ_j c$ as $abc$ in the proof of the following fact. \medskip

\noindent \textbf{Fact 3:} For each $s=1,...,k^{*}$, 

(i) $c_{s}^{b}\ne c_{s}^{a}$, and 

(ii) $c_{s}^{a}b_{3k^{*}+1}b_{3k^{*}}...b_{k^{*}+s+1}c_{s}^{b}$.\medskip

\noindent \textbf{Proof:} Note that by Fact 1, we immediately have $b_{3k^{*}+1}....b_{2}b_{1}$. We begin by showing (i).   Suppose by contradiction that $c_{s}^{a}=c_{s}^{b}\equiv c_{s}$ for some $s=1,...,k^{*}$. We argue that Fact 2 is violated. First, suppose that $c_{s}\succ_{j}b_{k^{*}+s+1}$.  Hence, 
\begin{align*}\label{f3csacsbi}
	c_{s}b_{k^{*}+s+1}b_{k^{*}+s}...b_{s+1}b_{s}.
\end{align*}
Since $c_{s}$ is ranked higher than $b_{s}$ by more than $k^{*}$ positions under $\succ_{j}$, this violates Fact 2(i). Now suppose that $b_{k^{*}+s+1}\succ_{j}c_{s}$. Therefore, 
\begin{align}
	a_{s+1}b_{4k^{*}+1}...b_{k^{*}+s+1}c_{s},
\end{align}
by Fact 1. However since $a_{s+1}$ is ranked higher than $c_{s}$ by more than $k^{*}$, this violates Fact 2(ii). Note also that we must have $c_{s}\ne b_{k^{*}+s+1}$, as otherwise, $c_{s}=b_{k^{*}+s+1}$, so that $b_{k^{*}+s+1}$ is the student unmatched by a college $j'\ne j$ that then matches $b_{s}$. This, however, violates Lemma \ref{DA_as_monotone_operator}, since the fact that $b_{k^{*}+s+1}$ is more than $k^{*}$ positions higher than $b_{s}$ under $\succ_{j}$ implies that college $j'$ is made worse off in the iteration of $T$ in which it matches $b_{s}$ and unmatches $b_{k^{*}+s+1}$. Since the preferences are strict, we arrive at a contradiction regarding the college $j$'s preference ordering between $b_{k^*+ s + 1}$ and $c_s$. Hence, the statement (i) in Fact 3 follows.  

We now show (ii). Fix any $s=1,...,k^{*}$.  By Fact 3(i), $c_{s}^{b}\ne c_{s}^{a}$.  First, we must have $c_{s}^{a}\succ_jb_{3k^{*}+1}$, as otherwise  $a_{s+1}b_{4k^{*}+1}...b_{3k^{*}+1}c_{s}^{a}$ by Fact 1, which violates Fact 2(ii). Lastly, we must have $c_{s}^b\ne b_{k^{*}+s+1}$, as otherwise we violate  Lemma \ref{DA_as_monotone_operator} as we just saw in the proof of (i). We also must have $b_{k^{*}+s+1}\succ_j c_{s}^{b}$, as otherwise, $c_{s}^{b}b_{k^{*}+s+1}...b_{s}$ by Fact 1 which violates Fact 2(i). Thus we have shown (ii). $\square$ \medskip

Now we are ready to complete the proof of the lemma. It is convenient to begin the recursive argument that follows by letting   $c_{s,0}:=c_{s}^{b}$ for each $s=1,...,k^{*}$.  By Fact 3, we have 
\begin{align}\label{step1_recursionsosm}
	\{c_{1}^{a},...,c_{k^{*}}^{a}\}b_{3k^{*}+1}...b_{2k^{*}+1}\{c_{1,0},...,c_{k^{*},0}\}.
\end{align}
Let $c_{s_1,0}$ be the worst student among $c_{1,0},...,c_{k^{*},0}$ according to $\succ_{j}$.  Let $c_{s_1,1}$ be the student unmatched by the college that then matches $c_{s_1,0}$. Then we must have $b_{2k^{*}+1}\succ_{j}c_{s_1,1}$, as otherwise the college that matches $c_{s_1,0}$ and unmatches $c_{s_1,1}$ is made worse off, violating Lemma \ref{DA_as_monotone_operator}. 

Next, we let $c_{s,1}:=c_{s,0}$ for all $s\in \{1,...,k^{*}\}\backslash\{ s_1\}$. By (\ref{step1_recursionsosm}) and $b_{2k^{*}+1}\succ_{j}c_{s_1,1}$, we must have
\begin{align}\label{step2_recursion}
	\{c_{1}^{a},...,c_{k^{*}}^{a}\}b_{3k^{*}+1}...b_{2k^{*}+1}\{c_{1,1},...,c_{k^{*},1}\}.
\end{align}
Let $c_{s_2,1}$ be the worst student among $c_{1,1},...,c_{k^{*},1}$ according to $\succ_{j}$. Let $c_{s_2,2}$ denote the student unmatched by the college that matches student $c_{s_2,1}$. By Lemma \ref{DA_as_monotone_operator}, this student must again satisfy $b_{2k^{*}+1}\succ_{j}c_{s_2,2}$.  

Now we denote $c_{s,2}:=c_{s,1}$ for all $s\in \{1,...,k^{*}\}\backslash\{s_2\}$. The next displaced student in the sequence must again be below $b_{2k^{*}+1}$. By  continuing to displace students recursively in this fashion, we find that we can never displace any student above $b_{2k^{*}+1}$. In particular, it follows that for any $s,s'\in \{1,...,k^{*}\}$,  we cannot have that student $c_{s}^{a}$ is displaced after student $c_{s'}^{b}$ under the iterations of $T$ that generate the sets $N_{j,1}$ and $N_{j,0}$ for college $j$.

Now fix any  $s=1,...,k^{*}$. Consider the iteration of $T$ on which student $a_{s}$ is matched to college $j$ (so that $a_{s+1}$ is not yet matched to $j$).  By Lemma  \ref{DA_as_monotone_operator}, we must have that either: (i) $b_{s}$  is unmatched by college $j$ on the same iteration of $T$ on which $a_{s}$ is matched to $j$, or (ii) $b_{s}$ is unmatched by college $j$ on a previous iteration of $T$ (before $a_{s}$ is matched to $j$). In either case, it follows that $c_{s}^{b}$ must be displaced before $c_{s}^{a}$, so that $a_{s+1}$ can then be matched with college $j$. By our previous arguments, however, we know that $c_{s}^{b}$ cannot be displaced before $c_{s}^{a}$ under the iterations of $T$ that ultimately generate the sets $N_{j,1}$ and $N_{j,0}$ for college $j$. Hence, we cannot have $a_{s+1}\in N_{j,1}$ for any $s=1,...,k^{*}$.  Therefore, we have shown that if  $|N_{j,0}|= t$ and $|N_{j,1}|= t$ for some integer $t>4k^{*}$, then Lemma \ref{DA_as_monotone_operator} is violated. $\blacksquare$\medskip

\begin{corollary}\label{cor:students_changing}
	Let $\boldsymbol{u} = (\boldsymbol{v},\boldsymbol{w}) \in\mathbb{U}$ be the preference profiles with $h(\boldsymbol{w}) =k$ for some $k = 0,1,...,n$. Then for each $i\in N$ and $\boldsymbol{u}\in\mathbb{U}$, we
	have 
	\begin{align*}
		|\{i'\in N\backslash\{i\}:\mu^*(i',\boldsymbol{u})\neq\mu_{i}^*(i',\boldsymbol{u}_{-i})\}|\leq 4m(k\vee1)+1.
	\end{align*}
	
\end{corollary}

\noindent \textbf{Proof: } By Lemma \ref{lemm:equilibration_theorem}, $\mu^*(\cdot,\boldsymbol{u})$ is obtained from $\mu_{i}^*(\cdot,\boldsymbol{u}_{-i})$ through the iterations of $T$. Observe that at most one student who is matched to a college under
$\mu_i^*(\cdot,\boldsymbol{u}_{-i})$ can be left unmatched as a result of the
iterations of $T$ process. That is, the set 
\[
A_i=\left\{i'\in N:\mu^*(i',\boldsymbol{u})=0\text{ and }i'\in\mu_{i}^{*-1}(j,\boldsymbol{u}_{-i})\text{ for some }j\text{\ensuremath{\in M}}\right\},
\]
is either a singleton, say, $\{i^{*}\}$, for some $i^{*}\in N\backslash\{i\}$, or an empty set. 
A student $i'\in N\backslash\{i,i^{*}\}$ satisfies $\mu^*(i',\boldsymbol{u})\neq\mu_{i}^*(i',\boldsymbol{u}_{-i})$,
if and only if, for exactly one college, say $j$, the student is a member of $\mu^{*-1}(j,\boldsymbol{u})$
but not $\mu_{i}^{*-1}(j,\boldsymbol{u}_{-i})$. Thus for each student $i'\in N\backslash\{i\}$ we have that
\begin{align*}
	1\{\mu^*(i',\boldsymbol{u})\neq\mu_{i}^*(i',\boldsymbol{u}_{-i})\}= \sum_{j\in M}1\{i'\in\mu^{*-1}(j,\boldsymbol{u})\backslash\mu_{i}^{*-1}(j,\boldsymbol{u}_{-i})\}+1\{i'=i^{*}\}.
\end{align*}
Therefore, summing over $i'\in N\backslash\{i\}$, we have 
\begin{align*}
	& |\{i'\in N\backslash\{i\}:\mu^*(i',\boldsymbol{u})\neq\mu_{i}^*(i',\boldsymbol{u}_{-i})\}|\\
	&\le \sum_{j\in M}|\mu^{*-1}(j,\boldsymbol{u})\backslash\mu_{i}^{*-1}(j,\boldsymbol{u}_{-i})|+\sum_{i'\in N\backslash\{i\}}1\{i'=i^{*}\}\le  4m(k\vee1)+1,
\end{align*}
where we used Lemma \ref{lem:change_at_college} for the last bound.  $\blacksquare$\medskip

\begin{lemma}\label{lemm:BDC sosm} Choose any $k=0,1,2,...,n$, and any $(\boldsymbol{u}',\boldsymbol{u})$ such that $\boldsymbol{u}$ and $\boldsymbol{u}'$ are generated by $(\tilde s, \tilde z) \in \mathcal{\tilde S} \times \mathcal{\tilde Z}$ and $(\tilde s', \tilde z) \in \mathcal{\tilde S} \times \mathcal{\tilde Z}$, where $(\tilde s,\tilde z)$ and $(\tilde s',\tilde z)$ satisfy the conditions (a),(b) and (c) in Lemma \ref{lemm:bounded_difference_condition} in the main text, with the chosen $k$.
	
Then, for any $j  \in M'$,
	\begin{align}
		\label{eq:5pt1ii2-1-1-2}
		&\left|\left\{i'\in N:1\{\mu^{*}(i',\boldsymbol{u}) = j\} \ne  1\{\mu^{*}(i',\boldsymbol{u}') = j\}\right\} \right|\leq 16(k\vee1) +1, \text{ and } \\\notag
		&\left|\{i' \in N:\mu^{*}(i',\boldsymbol{u})\neq\mu^*(i',\boldsymbol{u}')\}\right|\leq 8m(k\vee1)+3.
	\end{align}
\end{lemma}

\noindent \textbf{Proof}:
We begin by showing the second bound in (\ref{eq:5pt1ii2-1-1-2}). Choose $(\boldsymbol{u}',\boldsymbol{u})$ as in the lemma. Let $i$ be the student such that $\tilde s_i \ne \tilde s_i'$ in the condition (a) in Lemma \ref{lemm:bounded_difference_condition}. Then, we have
\begin{align*}
	\boldsymbol{u}_{-i} = \boldsymbol{u}_{-i}',
\end{align*}
by Assumption \ref{assump: index utility2}, because the elimination of student $i$ in the market does not alter the preference ordering between other students by any college. Hence
\begin{align}
	\left|\{i'\in N:\mu^*(i',\boldsymbol{u})\neq\mu^*(i',\boldsymbol{u}')\}\right| & =\sum_{i'\in N}1\{\mu^*(i',\boldsymbol{u})\ne\mu^*(i',\boldsymbol{u}')\}\label{eq:triangle_ineq}\\ \notag
	&\le \sum_{i'\in N}1\{\mu^*(i',\boldsymbol{u})\ne\mu^*(i',\boldsymbol{u}'),i' \ne i\} + 1\\
	& \leq \sum_{i'\in N}1\{\mu^*(i',\boldsymbol{u})\ne \mu_{i}^*(i',\boldsymbol{u}_{-i}), i' \ne i\}\nonumber \\
	& \quad +\sum_{i'\in N}1\{\mu_{i}^*(i',\boldsymbol{u}_{-i})\ne\mu^*(i',\boldsymbol{u}'), i' \ne i\} + 1.\nonumber 
\end{align}
By Corollary \ref{cor:students_changing},
\begin{align*}
	\sum_{i'\in N}1\{\mu^*(i',\boldsymbol{u})\ne \mu_{i}^*(i',\boldsymbol{u}_{-i}), i' \ne i\} & \leq|\{i'\in N\backslash\{i\}:\mu^*(i',\boldsymbol{u})\neq\mu_{i}^*(i',\boldsymbol{u}_{-i})\}|\\
	& \leq 4m(k\vee1)+1.
\end{align*}
Since the above bound is uniform over $\boldsymbol{u}\in\mathbb{U}$ and $\boldsymbol{u}_{-i} = \boldsymbol{u}_{-i}'$, the same bound applies to the last sum in (\ref{eq:triangle_ineq}). Thus we conclude that 
\begin{align*}
	\left|\{i'\in N:\mu^*(i',\boldsymbol{u})\neq\mu^*(i',\boldsymbol{u}')\}\right|  \leq8 m(k\vee 1) + 3,
\end{align*}
establishing the second bound in the lemma.

As for the first bound in (\ref{eq:5pt1ii2-1-1-2}), we again choose $(\boldsymbol{u}',\boldsymbol{u})$ as in the lemma so that $\boldsymbol{u}'$ and $\boldsymbol{u}$ differ by the quality of one student, $i$. Then for any $j\in M'$ we have
\begin{align}
	& \sum_{i'\in N} \left|1\{\mu^*(i',\boldsymbol{u})=j\}-1\{\mu^*(i',\boldsymbol{u}')=j\} \right|\nonumber \\
	& \le \sum_{i'\in N}1\{\mu^*(i',\boldsymbol{u})=j\}1\{\mu^*(i',\boldsymbol{u}')\neq j\}1\{i'\neq i\}\label{eq:college_j_bound}\\
	& +\sum_{i'\in N}1\{\mu^*(i',\boldsymbol{u})\neq j\}1\{\mu^*(i',\boldsymbol{u}')=j\}1\{i'\neq i\}+1\nonumber.
\end{align}
By multiplying each summand with $i'$ in the first sum on the right hand side of (\ref{eq:college_j_bound})
by $1=1\{\mu_{i}^*(i',\boldsymbol{u}_{-i})\neq j\}+1\{\mu_{i}^*(i',\boldsymbol{u}_{-i})=j\}$, we bound this first sum by 
\begin{align}
	\label{college_j_bound22}
	& \sum_{i'\in N}1\{\mu^*(i',\boldsymbol{u})=j\}1\{\mu_{i}^*(i',\boldsymbol{u}_{-i})\neq j\}1\{i'\neq i\}\\ \notag
	& +\sum_{i'\in N}1\{\mu^*(i',\boldsymbol{u}')\neq j\}1\{\mu_{i}^*(i',\boldsymbol{u}_{-i})=j\}1\{i'\neq i\}\leq  8(k\vee 1).
\end{align}
Note that if $j\in M$, the last inequality follows immediately by Lemma \ref{lem:change_at_college}. If, on the other hand,  $j=0$, then the last inequality follows by Lemma \ref{lemm:equilibration_theorem} and the definition of $T$.\footnote{To see this, consider the two terms on the left-hand side of (\ref{college_j_bound22}) in the case of $j=0$. The first term has an upper bound of 1, since at most one student $i'\in N\backslash\{i\}$ matched to some college under $\mu_{i}^{*}(\cdot,\boldsymbol{u}_{-i})$  becomes unmatched in the iterations of  $T(\cdot,\boldsymbol{u})$ that yield $\mu^{*}(\cdot,\boldsymbol{u})$ from $\mu_{i}^{*}(\cdot,\boldsymbol{u}_{-i})$. The second term is equal  to zero, since no student in $i'\in N\backslash\{i\}$  unmatched under $\mu_{i}^{*}(\cdot,\boldsymbol{u}_{-i})$ is matched in the iterations of $T(\cdot,\boldsymbol{u}')$ that yield  $\mu^{*}(\cdot ,\boldsymbol{u}')$ from $\mu_{i}^{*}(\cdot,\boldsymbol{u}_{-i}).$ } 

Since the same bound of $8(k\vee1)$ also holds for the second sum on the right hand side of (\ref{eq:college_j_bound}),
we conclude that 
\begin{align*}
	\left|\left\{ i'\in N:1\{\mu^*(i',\boldsymbol{u})=j\}\ne1\{\mu^*(i',\boldsymbol{u}')=j\}\right\} \right| & \leq 16(k\vee1)+1.
\end{align*}
$\blacksquare$ 

\subsection{Bounding the Distance Between SOSM and an Arbitrary Stable Matching}
 
For $\boldsymbol{u}\in \mathbb{U}$, and any matching $\mu: N \rightarrow M'$, we define
\begin{align*}	
	N^{H}(\mu)&\equiv\{i\in N:\mu(i)\ne \mu^{*}(i,\boldsymbol{u})\},\\
	N^{*}_{j,1}(\mu)&\equiv \mu^{-1}(j)\backslash  \mu^{*-1}(j,\boldsymbol{u}),\text{ and }
	N^{*}_{j,0}(\mu) \equiv \mu^{*-1}(j,\boldsymbol{u})\backslash \mu^{-1}(j).
\end{align*}
The following is a straightforward consequence of Theorem 5.12 of \cite{Roth/Sotomayor:90:TwoSidedMatching}. 

\begin{corollary}\label{basic_properties_student_optimal_to_other_stable} 
	
(i) For any $\boldsymbol{u}\in \mathbb{U}$ and any $\mu \in \mathcal{S}(N,M,\boldsymbol{u})$, 
	\begin{align}\label{i_changes_colleges_if_and_only_if}
    N^{H}(\mu) = \bigcup_{j_1, j_2 \in M: j_1 \ne j_2} \left(N^{*}_{j_{1},0}(\mu)\cap N^{*}_{j_{2},1}(\mu) \right).
   \end{align}

(ii) For any $\boldsymbol{u}\in \mathbb{U}$, any $\mu \in \mathcal{S}(N,M,\boldsymbol{u})$, and for any college $j\in M$,
\begin{align}\label{cardinality_constraint}
	|N^{*}_{j,1}(\mu)|=|N^{*}_{j,0}(\mu)|.
\end{align}
\end{corollary}

\noindent \textbf{Proof: } Let us first prove (i). Note that since both $\mu(\cdot)$ and $\mu^{*}(\cdot,\boldsymbol{u})$ are stable matchings under $\boldsymbol{u}$, the set of students matched to some college must be identical across the two matchings by Theorem 5.12 of \cite{Roth/Sotomayor:90:TwoSidedMatching}. The desired result comes from this immediately.

We now prove (ii).  By Theorem 5.12 of \cite{Roth/Sotomayor:90:TwoSidedMatching}, the set of filled positions must be identical across the stable matchings, $\mu(\cdot)$ and $\mu^{*}(\cdot,\boldsymbol{u})$.  Hence, 
\begin{align}
	|\mu^{-1}(j)|=|\mu^{*-1}(j,\boldsymbol{u})|\text{ for all }j\in M.
\end{align}
Thus, (ii) follows as in the proof of Corollary \ref{cor:relative_size_Nj0Nj1}. $\blacksquare$

\begin{corollary}\label{other_stable_matching_always_Better}
For any $\boldsymbol{u}\in \mathbb{U}$, any $\mu \in \mathcal{S}(N,M,\boldsymbol{u})$, and for any college $j\in M$,
	\begin{align}\label{eq:college_always_likes_students_at_other_stable_matching_better}
		i_{1}\succ_{j} i_{2} \text{ for any }i_{1}\in N^{*}_{j,1}(\mu) \text{ and }i_{2}\in N^{*}_{j,0}(\mu).
	\end{align}
\end{corollary}

\noindent \textbf{Proof:} Since $\mu^{*}(\cdot,\boldsymbol{u})$ is the college-worst stable
matching by Corollary 5.30 of \cite{Roth/Sotomayor:90:TwoSidedMatching}, we have $\mu(\cdot)\succsim\mu^{*}(\cdot,\boldsymbol{u})$ for any stable matching $\mu$. 
Therefore, the result follows from Lemma 5.25 of \cite{Roth/Sotomayor:90:TwoSidedMatching}.\footnote{Note that if $\mu(\cdot)=\mu^{*}(\cdot,\boldsymbol{u})$ then  $N^{*}_{j,1}(\mu) $ and $N^{*}_{j,0}(\mu) $ are both empty and  (\ref{eq:college_always_likes_students_at_other_stable_matching_better}) holds trivially. If $\mu(\cdot)\ne \mu^{*}(\cdot,\boldsymbol{u})$, then $\mu(\cdot)\succsim\mu^{*}(\cdot,\boldsymbol{u})$ implies that we must have $\bar{\mu}_{M}(c)\succ_{j}\bar{\mu}^{*}_{M}(c)$ at some position $c$ of college $j$, where $\bar{\mu}=(\bar{\mu}_{N},\bar{\mu}_{M})$ and $\bar{\mu}^{*}=(\bar{\mu}_{N}^{*},\bar{\mu}_{M}^{*})$ denote the stable-matchings corresponding to $\mu(\cdot)$ and $\mu^{*}(\cdot,\boldsymbol{u})$ in the related one-to-one market. Thus, we must have  (\ref{eq:college_always_likes_students_at_other_stable_matching_better}), since $\bar{\mu}_{M}(c)\succ_{j}\bar{\mu}^{*}_{M}(c)$ at some position $c$ of college $j$ implies that $\bar{\mu}_{M}(c')\succeq_{j}\bar{\mu}^{*}_{M}(c')$ for all positions $c'$ of college $j$ by Lemma 5.25 of \cite{Roth/Sotomayor:90:TwoSidedMatching}.} $\blacksquare$\medskip 

Let $\boldsymbol{u}\in \mathbb{U}$ and $\mu \in \mathcal{S}(N,M,\boldsymbol{u})$. It is helpful to formalize the notion of ``displacement'' of one student by another as we move from $\mu^{*}(\cdot, \boldsymbol{u})$ to $\mu(\cdot)$. Given a college $j\in M$, a student $i\in N$, and a set $A\subset N$, let
\begin{align}
r_{ji}(A)\equiv\sum_{i' \in A}1\{i'\succ_{j} i\}+1.
\end{align}
Thus, $r_{ji}(A)$ denotes the rank of student $i$ in the set $A$ according to $\succ_{j}$.\footnote{For example, if $A=\{i_{1},i_{2},i_{3}\}$ and $i_{1}\succ_{j}i_{3}\succ_{j}i_{2}$, then $ r_{ji_{3}}(A)=2$.}  Given two students $a,b\in N^{H}(\mu) $ and a college $j\in M$, we write $a\vartriangleright_{j}b$ if and only if 
\begin{align}
a\in N^{*}_{j,1}(\mu),\text{ } b\in  N^{*}_{j,0}(\mu) \text { and }r_{ja}(N^{*}_{j,1}(\mu))=r_{jb}(N^{*}_{j,0}(\mu)).
\end{align}
In this case, we say that student $a$ \bi{displaces} student $b$ from college $j$ (equivalently, student $b$ \bi{is displaced by} student $a$ from college $j$) as we move from $\mu^{*}$ to $\mu$.  For any two students $a,b\in N^{H}(\mu) $, we write $a\vartriangleright b$  if and only if $a\vartriangleright_{j}b$ for some $j\in M$ and say that student $a$ \bi{displaces} student $b$ (equivalently, student $b$ \bi{is displaced by} student $a$). For any students $a,b\in N^{H}(\mu) $ (not necessarily distinct), we also write $a\ntriangleright b$ if $a$ does not displace $b$. Similarly, we write $a\ntriangleright_{j} b$ if we wish to specify that $a$ does not displace college $b$ from college $j$.

We highlight some useful properties of $\vartriangleright$. First, $\vartriangleright$ must satisfy $a\ntriangleright a$ for any student $a\in N^{H}(\mu) $, since $N^{*}_{j,1}(\mu) \cap N^{*}_{j,0}(\mu) =\varnothing$ for any $j\in M$ from definitions.   Hence, $\vartriangleright$ is an irreflexive binary relation on $N^{H}$. In addition, we must also have  $b\ntriangleright_{j} a$ for any distinct pair of students $a,b\in N^{H}(\mu)$ satisfying that $a\vartriangleright_{j} b$, which again follows by $N^{*}_{j,1}(\mu) \cap N^{*}_{j,0}(\mu) =\varnothing$. Note also that we must have $a\succ_{j}b$ whenever  $a\vartriangleright_{j} b$, which is a consequence of Corollary \ref{other_stable_matching_always_Better}.

\medskip

\begin{lemma}\label{every_student_displaces_and_is_displaced}
For any $\boldsymbol{u}\in \mathbb{U}$, any $\mu \in  \mathcal{S}(N,M,\boldsymbol{u})$, and any $i_{1}\in N^{H}(\mu)$, we have the following.

\noindent (i) Student $i_{1}$ is displaced by one and only one student, say, $i_{0}$, and student $i_{1}$ displaces one and only one student, say, $i_{2}$.\\
\noindent (ii)  Let $i_{0}$ and $i_{2}$ be the students in (i), and suppose that $i_{0}\ne i_{2}$. Then $i_{0},i_{1},i_{2}$ are distinct students in $N^{H}$ satisfying 
\begin{align}
	i_{0}\vartriangleright_{j_{1}} i_{1}\vartriangleright_{j_{2}} i_{2},
\end{align}
for some distinct colleges $j_{1},j_{2}\in M$.

\end{lemma}
\noindent \textbf{Proof: } (i) Since $i_{1}\in N^{H}(\mu)$, we have by Corollary \ref{basic_properties_student_optimal_to_other_stable}(i) that   $i_{1}\in N^{*}_{j_{1},0}(\mu)$ and $i_{1}\in N^{*}_{j_{2},1}(\mu)$ for two distinct colleges, $j_{1},j_{2}\in M$. By Corollary \ref{basic_properties_student_optimal_to_other_stable}(ii), the sets $N^{*}_{j_{1},0}(\mu)$ and $N^{*}_{j_{1},1}(\mu)$ have the same cardinality. Since $N^{*}_{j_{1},0}(\mu)$ and $N^{*}_{j_{1},1}(\mu)$ are also disjoint, we conclude that there must be a single student $i_{0}\ne i_{1}$ belonging to the set $N^{*}_{j_{1},1}(\mu)$ with the same rank in the set $N^{*}_{j_{1},1}(\mu)$ according to $\succ_{j_{1}}$ that $i_{1}$ has in the set $N^{*}_{j_{1},0}(\mu)$ according to $\succ_{j_{1}}$.  By applying Corollary \ref{basic_properties_student_optimal_to_other_stable}(ii) and the same logic, we can also find a single student $i_{2}\ne i_{1}$ belonging to the set $N^{*}_{j_{2},0}(\mu)$ with the same rank in $N^{*}_{j_{2},0}(\mu)$ according to  $\succ_{j_{2}}$  that $i_{1}$ has in the set $N^{*}_{j_{2},1}(\mu)$ according to  $\succ_{j_{2}}$. So we have (i).  

(ii) Let $i_{0}$ and $i_{2}$ be the students in (i), so that $i_{0}\vartriangleright_{j_{1}} i_{1}\vartriangleright_{j_{2}} i_{2}$ for some colleges  $j_{1},j_{2}\in M$, where we also assume that $i_{0}\ne i_{2}$. First, note that $i_{0}\ne i_{1}$ and $i_{1}\ne i_{2}$ by the proof of (i). Hence $i_{0},i_{1},i_{2}$ are mutually distinct. Lastly, to see that $j_{1}$ and $j_{2}$ are distinct, suppose by contradiction that $j_{1}=j_{2}\equiv j$. Then since $i_{0}\vartriangleright_{j} i_{1}\vartriangleright_{j} i_{2}$, we must have $i_{1}\in N^{*}_{j,0}(\mu) \cap N^{*}_{j,1}(\mu)$. However, this is not possible because $N^{*}_{j,0}(\mu) \cap N^{*}_{j,1}(\mu)=\varnothing$. Thus, we must have $j_{1}\ne j_{2}$. This establishes (ii). $\blacksquare$

\begin{lemma}\label{any_student_is_part_of_cycle}
	Let $\boldsymbol{u}\in \mathbb{U}$ and $\mu \in  \mathcal{S}(N,M,\boldsymbol{u})$. Let $i_{0},i_{1},i_{2},i_{3} \in N^{H}(\mu)$ be any four distinct students satisfying $	i_{0}\vartriangleright i_{1}\vartriangleright i_{2}\vartriangleright i_{3}$. Then either of the following two cases holds.
	
Case 1:
	\begin{align}\label{any_student_is_part_of_cycle_2}
		i_{0}\vartriangleright i_{1}\vartriangleright i_{2}\vartriangleright i_{3}\vartriangleright i_{0},
	\end{align}

Case 2: For some finite $r\geq 1$, there exist students $i_{4},...,i_{3+r}\in N^{H}(\mu)$ satisfying that
	\begin{align}\label{any_student_is_part_of_a_cycle_3}
		i_{0}\vartriangleright i_{1}\vartriangleright i_{2}\vartriangleright i_{3}\vartriangleright i_{4}\vartriangleright ...\vartriangleright i_{3+r} \vartriangleright i_{0},
	\end{align}
where $i_{0},i_{1},i_{2},i_{3},i_{4},...,i_{3+r}$ are all distinct students.
\end{lemma}
\noindent \textbf{Proof: }
	 Let $i_{0},i_{1},i_{2},i_{3} \in N^{H}(\mu)$ be any four students satisfying 
	\begin{align}\label{displacement_sequence}
	i_{0}\vartriangleright i_{1}\vartriangleright i_{2}\vartriangleright i_{3}, \text{ with }i_{0},i_{1},i_{2},i_{3}\text{ distinct.}
\end{align}
	Consider the student $i_{3}$. By Lemma \ref{every_student_displaces_and_is_displaced}, there is one and only one student, say, $i_{4}$, that satisfies $i_{3}\vartriangleright i_{4}$ for $i_{3}$. By Lemma \ref{every_student_displaces_and_is_displaced} and   (\ref{displacement_sequence}), we have $i_{4} \in N^{H}(\mu)\backslash\{i_{1},i_{2},i_3\}$.\footnote{To see that $i_4 \notin \{i_1,i_2,i_3\}$, suppose that $i_4=i_1$. Then since $i_3 \vartriangleright i_4$ and $i_0 \vartriangleright i_1$, then we must have $i_3=i_0$ because $i_1$ is displaced by one and only one student. However, this violates our assumption that $i_3$ and $i_0$ are distinct. If $i_4=i_2$, then $i_3=i_1$ (for the same reason as above), this time violating our assumption that $i_3$ and $i_1$ are distinct. Finally, by the irreflexivity of $\vartriangleright$, we cannot have $i_4=i_3$.} If $i_{4}=i_{0}$, we have (\ref{any_student_is_part_of_cycle_2}), and Case 1 is immediately satisfied. 

For the remainder of the proof, we will show that Case 2 must hold under the assumption that $i_{4}\ne i_{0}$. So suppose that $i_{4}\ne i_{0}$. By Lemma \ref{every_student_displaces_and_is_displaced} and (\ref{displacement_sequence})  (with $i_{3}\vartriangleright i_{4}$ and $i_{4}\ne i_{0}$) we must have 
		\begin{align}\label{displacement_sequence_1_all_distinct}
	i_{0}\vartriangleright i_{1}\vartriangleright i_{2}\vartriangleright i_{3}\vartriangleright i_{4}, \text{ with }i_{0},i_{1},i_{2},i_{3},i_{4}\text{ distinct.}
\end{align}
 By  Lemma \ref{every_student_displaces_and_is_displaced}, there exists one and only one student, satisfying $i_{4}\vartriangleright i_{5}$ for $i_{4}$. By Lemma \ref{every_student_displaces_and_is_displaced} and (\ref{displacement_sequence_1_all_distinct}), we have $i_{5}\in N^{H}(\mu) \backslash \{i_{1},i_{2},i_{3},i_{4}\}$. If $i_{5} = i_{0}$, then we have (\ref{any_student_is_part_of_a_cycle_3}), so that Case 2 holds with $r=1$.  If $i_{5}\ne i_{0}$, then by Lemma \ref{every_student_displaces_and_is_displaced} and (\ref{displacement_sequence_1_all_distinct}) we have that 
\begin{align}\label{displacement_sequence_2_all_distinct}
	i_{0}\vartriangleright i_{1}\vartriangleright i_{2}\vartriangleright i_{3}\vartriangleright i_{4}\vartriangleright i_{5}, \text{ with }i_{0},i_{1},i_{2},i_{3},i_{4},i_{5} \text{ distinct. } 
\end{align}
 We then consider whether or not $i_{6}$, the unique student satisfying $i_{5}\vartriangleright i_{6}$ for $i_{5}$ is equal to $i_{0}$. If $i_{6} = i_{0}$, then we have (\ref{any_student_is_part_of_a_cycle_3}) with all distinct students, and we have Case 2 with $r=2$. Otherwise, we go on to the next student. Since there are only finitely many students in the set $N$, it follows that sequence of students $i_{0},i_{1},i_{2},i_{3},i_{4},...$ (who are all distinct) must eventually terminate with $i_{3+r'} = i_{0}$ for some $r'\geq 1$. Hence, we conclude that when (\ref{displacement_sequence}) holds with the student $i_{4}$ displaced by $i_{3}$ satisfying $i_{4}\ne i_{0}$, then Case 2 must hold. $\blacksquare$

\begin{lemma}\label{lemm:bounded_between_two_stable_preliminary} Let $\boldsymbol{u} = (\boldsymbol{v},\boldsymbol{w}) \in\mathbb{U}$ be the preference profiles with $h(\boldsymbol{w}) =k$ for some $k =0,...,n$. Let $\mu \in  \mathcal{S}(N,M,\boldsymbol{u})$. Then for each $j\in M$,
	\begin{align}\label{eq:bounded_between_two_stable_preliminary_equation}
		|N^{*}_{j,1}(\mu)|\leq 4(k\vee 1)\text{  and }  	|N^{*}_{j,0}(\mu)|\leq 4(k \vee 1).
	\end{align}
\end{lemma}

\noindent \textbf{Proof: } Let $k^{*}\equiv k\vee 1$. Suppose by contradiction that $|N^{*}_{j,0}(\mu)|= t$ and $|N^{*}_{j,1}(\mu)|= t$ for some integer $t>4k^{*}$. Let us enumerate
\begin{align*}
	N^{*}_{j,0}(\mu) = \{b_{1},b_{2},...,b_{t-1},b_{t}\},
\end{align*}
so that a student with a lower index is worse according to college $j$'s preference. Furthermore, let us enumerate 
\begin{align*}
	N^{*}_{j,1}(\mu) = \{a_{2},...,a_{t-1},a_{t},a_{t+1}\},
\end{align*}
so that for each $s=1,...,t$, $a_{s+1}$ denotes the unique student  satisfying $a_{s+1}\vartriangleright_{j}b_{s}$ for student $b_{s}$ (which exists by Lemma  \ref{every_student_displaces_and_is_displaced}). Thus $a_{s+1}$ represents the student that displaces $b_{s}$ from college $j$. By Corollary \ref{other_stable_matching_always_Better} and the above ordering convention, we immediately obtain the following fact. \medskip 

\noindent \textbf{Fact I: } For each $s=1,...,4k^{*}+1$, $a_{s+1}\succ_{j}b_{4k^{*}+1}\succ_{j}b_{4k^{*}}\succ_{j}...\succ_{j}b_{1}$. \medskip 

Next, for each $s=1,...,k^{*}$,  let $c_{s}^{b}$ denote the unique student satisfying $b_{s}\vartriangleright c_{s}^{b} $  and let $c_{s}^{a}$ denote the unique student satisfying $c_{s}^{a}\vartriangleright a_{s+1}$. Thus, for any $s=1,...,k^{*}$,
\begin{align}\label{displacement_ordering_s}
	c_{s}^{a}\vartriangleright a_{s+1}\vartriangleright b_{s}\vartriangleright 	c_{s}^{b}.
\end{align}
Since $a_{s+1}$ displaces $b_{s}$ from college $j$ we have by Lemma \ref{every_student_displaces_and_is_displaced}(ii) that $c_{s}^{b}$ represents a student that is unmatched by some college (not $j$)  which then matches $b_{s}$, and $c_{s}^{a}$ represents a student matched to some college (not $j$) which then unmatches $a_{s+1}$ (before $a_{s+1}$ matches with college $j$).\medskip 

\noindent \textbf{Fact II: } For each $s=1,...,k^{*}$,

(i) $c_{s}^b$ is not ranked higher than $b_{s}$ by more than $k^{*}$ positions under $\succ_{j}$, and

(ii) $c_{s}^a$ is not ranked lower than $a_{s+1}$ by more than $k^{*}$ positions under $\succ_{j}$.\medskip

\noindent\textbf{Proof:} The result follows as in the proof of Fact 2 in Lemma \ref{lem:change_at_college}, with Corollary \ref{other_stable_matching_always_Better} and $\vartriangleright$ taking the place of Lemma \ref{DA_as_monotone_operator} and iterations of $T$. $\square$ \medskip 

\noindent \textbf{Fact III:} For each $s=1,...,k^{*}$, 

(i) $c_{s}^{b}\ne c_{s}^{a}$, and 

(ii) $c_{s}^{a}b_{3k^{*}+1}b_{3k^{*}}...b_{k^{*}+s+1}c_{s}^{b}$.\medskip

\noindent \textbf{Proof: } We have the result by following the argument used in the proof of Fact 3 from Lemma \ref{lem:change_at_college}, taking Facts I,  II and $\vartriangleright$ in place of Facts 1, 2 and iterations of $T$. $\square$ \medskip

In light of Fact III, it follows that for each $s=1,...,k^{*}$, the students $c_{s}^{b}$, $a_{s+1}$, $b_{s}$, $c_{s}^{a}$ are mutually distinct.\footnote{We have $c_{s}^{a}\ne a_{s+1}$, $a_{s+1}\ne b_{s}$, and  $b_{s}\ne c_{s}^{b}$, by irreflexivity of $\vartriangleright$. By Fact III(i),  we have $c_{s}^{b}\ne c_{s}^{a}$. By Fact III(ii) and Fact I, we have $c_{s}^{a}\ne b_{s}$, and $a_{s+1}\ne c_{s}^{b}$.} Thus, by  Lemma \ref{any_student_is_part_of_cycle} and (\ref{displacement_ordering_s}), we have the following displacement ordering over students for any $s=1,...,k^{*}$:
\begin{align}
	\label{T_operator_analogue_3}
	c_{s}^{a}&\vartriangleright a_{s+1}\vartriangleright b_{s}\vartriangleright 	c_{s}^{b} \vartriangleright c_s^a,
\end{align}
or
\begin{align}\label{T_operator_analogue_2}
	c_{s}^{a}&\vartriangleright a_{s+1}\vartriangleright b_{s}\vartriangleright 	c_{s}^{b}\vartriangleright i_{s,1}'\vartriangleright...\vartriangleright i_{s,r_{s}}' \vartriangleright c_{s}^{a},
\end{align}
for some $r_{s}\geq 1$, with all students distinct.

As in the proof of Lemma \ref{lem:change_at_college}, we use a recursive argument.  Let $c_{s,0}:=c_{s}^{b}$ for each $s=1,...,k^{*}$. By Fact III, we have 
\begin{align}\label{step1_recursion}
	\{c_{1}^{a},...,c_{k^{*}}^{a}\}b_{3k^{*}+1}...b_{2k^{*}+1}\{c_{1,0},...,c_{k^{*},0}\}.
\end{align}
Let $c_{s_1,0}$ be the worst student among $c_{1,0},...,c_{k^{*},0}$ according to $\succ_{j}$.  Let $c_{s_1,1}$ denote the unique student displaced by $c_{s_1,0}$, i.e., satisfying that $c_{s_1,0}\vartriangleright c_{s_1,1}$. Note that we must have $b_{2k^{*}+1}\succ_{j}c_{s_1,1}$ in (\ref{step1_recursion}), as otherwise $c_{s_1,1} \succ_j b_{2k^{*}+1}$ means that the rank of $c_{s_1,1}$ is higher than that of $c_{s_1,0}$ by more than $k^*$ according to $\succ_j$, as in the proof of Fact II, so that $c_{s_1,1} \succ_{j'} c_{s_1,0}$ and yet $c_{s_1,0}\vartriangleright_{j'} c_{s_1,1}$ for some college $j'$, which violates Corollary \ref{other_stable_matching_always_Better}.

Next, we let $c_{s,1}:=c_{s,0}$ for all $s\in \{1,...,k^{*}\}\backslash\{ s_1\}$. By (\ref{step1_recursion}) and $b_{2k^{*}+1}\succ_{j}c_{s_1,1}$, we must have
\begin{align}\label{step2_recursion}
	\{c_{1}^{a},...,c_{k^{*}}^{a}\}b_{3k^{*}+1}...b_{2k^{*}+1}\{c_{1,1},....,c_{k^{*},1}\}.
\end{align}
Let $s_2$ denote the index  $s\in \{ 1,...,k^{*}\}$ satisfying that $c_{s,1}$ is the worst student among $c_{1,1},...,c_{1,k^{*}}$ according to $\succ_{j}$ and $c_{s_2,2}$ denote the unique student  satisfying $c_{s_2,1}\vartriangleright c_{s_2,2}$. By Corollary \ref{other_stable_matching_always_Better}, this student must again satisfy $b_{2k^{*}+1}\succ_{j}c_{s_2,2}$.   Now we denote $c_{s,2}:=c_{s,1}$ for all $s\in \{1,...,k^{*}\}\backslash\{s_2\}$.  The next displaced student in the sequence must again be below $b_{2k^{*}+1}$ in the preference of college $\succ_j$. By continuing in this fashion, we find that we can never displace any student above $b_{2k^{*}+1}$. In particular, we can never displace $c_{s}^{a}$ after $c_{s'}^{b}$ for any $s,s'\in \backslash\{1,...,k^{*}\}$. This violates both (\ref{T_operator_analogue_3}) and (\ref{T_operator_analogue_2}). Thus, we cannot have $|N_{j,0}^*(\mu)|= t$ and $|N_{j,1}^*(\mu)|= t$ for some integer $t>4k^{*}$. $\blacksquare$

\begin{lemma}\label{lemm:bounded_between_two_stable} Let $\boldsymbol{u} = (\boldsymbol{v},\boldsymbol{w}) \in\mathbb{U}$ be any preference profile with $h(\boldsymbol{w}) =k$ for some $k =0,...,n$. Let $\mu \in  \mathcal{S}(N,M,\boldsymbol{u})$. Then for any $j\in M'$:
	\begin{align}
		\label{eq:5pt1ii2-1-1-3}
		\left|\left\{i\in N:1\{\mu(i) = j\} \ne  1\{\mu^{*}(i,\boldsymbol{u}) = j\}\right\} \right|\leq 8(k\vee1).
	\end{align}
\end{lemma}

\noindent \textbf{Proof: } For any $j\in M'$ we have
\begin{align}
	\label{eq:college_j_bound2}
	& \sum_{i\in N} \left|1\{\mu(i)=j\}-1\{\mu^{*}(i,\boldsymbol{u})=j\} \right| 
	\nonumber \\
	& \le \sum_{i\in N}1\{\mu(i)=j\}1\{\mu^{*}(i,\boldsymbol{u})\neq j\} +\sum_{i'\in N}1\{\mu(i)\neq j\}1\{\mu^{*}(i,\boldsymbol{u})=j\}.\nonumber
\end{align}
For $j\in M$, the sum on the right hand side is bounded by $8(k \vee 1)$ by Lemma \ref{lemm:bounded_between_two_stable_preliminary}, and for $j=0$, it is bounded by zero by Theorem 5.12 of \cite{Roth/Sotomayor:90:TwoSidedMatching}. $\blacksquare $

\subsubsection{Proof of Lemma \ref{lemm:bounded_difference_condition}}

Let $\mu^{*}$ be the SOSM mechanism and let $\mu(\cdot; \alpha(\tilde s, \tilde z))$ and $\mu(\cdot; \alpha(\tilde s', \tilde z))$ be stable matchings as given in the lemma. Fix any $k=0,1,...,n$ and choose $j \in M$. The preference profiles $\boldsymbol{u}$ and $\boldsymbol{u}'$ are generated from $(\tilde s, \tilde z)$ and $(\tilde s', \tilde z)$ such that $h(\tilde s,\tilde z) = h(\tilde s',\tilde z) = k$. By the triangle inequality,
\begin{align*}
	&\left|\left\{i\in N:1\{\mu(i;\alpha(\tilde s, \tilde z)) = j\} \ne  1\{\mu(i;\alpha(\tilde s', \tilde z)) = j\}\right\} \right| \\
	&\quad  \leq \left|\left\{i\in N:1\{\mu(i;\alpha(\tilde s, \tilde z)) = j\} \ne  1\{\mu^{*}(i,\boldsymbol{u}) = j\}\right\} \right|\\
	&\quad \quad +\left|\left\{i\in N:1\{\mu^{*}(i,\boldsymbol{u}) = j\} \ne  1\{\mu^{*}(i,\boldsymbol{u}') = j\}\right\} \right|\\
	&\quad \quad  +\left|\left\{i\in N:1\{\mu^{*}(i,\boldsymbol{u}') = j\} \ne  1\{\mu(i;\alpha(\tilde s', \tilde z)) = j\}\right\} \right|.
\end{align*}
By Lemma \ref{lemm:bounded_between_two_stable}, the first and the third terms on the right-hand side of the above display are each bounded by $8(k\vee1)$. By Lemma \ref{lemm:BDC sosm}, the second term is bounded by $16(k\vee1)+1$. $\blacksquare$

\putbib[LLN]
\end{bibunit}

\end{document}